\documentclass[12pt]{article}

\usepackage{cite}
\usepackage{times}
\usepackage{units}
\usepackage{graphicx}
\usepackage{bm}
\usepackage{amsmath}
\usepackage[pdffitwindow=false,bookmarks=true,pdfauthor={RT},colorlinks=true,citecolor=blue,linkcolor=blue]{hyperref}
\usepackage{hypcap}
\usepackage{multirow}
\usepackage{bigdelim}
\usepackage{array}
\usepackage{tabularx}
\usepackage{ulem}
\usepackage{color}
\usepackage{xspace}
\usepackage{xcolor}
\usepackage{amsmath}
\usepackage{amssymb}
\usepackage{mathrsfs}

\newcommand*{\Vneg}{\ensuremath{V^-}\xspace}

\newcommand*{\Vpos}{\ensuremath{V^+}\xspace}

\newcommand*{\Vpm}{\ensuremath{V^{\pm}}\xspace}

\newcommand*{\fm}{\ensuremath{f_\text{m}}\xspace}

\newcommand*{\df}{\ensuremath{\Delta f}\xspace}

\newcommand*{\Vb}{\ensuremath{V_\text{b}}\xspace}

\newcommand*{\rpar}{\ensuremath{\boldsymbol{\text{r}}_{||}}\xspace}

\newcommand*{\rparprime}{\ensuremath{\boldsymbol{\text{r}}_{||}'}\xspace}

\newcommand*{\rpardprime}{\ensuremath{\boldsymbol{\text{r}}_{||}''}\xspace}

\newcommand*{\rparprimeZero}{\ensuremath{\boldsymbol{\text{r}}_{||0}'}\xspace}

\newcommand*{\rparprimePhi}{\ensuremath{\boldsymbol{\text{r}}_{||\phi}'}\xspace}

\newcommand*{\rr}{\ensuremath{\boldsymbol{\text{r}}}\xspace}

\newcommand*{\RR}{\ensuremath{\boldsymbol{\text{R}}}\xspace}

\newcommand*{\rrprime}{\ensuremath{\boldsymbol{\text{r}}'}\xspace}

\newcommand*{\rparidprime}{\ensuremath{\boldsymbol{\text{r}}_{||i}''}\xspace}

\newcommand*{\PhiQD}{\ensuremath{\Phi_\text{QD}}\xspace}

\newcommand*{\PhiTip}{\ensuremath{\Phi_\text{T}}\xspace}

\newcommand*{\Phis}{\ensuremath{\Phi_\text{s}}\xspace}

\newcommand*{\Phit}{\ensuremath{\Phi_\text{t}}\xspace}

\newcommand*{\Phistar}{\ensuremath{\Phi^*}\xspace}

\newcommand*{\GS}{\ensuremath{G_\mathscr{S}}\xspace}

\newcommand*{\GSD}{\ensuremath{G^\mathrm{D}_\mathscr{S}}\xspace}

\newcommand*{\GTD}{\ensuremath{G^\mathrm{D}_\mathscr{T}}\xspace}

\newcommand*{\GSN}{\ensuremath{G^\mathrm{N}_\mathscr{S}}\xspace}

\newcommand*{\FS}{\ensuremath{F_\mathscr{S}}\xspace}

\newcommand*{\FT}{\ensuremath{F_\mathscr{T}}\xspace}

\newcommand*{\surf}{\ensuremath{\mathscr{S}}\xspace}

\newcommand*{\surfT}{\ensuremath{\mathscr{T}}\xspace}

\newcommand*{\vol}{\ensuremath{\mathscr{V}}\xspace}

\newcommand*{\gammastar}{\ensuremath{\gamma^*}\xspace}

\newcommand*{\gammaplain}{\ensuremath{\gamma}\xspace}

\newcommand*{\gammaTopo}{\ensuremath{\gamma_\text{topo}}\xspace}

\newcommand*{\gammaPP}{\ensuremath{\gamma_\text{pp}}\xspace}

\newcommand*{\gammaAx}{\ensuremath{\gamma_\text{axial}}\xspace}

\newcommand*{\zt}{\ensuremath{z_\text{t}}\xspace}

\newcommand*{\zz}{\ensuremath{z}\xspace}

\newcommand*{\Vstar}{\ensuremath{V^*}\xspace}

\newcommand*{\Pperp}{\ensuremath{P_\perp}\xspace}

\newcommand*{\piPerp}{\ensuremath{\Pi_\perp}\xspace}

\newcommand*{\teff}{\ensuremath{t_{\text{d}}}\xspace}

\newcommand*{\tBar}{\ensuremath{\bar{t}}\xspace}

\newcommand*{\alpharel}{\ensuremath{\alpha_{\text{rel}}}\xspace}

\newcommand\numberthis{\addtocounter{equation}{1}\tag{\theequation}}

\newcommand*{\hatGamma}{\ensuremath{\hat{\Gamma}}\xspace}

\newcommand*\colvec[3][]{
	\begin{pmatrix}\ifx\relax#1\relax\else#1\\\fi#2\\#3\end{pmatrix}
}

\title{The Theory of Scanning Quantum Dot Microscopy}
\author{
	Christian~Wagner$^{1,2\ast}$ and
	F.~Stefan~Tautz$^{1,2,3}$
	\\
	\\
	\normalsize{$^{1}$Peter Gr\"unberg Institut (PGI-3), Forschungszentrum J\"ulich, 52425 J\"ulich, Germany}\\
	\normalsize{$^{2}$J\"ulich Aachen Research Alliance (JARA)-Fundamentals of Future Information Technology,}\\
	\normalsize{52425 J\"ulich, Germany}\\
	\normalsize{$^{3}$Experimentalphysik IV A, RWTH Aachen University, Otto-Blumenthal-Stra\ss{}e,}\\
	\normalsize{52074 Aachen, Germany}\\
	\normalsize{$^\ast$To whom correspondence should be addressed; E-mail:  c.wagner@fz-juelich.de.}
}
\date{ }

\begin{document}
\maketitle

\sffamily
\small
\noindent
\textbf{Electrostatic forces are among the most common interactions in nature and omnipresent at the nanoscale. Scanning probe methods represent a formidable approach to study these interactions locally. The lateral resolution of such images is, however, often limited as they are based on measuring the force (gradient) due to the entire tip interacting with the entire surface. Recently, we developed scanning quantum dot microscopy (SQDM), a new technique for the imaging and quantification of surface potentials which is based on the gating of a nanometer-size tip-attached quantum dot by the local surface potential and the detection of charge state changes via non-contact atomic force microscopy. Here, we present a rigorous formalism in the framework of which SQDM can be understood and interpreted quantitatively. In particular, we present a general theory of SQDM based on the classical boundary value problem of electrostatics, which is applicable to the full range of sample properties (conductive \textit{vs} insulating, nanostructured \textit{vs} homogeneously covered). We elaborate the general theory into a formalism suited for the quantitative analysis of images of nanostructured but predominantly flat and conductive samples.}

\normalsize 
\bigskip
\normalfont

\newpage
\tableofcontents
\baselineskip24pt
\newpage

\section{Introduction to SQDM}

\subsection{Electrostatic potentials at the nanoscale and their measurement}

Electrostatic forces are among the most common interactions in nature. While they appear in the macroscopic world only when excess charges are present, they are omnipresent at the nanoscale because the constituents of matter, electrons and nuclei, carry discrete charges. These fields significantly influence microelectromechanical systems \cite{Dong2007} as well as nanoelectronic components, for example as built-in interface potentials \cite{Campbell1997,Matyba2009} or unwanted background charges \cite{Deng2010}. 

A common method to image and quantify electric potentials on surfaces is Kelvin probe force microscopy (KPFM) which measures the potential difference between a surface and a probe \cite{Nonnenmacher1991,Kikukawa1995,Jacobs1998,Kitamura1998,Glatzel2003,Zerweck2005,Gross2009a,Sadewasser2009,Baier2012,Sadewasser2012,Schuler2014,Albrecht2015,Rahe2016,Schulz2018}. The measurement principle of KPFM is derived from the classical Kelvin probe insofar as the electrostatic interaction between two objects, tip and surface, is minimized by application of a compensating bias. The lateral resolution of KPFM arises from the distance dependence of the electrostatic force \cite{Jacobs1998,Zerweck2005,Cohen2013}: The forces at the closest distance between sample and probe, i.e., close to the tip apex, outweigh the forces between the probe and more remote surface areas. Therefore, the potential difference in the vicinity of the tip apex is measured when nulling the overall electrostatic force. Consequently, surface potential variations can be imaged when scanning the probe over the surface. 

However, the lateral resolution of KPFM is intrinsically limited by this working principle in which the measurement signal is determined by the entire tip interacting with the entire surface. Different tip shapes have been explored but even ultimately sharp tips are not optimal because a shrinking (relevant) interaction area likewise reduces the signal-to-noise ratio \cite{Jacobs1998}. To obtain maximal resolution, the tip-surface distance has to be decreased to a point where chemical forces between tip and sample start acting \cite{Sadewasser2009}. A rather recent development in this direction is the passivation of a standard metal probe with a single CO molecule \cite{Schuler2014,Albrecht2015}. However, at these small distances additional effects besides the contact potential difference appear and hamper the quantitative interpretation of the data. Overcoming these limitations is an active field of research \cite{Sadewasser2009,Albrecht2015}. 

Recently we developed scanning quantum dot microscopy (SQDM), a new technique for the imaging and quantification of surface potentials \cite{Wagner2015,Green2016,Wagner2019}. SQDM is based on the gating of a nanometer-size tip-attached quantum dot (QD) by the local surface potential. The technique detects changes in the charge state of the QD via non-contact atomic force microscopy (NC-AFM) and (like KPFM) applies and adjusts a compensation voltage \Vb across the tip-surface junction to maintains the gating condition where the QD charge state changes. Since local surface potential variations contribute to the gating, their effect has to be compensated by the respective adjustments of \Vb. The introduction of a dedicated controller for SQDM imaging \cite{Maiworm2018} reduced the effort of SQDM and enabled the imaging of large surface areas \cite{Wagner2019}. Finally, an efficient image deconvolution technique enabled the interpretation of SQDM images in terms of electric potentials in the sample surface plane \cite{Wagner2019}.

Here, we derive the rigorous formalism in the framework of which SQDM can be understood and interpreted quantitatively. In particular, we present a general theory of SQDM based on the classical boundary value problem of electrostatics, which is applicable to the full range of sample properties (conductive \textit{vs} insulating, nanostructured \textit{vs} homogeneously covered). We elaborate the general theory into a formalism suited for the quantitative analysis of images of nanostructured but predominantly flat and conductive samples. This formalism is applicable to, e.g., atomically ordered metal surfaces with point or line defects, including adsorbates such as isolated molecules, molecular films or 2D materials. 

\subsection{The primary and secondary measurands of SQDM}

\label{primaryMeasurands}

\begin{figure}[!ht]
	\centering
	\includegraphics[width=12cm]{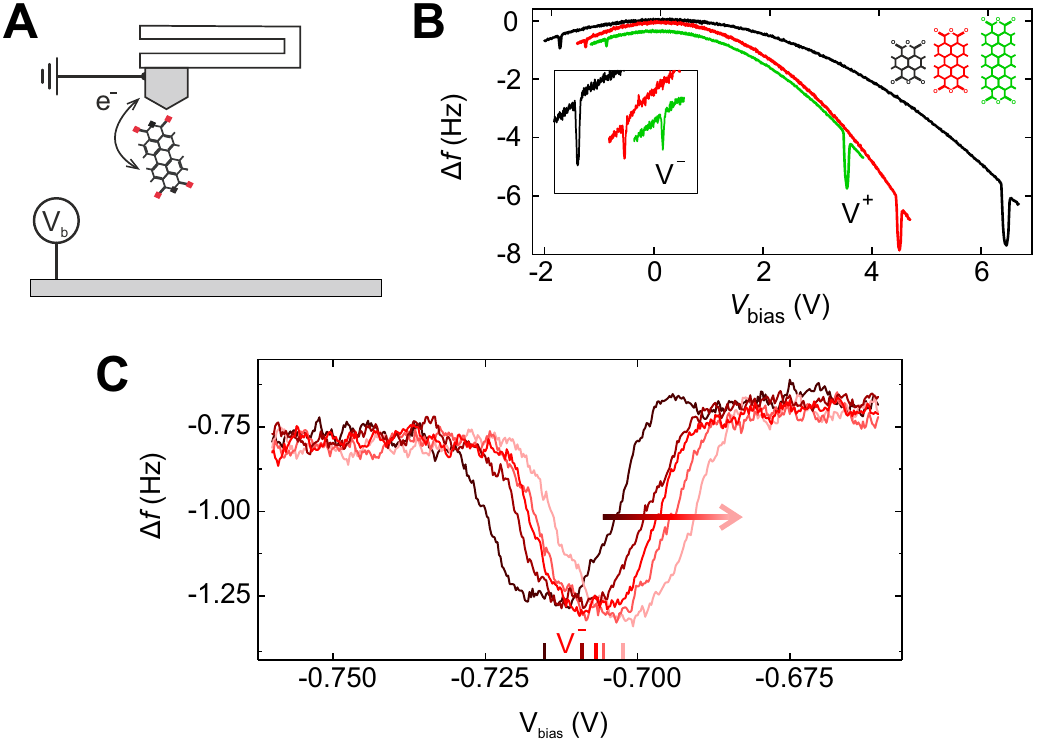}
	\sffamily
	\caption{\footnotesize  \textbf{The principle of SQDM} \textbf{(A)} A single molecule (here PTCDA) is attached to the SPM tip via a stable directional bond such that the molecule remains vertical and its electronic states barely overlap with the tip. Then, the molecule acts as a QD and can be gated and charged by individual electrons by applying a bias to the sample. \textbf{(B)} The charging events can be detected in $\Delta f(\Vb)$ curves as sharp dips superimposed on the usual Kelvin parabola. Shown are $\df(\Vb)$ curves for three different sensor QD molecules recorded with a qPlus-type NC-AFM/STM. \textbf{(C)} Series of $\df(\Vb)$ spectra around \Vneg for a decreasing surface potential beneath the tip+QD sensor during scanning. As a result \Vneg shifts to more positive values such that the sum of surface potential and applied potential \Vb remains (approximately) constant \cite{Wagner2015}.\normalsize}
	\normalfont 
	\label{Fig_SQDM_Principle}
\end{figure}

Before presenting an extensive and rigorous discussion of the theoretical foundation of SQDM, we introduce the basic principles and quantities relevant for SQDM. The working principle of SQDM is illustrated in Fig.~\ref{Fig_SQDM_Principle}. SQDM requires a QD firmly attached to the tip apex of a non-contact atomic force microscope (NC-AFM) (Fig.~\ref{Fig_SQDM_Principle}A). The levels of the QD are electronically decoupled from tip, and at least one of them is positioned near the Fermi energy of the tip $E_{\text{F}}$. Then, the charge state of that level can be controlled by gating via the sample bias \Vb. Changes in the charge state of the QD can be detected as abrupt steps in the tip-sample force and thus as dips in the measured frequency shift signal \df (Fig.~\ref{Fig_SQDM_Principle}B). 

One experimental realization of SQDM \cite{Wagner2015} is the case of a single molecule as the QD with an initially singly occupied level which can either be emptied or doubly occupied by gating. The bias voltages (center of each dip) at which one electron is added (\Vpos) or removed (\Vneg) are specific properties of the tip + QD system (referred to as sensor below) and generally increase with decreasing QD size (Fig.~\ref{Fig_SQDM_Principle}B) \cite{Green2016,Temirov2018}. 

For a given sensor the change in charge state happens at a specific well-defined potential $\PhiQD^\pm$ of the QD measured with respect to the grounded tip. If the tip is scanned across the surface, the laterally varying surface potential leads to variations in \PhiQD. This shifts the charging voltages \Vpm (Fig.~\ref{Fig_SQDM_Principle}C) \cite{Wagner2015}. Moreover, lateral variations of the gating efficiency which stem from the surface topography will affect the relation between \PhiQD and \Vb and thus also cause $\Vpm$ to change. \Vpos and \Vneg are the primary measurands of SQDM. 

Next, we derive expressions which connect these primary measurands with properties of the sample surface that are of interest. The general expression for the potential at the QD is
\begin{equation}
\PhiQD = \Phistar+\alpha\Vb+\PhiTip,
\label{EqPhiQD1}
\end{equation}
where \Phistar denotes the contribution from the electrostatic  potential distribution on the sample surface, $\alpha\Vb$ the contribution from the applied bias and \PhiTip the contribution from local potentials on the tip surface. The latter is constant for a given sensor. The gating efficiency is defined as 
\begin{equation}
\alpha \equiv \frac{d\PhiQD}{d\Vb}.
\label{EqAlpha1}
\end{equation}

Equation~\ref{EqPhiQD1} will be rigorously derived in Section.~\ref{sec:Dirichlet_SQDM} (Eq.~\ref{PhiGF3}).

Similar to KPFM, SQDM also measures \textit{changes} in surface potential and no absolute surface potentials (work functions). Hence, one has to select a reference position $\rr_0$ of the sensor above a region of the sample surface in which the surface potential is constant. At this reference position we define the surface potential as zero which likewise defines $\Phistar(\rr_0) \equiv 0$. We denote the primary measurands at $\rr_0$ as $\Vpm_0 \equiv \Vpm(\rr_0)$, and the gating efficiency at $\rr_0$ as $\alpha_0 \equiv \alpha(\rr_0)$. It is moreover helpful to define a relative gating efficiency as
\begin{equation}
\alpharel \equiv \frac{\alpha}{\alpha_0}.
\label{EqAlphaRel1}
\end{equation}
With these definitions we now equate two versions of Eq.~\ref{EqPhiQD1}, one at $\rr_0$ and one at any other sensor position \rr. Moreover, we do so for the positive and negative charging events, which yields
\begin{equation}
\alpha_0 \Vpos_0 = \alpha(\rr) \Vpos(\rr) + \Phistar(\rr)
\label{EqEquatePos}
\end{equation}
\begin{equation}
\alpha_0 \Vneg_0 = \alpha(\rr) \Vneg(\rr) + \Phistar(\rr),
\label{EqEquateNeg}
\end{equation}
respectively. 

We can derive the following two expressions via a series of straightforward algebraic transformation from Eqs.~\ref{EqAlphaRel1}, \ref{EqEquatePos} and \ref{EqEquateNeg}, leaving out the variable \rr for simplicity

\begin{equation}
\alpharel = \frac{\Vpos_0-\Vneg_0}{\Vpos-\Vneg}
\label{EqAlphaRel2}
\end{equation}

\begin{equation}
\Phistar = \alpha\left( \frac{\Vneg_0}{\alpharel}-\Vneg\right) = \alpha\left( \frac{\Vpos_0}{\alpharel}-\Vpos\right)
\label{EqPhistar1}
\end{equation}

With these equations we have quantified two main parameters of Eq.~\ref{EqPhiQD1} in terms of measured \Vpm and $\Vpm_0$. It is intuitively clear that the two quantities \alpharel and \Phistar are related to the surface potential and topography. A substantial part of this paper is dedicated to the derivation of this explicit relation. Looking at Eq.~\ref{EqPhistar1}, we observe that it only specifies $\Vstar \equiv \frac{\Phistar}{\alpha}$. The reason is that \Phistar is measured by compensation and the applied compensation voltage acts on \PhiQD via $\alpha$.

We will discuss in the following sections how the secondary measurands of SQDM, \Vstar and \alpharel, can be derived in a rigorous theoretical framework and how they are related to actually relevant surface properties like the local surface potential $\Phis(\rr)$.

\subsection{The imaging formalism of SQDM}
\label{SecImagingFormalism}

SQDM is a scanning probe microscopy. The probe, in our case the sensor consisting of tip and quantum dot, is scanned at a certain height $z$ across the sample surface, with the aim to learn something about the sample surface. Conceptually, it is opportune to distinguish between the \textit{imaging plane} at $z$ and the \textit{object surface}, which is the sample surface itself. Note that since the imaging plane is a plane, while the object surface is an arbitrary surface, the vertical distance between the two is not the same at every lateral position of the probe. 

Abstractly, the problem of SQDM can be formulated as follows: Knowing the imaged quantity $I$ in the imaging plane from the measurement, we want to learn about the object quantity $O$ in the object surface. Specifically, the following questions arise: What are the image and object quantities, what is the mathematical relation between $I$ and $O$, and how can $O$ be determined once $I$ has been measured? We thus have to identify $I$ and $O$ and their relation 
\begin{equation}
O(\rrprime)=\hat{O}[I(\rr)],
\label{I_O}
\end{equation}
where $\hat{O}$ is a functional and $O(\rrprime)$ and $I(\rr)$ are functions defined in the object surface and imaging planes, respectively. In Chapter~\ref{sec:Dirichlet_SQDM} we prove that $\PhiQD(\rr)$, the electrostatic potential at the position of the QD, can be written as 
\begin{equation}
\PhiQD(\rr) = \iint\limits_{\text{surface}}\gammaplain(\rr,\rrprime)[\Phis(\rrprime)+\Vb]d^2\rrprime + \PhiTip.
\label{EqPhiQD2}
\end{equation}
$\Phis(\rrprime)$ is the electrostatic potential in the object surface at \rrprime, \Vb is the bias voltage applied to the sample (object), \PhiTip is the constant contribution from the potential distribution on the tip surface, and \gammaplain is an integral kernel to be determined below. The integral is carried out over the entire object surface. Eq.~\ref{EqPhiQD2} corresponds to an electrostatic boundary value problem $\hat{B}$, which can abstractly be written as 
\begin{equation}
\PhiQD(\rr)=\hat{B}[O(\rrprime)-I]
\label{boundary_problem_conceptual}
\end{equation}
with the object function $O(\rrprime)=\Phis(\rrprime)$ and the parameter $I=\Vb$. Within the measurement protocol of SQDM, $\hat{B}$ is in fact executed, in the sense that the total surface potential $O(\rrprime)-I$ is adjusted by the parameter $I$ in order to realize a specific constant value of $\PhiQD$ (corresponding to the charging potential $\PhiQD^\pm$) for all \rr. From the condition 
\begin{equation}
\hat{B}[O(\rrprime)-I]=\mathrm{const.}
\label{boundary_problem_conceptual_constant}
\end{equation}
the image function $I(\rr)$ is generated in the experiment. Mathematically, $I(\rr)=\Vstar(\rr)$ is given by
\begin{equation}
\Vstar(\rr) = \iint\limits_{\text{surface}}\gammastar(\rr,\rrprime)\Phis(\rrprime)d^2\rrprime.
\label{equation3}
\end{equation}
Eq.~\ref{equation3}, which can be derived straightforwardly from Eq.~\ref{EqPhiQD2} as we will show later (Eq.~\ref{EqVstar2}), has the form
\begin{equation}
I(\rr)=\hat{O}^{-1}[O(\rrprime)].
\label{inv_I_O}
\end{equation}
Hence, the inverse of Eq.~\ref{equation3} is the sought-after relation Eq.~\ref{I_O} and thus provides a solution to the problem of obtaining the object function from the measured image function of SQDM. In other words, SQDM corresponds to the inversion of a boundary value problem. For a certain class of samples this inversion corresponds to a deconvolution with a kernel \gammastar.

The situation is different for the secondary measurand \alpharel. With respect to \alpharel as image function $I(\rr)$, Eq.~\ref{inv_I_O} reads (Eq.~\ref{integralalpha})
\begin{equation}
\alpharel(\rr) = \frac{1}{\alpha_0} \iint\limits_\text{surface} \gammaplain(\rr,\rrprime)d^2\rrprime
\label{integralalpha1}.
\end{equation}
The corresponding object function $O(\rrprime)$ turns out to be the surface topography which determines the function $\gammaplain(\rr,\rrprime)$. However, because the relation between topography and \gammaplain is complex, a series of approximations must be made to retrieve the topography from \alpharel (Sec.~\ref{sec:SpecificSolutions}).

\newpage
\section{SQDM as a boundary value problem of electrostatics}
\label{sec:BoundaryValueProblem}

\subsection{Poisson equation and Green's function formalism}
\label{sec:Poisson}

\begin{figure}[!ht]
	\centering
	\includegraphics[width=6.5cm]{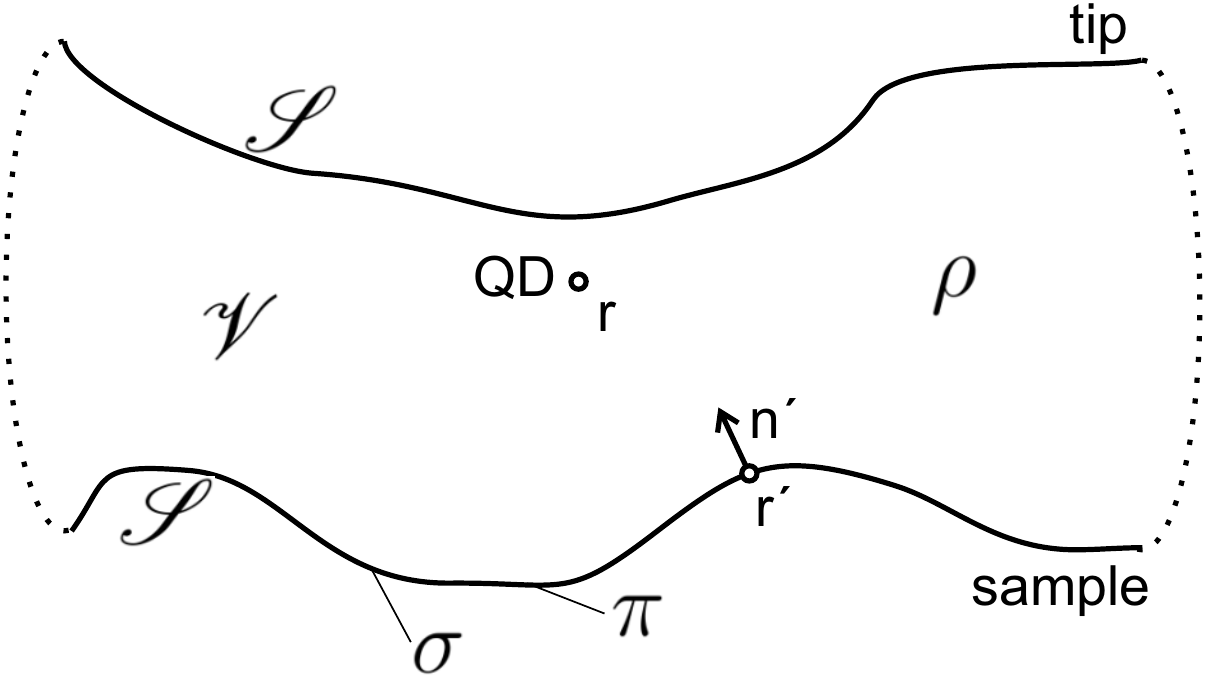}
	\sffamily
	\caption{\footnotesize  \textbf{Abstraction of the tip-QD-sample junction as an electrostatic boundary value problem.} The potential inside the volume \vol is defined by a set of boundary conditions on the enclosing surface \surf. The symbols are explained in the text.\normalsize}
	\normalfont 
	\label{Fig_Cavity1}
\end{figure}

\noindent As an essential abstraction for the forthcoming discussion we assume that the QD is point-like and that the tip is conductive and electrically connected to the sample at infinity such that tip and sample surface enclose a volume \vol (Fig.~\ref{Fig_Cavity1}). Under these assumptions the SQDM setup and the imaging problem outlined above (Eq.~\ref{boundary_problem_conceptual}) can be formulated as a boundary value problem of electrostatics.

The electrostatic potential $\Phi$ always and everywhere fulfills the Poisson equation 
\begin{equation}
	\Delta' \Phi(\rrprime)= -\frac{\rho(\rrprime)}{\epsilon_0}.   
		\label{Poisson}
\end{equation}
With the help of Green's second identity 
\begin{equation}
	\iiint\limits_{\mathscr{V}} (\phi \Delta \psi - \psi \Delta \phi)dV=\oint\limits_{\mathscr{S}} \bigg[\phi\frac{\partial \psi}{\partial \boldsymbol{\text{n}}}-\psi\frac{\partial \phi}{\partial \boldsymbol{\text{n}}}\bigg] \cdot d\boldsymbol{\text{a}},
	\label{Greensidentity_phi_psi}
\end{equation}
where \surf is a closed surface and \vol the enclosed volume, setting $\psi(\rrprime)=1/|\rr-\rrprime|$ and $\phi(\rrprime)=\Phi(\rrprime)$, as well as using the mathematical identity 
\begin{equation}
	\Delta' \Big(\frac{1}{|\rr-\rrprime|}\Big)=-4\pi \delta(\rr-\rrprime),
	\label{Delta_one_over_r}
\end{equation}
Eq.~\ref{Poisson} and can be turned into an integral equation for the potential $\Phi(\rr)$, namely 
\begin{equation}
\Phi (\rr)= \iiint\limits_\mathscr{V}\frac{\rho(\rrprime)}{4\pi\epsilon_0|\rr-\rrprime|}d^3\rrprime + \frac{1}{4\pi\epsilon_0} \oint\limits_\mathscr{S}\bigg[\frac{1}{|\rr-\rrprime|}\epsilon_0\frac{\partial \Phi(\rrprime)}{\partial \boldsymbol{\text{n}}'}-\epsilon_0\Phi(\rrprime)\frac{\partial}{\partial \boldsymbol{\text{n}}'} \Big(\frac{1}{|\rr-\rrprime|}\Big)\bigg]d^2\rrprime.   
		\label{Poissonintegral}
\end{equation}
The normal derivatives are defined by
\begin{equation}
\frac{\partial}{\partial \boldsymbol{\text{n}}'}\equiv \boldsymbol{\text{n}}' \cdot \nabla',
\label{defintion_normal_derivative}
\end{equation}
where $\boldsymbol{\text{n}}'$ is the unit vector normal to \surf and pointing outward, i.e.~away from \vol. 

Although Eq.~\ref{Poissonintegral} expresses the potential $\Phi(\rr)$ at \rr in the volume \vol as the integral over the charge density $\rho(\rrprime)$ in \vol and the potential $\Phi(\rrprime)$ and its normal derivative $\frac{\partial \Phi(\rrprime)}{\partial \boldsymbol{\text{n}}'}$ on the boundary \surf of \vol (Fig.~\ref{Fig_Cavity1}), this equation does not define a valid boundary value problem for $\Phi$, because specifying arbitrary $\Phi$ \textit{and} $\frac{\partial \Phi}{\partial \boldsymbol{\text{n}}'}$ on \surf does not yield a solution to Eqs.~\ref{Poisson} or \ref{Poissonintegral}, as solutions are already uniquely defined by \textit{either} specifying  $\Phi$  (Dirichlet boundary conditions) \textit{or} $\frac{\partial \Phi}{\partial \boldsymbol{\text{n}}'}$ (Neumann boundary conditions) on \surf \cite{Jackson1999}. Specifying both independently overdetermines $\Phi$. 

Eq.~\ref{Poissonintegral} is instructive, however, because it shows that the potential $\Phi$ at point \rr is determined by three contributions (Fig.~\ref{Fig_Cavity1}): (1) the charge density $\rho$ in \vol, i.e., each point charge at $\rrprime$ that contributes to $\rho$ being the source of a Coulomb potential, plus (2) the potential due to a surface charge density $\sigma=\epsilon_0\frac{\partial \Phi}{\partial \boldsymbol{\text{n}}'}$ on \surf and (3) the potential due to a dipole moment density $\Pi=-\epsilon_0\Phi$ of a double layer on \surf. It is easy to show that the potential of a double layer of equal and opposite surface charge densities gives rise to a potential \cite{Jackson1999}
\begin{equation}
\Phi(\rr)= \frac{1}{4\pi\epsilon_0}\iint \Pi(\rrprime)\boldsymbol{\text{n}} \cdot \nabla' \Big(\frac{1}{|\rr-\rrprime|}\Big)da.
\label{doublelayerpotential}
\end{equation}
where $\Pi(\rrprime)$ is the dipole density. Equation~\ref{Poissonintegral} applied to the potential at the position of the QD thus shows that there is a relation of either charge density or dipole density and SQDM measurands. 

Valid electrostatic boundary value problems for solving Laplace or Poisson equations can be formulated by the Green's function formalism \cite{Jackson1999}. Thereby, either Dirichlet or Neumann boundary conditions can be used. In a Dirichlet problem, the potential $\Phi$ is given on the boundary \surf of volume \vol (which corresponds to specifying a surface dipole density), while in a Neumann problem the normal derivative of the potential (surface charge density) is given on \surf. Since the charge density defines an electric field, which is the \textit{gradient} of the potential, an integration constant in the form of a global potential offset is additionally required in the Neumann problem. In the following we will discuss the implications arising from the two types of boundary conditions.

The Green's function specifies the contribution of each point on the surface \surf to the potential at any point in the volume \vol. A Green's function for the Laplace or Poisson equations in the volume \vol bordered by the surface \surf is a function $\GS(\rr,\rrprime)$ that satisfies the equation  
\begin{equation}
	\Delta' \GS(\rr,\rrprime)= -\frac{e}{\epsilon_0}\delta(\rr-\rrprime) \quad \quad \text{for all} \quad \quad \rrprime \in \mathscr{V}
		\label{Laplace_delta},
\end{equation}
Comparing to Eq.~\ref{Delta_one_over_r}, one sees that $\frac{e}{4\pi\epsilon_0|\rr-\rrprime|}$, i.e. the electric potential at \rr of a point charge $+e$ at $\rrprime$, is a possible Green's function. However, the most general expression for \GS is 
\begin{equation}
	\GS(\rr,\rrprime)=\frac{e}{4\pi\epsilon_0|\rr-\rrprime|} + \FS(\rr,\rrprime)
		\label{G_one_over_R_plus_F},
\end{equation}
where \FS satisfies the Laplace equation 
\begin{equation}
\Delta' \FS(\rr,\rrprime)= 0 \quad \quad \text{for all} \quad \quad \rrprime \in \mathscr{V}.
		\label{Laplace_delta_for_F}
\end{equation}
The freedom which \FS offers can now be used to implement Dirichlet or Neumann boundary conditions as well as specify the shape of the boundary. With $\psi(\rrprime)=\GS(\rr,\rrprime)$ and $\phi(\rrprime)=\Phi(\rrprime)$, Green's second identity Eq.~\ref{Greensidentity_phi_psi} becomes 
\begin{equation}
\begin{aligned}
\iiint\limits_\mathscr{V} &\big[\Phi(\rrprime) \Delta'  \GS(\rr, \rrprime)-\GS(\rr, \rrprime)\Delta' \Phi(\rrprime)\big] d^3 \rrprime
\\
= &\oint\limits_\mathscr{S} \bigg[\Phi(\rrprime)\frac{\partial \GS(\rr, \rrprime) }{\partial \boldsymbol{\text{n}}'}-\GS(\rr, \rrprime)\frac{\partial\Phi(\rrprime)}{\partial \boldsymbol{\text{n}}'}\bigg]d^2\rrprime
		\label{Greensidentitygeneral}.
\end{aligned}		
\end{equation}
With Eq.~\ref{Laplace_delta} this becomes
\begin{equation}
\begin{aligned}
\forall\,  \rr \in \mathscr{V}: \quad \Phi(\rr)=
\iiint\limits_\mathscr{V} &\GS(\rr, \rrprime) \frac{\rho(\rrprime)}{e}d^3\rrprime \\&+ \frac{\epsilon_0}{e}\oint\limits_\mathscr{S} \bigg[\GS(\rr, \rrprime)\frac{\partial\Phi(\rrprime)}{\partial \boldsymbol{\text{n}}'}-\Phi(\rrprime)\frac{\partial \GS(\rr, \rrprime) }{\partial \boldsymbol{\text{n}}'}\bigg]d^2\rrprime.
\label{Phi}
\end{aligned}		
\end{equation}

\subsection{Dirichlet problem for a metallic surface}
\label{sec:Dirichlet}

It is possible to make the first term in the surface integral in Eq.~\ref{Phi} zero by choosing \FS such that
\begin{equation}
		\forall \,\rr \in \mathscr{V},\, \forall \,\rrprime \in \mathscr{S}: \quad \GSD(\rr,\rrprime)= 0 
			\label{boundarycondition}.
\end{equation}
This is the Dirichlet boundary condition, the superscript D stands for "Dirichlet". The physical meaning of Dirichlet boundary conditions Eq.~\ref{boundarycondition} is clear: The potential $\GSD(\rr,\rrprime)$ at \rr of a point test charge located at $\rrprime$ on the surface \surf is forced to be zero everywhere in \vol and on \surf. This is achieved by external charges outside \vol which provide an additional potential $F(\rr,\rrprime)$ at \rr such that the total potential, including the potential of the test charge, is zero in \vol and on \surf. Since the external charge distribution which makes the potential vanish must depend on the position of the test charge, $F(\rr,\rrprime)$ depends parametrically on $\rrprime$. If \surf is the surface of a metal, then $F$ can be interpreted as arising from a surface charge density located on \surf and induced by the test charge at $\rrprime$.

With Eq.~\ref{boundarycondition}, Eq.~\ref{Phi} becomes
\begin{equation}
\begin{aligned}
 \forall\,  \rr \in \mathscr{V}: \quad \Phi(\rr)=
\iiint\limits_\mathscr{V} &\GSD(\rr, \rrprime) \frac{\rho(\rrprime)}{e}d^3\rrprime - \frac{\epsilon_0}{e}\oint\limits_\mathscr{S} \Phi(\rrprime)\frac{\partial \GSD(\rr,\rrprime) }{\partial \boldsymbol{\text{n}}'}d^2\rrprime.
\label{PhiDirichlet}
\end{aligned}		
\end{equation}
If moreover no charges are located in \vol, we obtain
\begin{equation}
\begin{aligned}
 \forall\,  \rr \in \mathscr{V}: \quad \Phi(\rr)=
 - \frac{\epsilon_0}{e}\oint\limits_\mathscr{S} \Phi(\rrprime)\frac{\partial \GSD(\rr, \rrprime) }{\partial \boldsymbol{\text{n}}'}d^2\rrprime.
\label{PhiGF}
\end{aligned}		
\end{equation}
The essence of Eq.~\ref{PhiGF} is as follows: If one puts a test charge on a closed and grounded metal surface \surf, the freely movable charge of the metal will redistribute to zero the potential everywhere on the surface and in the enclosed volume \vol. This means that  $\GSD(\rr,\rrprime)=0$ for all $\rrprime \in \mathscr{S}$. If now the test charge is moved away from the surface along the surface normal into the volume \vol, an image charge in the metal is created just below the test charge. This image charge becomes part of the charge distribution outside \vol that creates the potential $F$. Evidently, the now nonzero potential $\GSD(\rr,\rrprime)$ at \rr is the potential of a point dipole at the position $\rrprime$. For this reason, the integrand in Eq.~\ref{PhiGF}, which gives the total potential at \rr of an arbitrary surface potential distribution on \surf, is given by $\frac{1}{e}\frac{\partial \GSD}{\partial \boldsymbol{\text{n}}'}\times \epsilon_0 \Phi$, which is the potential $\GSD(\rr,\rrprime)=\GSD(\rr,\rrprime\in \mathscr{S})+\boldsymbol{\text{n}}'\cdot \nabla'\GSD(\rr,\rrprime)$ of the point dipole per unit charge, times the actual size $\epsilon_0 \Phi$ of the point dipole at $\rrprime$ (see Eq.~\ref{doublelayerpotential}). Note that any surface potential can be understood as the areal dipole density of a double layer, because at a double layer the potential always has a step ($\Phi_2-\Phi_1= \frac{\Pi}{\epsilon_0}$). The minus sign in Eq.~\ref{PhiGF} follows from the fact that the derivative is taken in the direction of the normal that points away from \vol, while we have moved the test charge at $\rrprime$ into \vol.   

\begin{figure}[!ht]
	\centering
	\includegraphics[width=8cm]{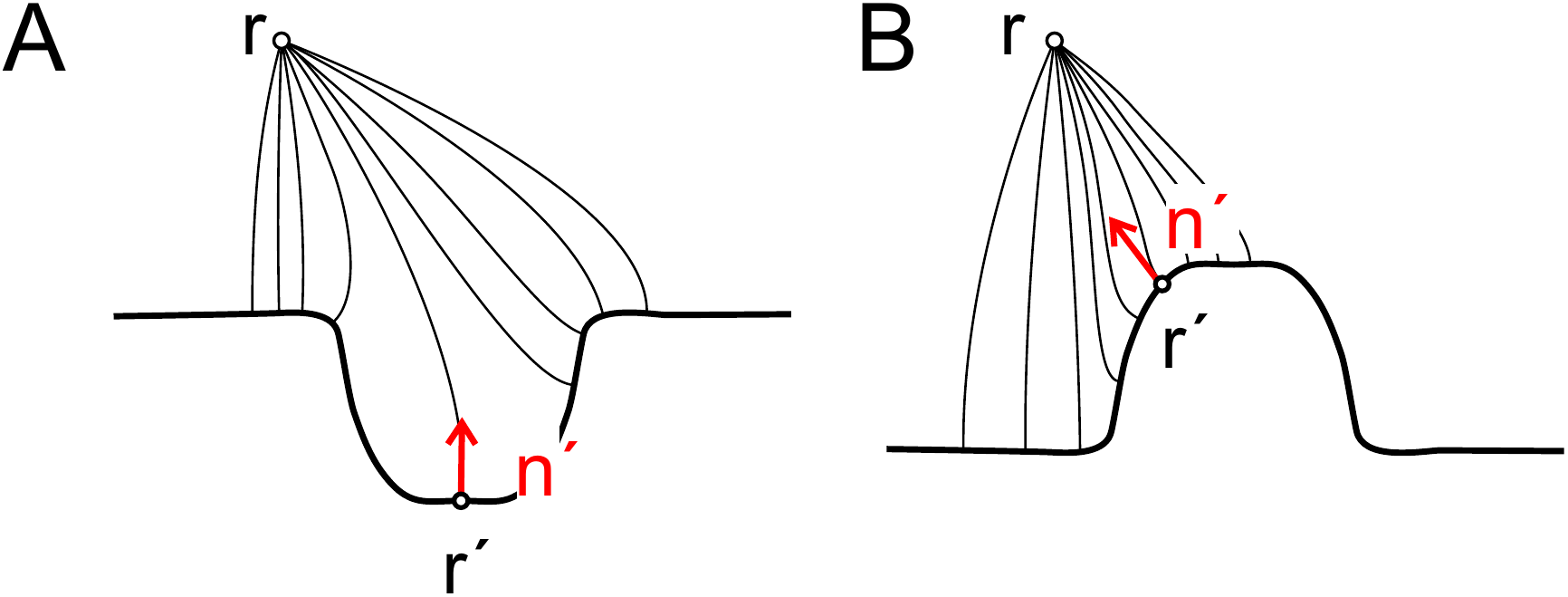}
	\sffamily
	\caption{\footnotesize  \textbf{Electric field lines at a conductive surface.} A test charge at \rr causes an electric field. \textbf{(A)} Within a depression of the conductive surface the electric field is weakened due to screening as illustrated by the sparsity of electric field lines inside the depression. \textbf{(B)} By the same token, the field is enhanced on top of a protrusion. The gradient $-\frac{\partial \GSD(\rr, \rrprime) }{\partial \boldsymbol{\text{n}}'}$ is correspondingly small in panel A and large in panel B.\normalsize}
	\normalfont 
	\label{Fig_SI_A}
\end{figure}

We note that for Dirichlet boundary conditions the Green's function is symmetric, i.e. $\GSD(\rr,\rrprime)=\GSD(\rrprime, \rr)$ \cite{Jackson1999}. This means that source point and field point of the test charge can be exchanged. Hence, an alternative interpretation of Dirichlet boundary conditions is possible (Fig.~\ref{Fig_SI_A}): If a test charge is placed at \rr in \vol, then the total potential, i.e. the potential due to the point charge itself and the potential $F$ due to the external charges, is zero for all $\rrprime$ on \surf. $-\frac{\partial \GSD(\rr, \rrprime) }{\partial \boldsymbol{\text{n}}'}$ then gives the normal component of the electric field on \surf, which is small near surface depressions, and large near protrusions (Fig.~\ref{Fig_SI_A}). In this view, the parametric dependence of $F$ on \rr arises because when the test charge changes its position in \vol, the external charges have to adjusted in order to provide zero potential on \surf. If \surf is the surface of a metal, $F$ can be interpreted as the potential due to image charges.

\subsection{Neumann problem for a dielectric surface}
\label{sec:Neumann}  

Next, we analyze the situation of a dielectric rather than a metallic surface.  In this case, Neumann boundary conditions, formulated in terms of surface charge densities instead of surface potentials, are appropriate. In the present section we recall the case of Neumann conditions on a closed surface \surf. 

For Neumann conditions, $F$ in Eq.~\ref{G_one_over_R_plus_F} is adjusted such that  
\begin{equation}
\forall\, \rr \in \mathscr{V},\,\forall\, \rrprime \in \mathscr{S}: \quad \frac{\partial \GSN(\rr, \rrprime)}{\partial \boldsymbol{\text{n}}'}=-\frac{e}{\epsilon_0 S},
\label{Neumann_condition}
\end{equation}
where $S$ is the total surface area of \surf. While it might seem more natural to assume that $\frac{\partial \GSN(\rr, \rrprime)}{\partial \boldsymbol{\text{n}}}=0$, this is inconsistent with Gauss' law, which for $\nabla' \GS(\rr,\rrprime)$ is
\begin{equation}
\int\limits_{\mathscr{V}} \nabla' \cdot \nabla' \GS(\rr,\rrprime) d^3 \rrprime = -\frac{e}{\epsilon_0}=\oint\limits_{\mathscr{S}} \boldsymbol{\text{n}} \cdot  \nabla' \GS(\rr,\rrprime) d^2 \rrprime= \oint\limits_{\mathscr{S}} \frac{\partial \GS(\rr, \rrprime)}{\partial \boldsymbol{\text{n}}'}d^2 \rrprime.
\label{Gauss_law_nabla_G}
\end{equation}
Here we have used Eq.~\ref{Laplace_delta} and \ref{defintion_normal_derivative}. If $\frac{\partial \GSN(\rr, \rrprime)}{\partial \boldsymbol{\text{n}}'}$ is constant on \surf, then Eq.~\ref{Neumann_condition} follows from Eq.~\ref{Gauss_law_nabla_G}. If Eq.~\ref{Neumann_condition} is inserted into Eq.~\ref{Phi}, then one obtains
\begin{equation}
\begin{aligned}
\forall\,  \rr \in \mathscr{V}: \quad \Phi(\rr)=
\iiint\limits_\mathscr{V} &\GSN(\rr, \rrprime) \frac{\rho(\rrprime)}{e}d^3\rrprime +\langle\Phi\rangle_\mathscr{S}\\&+ \frac{\epsilon_0}{e}\oint\limits_\mathscr{S} \GSN(\rr, \rrprime)\frac{\partial\Phi(\rrprime)}{\partial \boldsymbol{\text{n}}'}d^2\rrprime,
\label{Neumann_phi_1}
\end{aligned}		
\end{equation}
where we have used
\begin{equation}
	-\frac{\epsilon_0}{e}\oint\limits_\mathscr{S}\Phi(\rrprime)\frac{\partial \GSN(\rr, \rrprime) }{\partial \boldsymbol{\text{n}}'}d^2\rrprime=\frac{\epsilon_0}{e}\oint\limits_\mathscr{S}\Phi(\rrprime)\frac{e}{\epsilon_0 S} d^2\rrprime=\langle\Phi\rangle_\mathscr{S}.
\end{equation}
If there are no charges in \vol, one has with Neumann boundary conditions
\begin{equation}
\forall\,  \rr \in \mathscr{V}: \quad \Phi(\rr)=
\langle\Phi\rangle_\mathscr{S}+ \frac{\epsilon_0}{e}\oint\limits_\mathscr{S} \GSN(\rr, \rrprime)\frac{\partial\Phi(\rrprime)}{\partial \boldsymbol{\text{n}}'}d^2\rrprime.
\label{Neumann_phi_2}		
\end{equation}
The physical meaning of Neumann boundary conditions Eq.~\ref{Neumann_condition} can be understood in the following way: One puts a point test charge on \surf at $\rrprime$ and then adjusts a distribution of additional charges outside \vol such that the total potential $G^\mathrm{N}_\mathscr{V}(\rr,\rrprime)= \frac{e}{4\pi\epsilon_0|\rr-\rrprime|}+ F(\rr,\rrprime)$  at \rr inside \vol fulfils Eq.~\ref{Neumann_condition}, i.e.~changes with a fixed (independent of $\rrprime$) slope as the test charge is moved along the surface normal $\boldsymbol{\text{n}}'$. Moreover, this fixed slope tends to zero as \surf becomes large. Eq.~\ref{Neumann_condition} is evidently fulfilled if \surf is a dielectric surface that has, except at $\rrprime$, a specific areal charge density $\sigma$ sitting below it, just outside \vol. The $\sigma$ is constructed such that moving the test charge from its position on \surf to just outside \vol, that is into line with $\sigma$, creates the situation of constant potential (zero electric field) everywhere inside \vol ($\sigma$ is chosen accordingly; if, e.g., \surf is a sphere, $\sigma$ is a constant surface charge density). Before, with the test charge at its original position on \surf (i.e.~slightly inside $\sigma$), the electric field is not precisely zero in \vol, and hence there are small changes $\frac{\partial \GSN(\rr, \rrprime)}{\partial \boldsymbol{\text{n}}'}$ (Eq.~\ref{Neumann_condition}) of the potential at \rr as the test charge is moved along $\boldsymbol{\text{n}}'$. However, since for increasing area of \surf the contribution of the one slightly "misaligned" charge at $\rrprime$ to the total potential in \vol becomes negligible, the potential change $\frac{\partial \GSN(\rr, \rrprime)}{\partial \boldsymbol{\text{n}}'}$ at \rr also converges to zero as \surf grows. 

With this interpretation of Eq.~\ref{Neumann_condition}, it is straightforward to understand the physical essence of Eq.~\ref{Neumann_phi_2}. The integral Eq.~\ref{Neumann_phi_2} gives the total potential at \rr of an arbitrary surface charge distribution applied on \surf (not to be confused with the charge density outside \vol that produces $F$). Its integrand is therefore given by $\frac{1}{e}\GSN$, which is the potential of the surface charge density element at $\rrprime$ per unit charge, times $\epsilon_0\frac{\partial \Phi}{\partial \boldsymbol{\text{n}}'}$, the actual size of the applied surface charge density at $\rrprime$. The surface charge density is given by the latter expression, because $-\frac{\partial \Phi}{\partial \boldsymbol{\text{n}}'}$ is the normal component of the electric field at \surf, and generally steps in the normal electric field at surfaces are related to surface charge densities ($\boldsymbol{\text{E}}_2-\boldsymbol{\text{E}}_1= \frac{\sigma}{\epsilon_0}$). The integral in Eq.~\ref{Neumann_phi_2} is positive, because negative signs due to the definition of the electric field and due to taking the normal derivative in $-\boldsymbol{\text{n}}'$ direction cancel. In addition to the integral, the term $\langle \Phi\rangle_\mathscr{S}$ appears in Eq.~\ref{Neumann_phi_2}. It arises because, even in absence of the applied charge density $\epsilon_0\frac{\partial \Phi}{\partial\boldsymbol{\text{n}}'}$, the potential at \rr, given by $\GSN(\rr,\rrprime)$, does not remain constant as the test charge moves along $-\boldsymbol{\text{n}}'$; as an immediate consequence, the  surface integral in Eq.~\ref{Phi} with $\frac{\partial \GSN(\rr,\rrprime)}{\partial\boldsymbol{\text{n}}'}$ in the integrand is not zero. As we have seen in the Dirichlet case, this integral describes the effect of a double layer at \surf on the potential at \rr. In the present case of Neumann boundary conditions, this double layer arises from the polarization of the dielectric at its surface \surf by the charge density that sets up $F$. Because by construction (Eq.~\ref{Neumann_condition}) $\frac{\partial \GSN(\rr,\rrprime)}{\partial\boldsymbol{\text{n}}'}$ is constant, only the average of the double layer potential on \surf contributes to the potential at \rr, although the double layer may well be inhomogeneous on \surf.

With this, we conclude our discussion of the formalism for the description of Dirichlet and Neumann boundary conditions. We now turn to the application of this formalism to the specific case of SQDM.

\newpage

\section{Solution of the SQDM boundary value problem}
\label{sec:ApplicationToSQDM}

\subsection{Introduction}

As stated earlier, the central task of SQDM image interpretation is finding the relation between the potential \PhiQD at the QD and the properties of the boundary, i.e., tip and sample surface. In this respect, the goal is to actually determine the Green's function for a given measurement situation and invert the corresponding boundary value problem. We will present a formal solution below. However, it is practically always necessary to introduce simplifying assumptions about the nature of the boundary, particularly at the object surface, to solve this task. Clearly, there is not one single preferable approach, but rather a hierarchy of simplifications which can only partially explored in this paper.

The boundary at the object surface is fully specified by the boundary shape and the material of which the boundary consists. Depending on the material, appropriate boundary conditions can be chosen: Dirichlet conditions for a conductive surface, Neumann conditions for a dielectric surface. We outline the general case of a boundary which has metallic and dielectric parts in in Section~\ref{sec:Mixed}. In Section~\ref{sec:Approximativesolution} we introduce the first simplification, namely eliminating Neumann conditions at partially dielectric surfaces via the introduction on an effective \textit{dielectric topography} of metallic nature. In Section~\ref{sec:SpecificSolutions} we then apply some restrictions to the shape of the boundary, the strongest one being the assumption of a completely flat surface. Subsequently, we relax this assumption and sketch a possible approach to obtain the Green's functions for non-planar surfaces.

\subsection{Formal solution}
\label{sec:Mixed}

\begin{figure}[!ht]
	\centering
	\includegraphics[width=7cm]{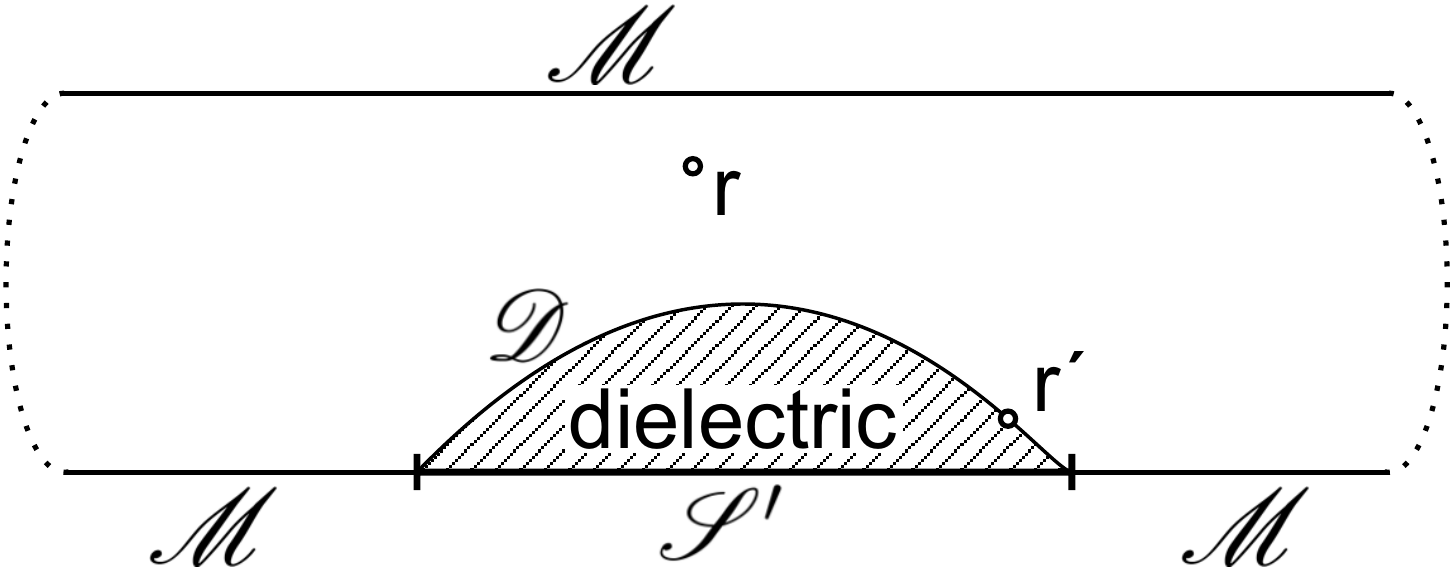}
	\sffamily
	\caption{\footnotesize  \textbf{Boundary value problem for a mixed surface.} The enclosing surface \surf consists of metallic ($\mathscr{M}$) and dielectric ($\mathscr{D}$) parts. A second, purely metallic surface $\surf'$ can be defined which coincides with \surf on $\mathscr{M}$ but differs in the regions $\mathscr{D}$ where it forms the metallic support beneath the dielectric nanostructure. \normalsize}
	\normalfont 
	\label{Fig_SI_B} 
\end{figure}

In this section we consider the most general situation that may be encountered in SQDM. While the tip is always metallic, the sample surface may consist of several elements:  
\begin{enumerate}
\item an arbitrary geometric topography of a metallic sample surface,
\item a dielectric layer of arbitrary dielectric properties and topography (not necessarily closed) on top of the metal surface; this includes the case of a thick dielectric on a metallic sample plate. 
\end{enumerate}
An example is given in Fig.~\ref{Fig_SI_B}. We split the closed surface \surf into two parts, the dielectric portion $\mathscr{D}$ and the metallic portion $\mathscr{M}$. The latter also comprises the tip. In Fig.~\ref{Fig_SI_B}, we also could consider the surface $\surf'$ which is fully metallic. However, then in Eq.~\ref{PhiDirichlet} the volume integral is not zero. Since in SQDM we must solve a problem of the kind defined in Eq.~\ref{inv_I_O} for 2D image and object functions, it is therefore more advantageous to consider the inner surface \surf which consists of $\mathscr{D}$ and $\mathscr{M}$. This situation can be handled by choosing Dirichlet boundary conditions on the metal, and Neumann boundary conditions on the dielectric. 

Mixed Dirichlet/Neumann boundary conditions have a unique solution \cite{Jackson1999}, such that the potential at point \rr is given by
\begin{equation}
\forall\,  \rr \in \mathscr{V}: \quad \Phi(\rr)=
\langle\Phi\rangle_\mathscr{D}+ \frac{\epsilon_0}{e}\int\limits_\mathscr{D} G^\mathrm{M}_\mathscr{S}(\rr, \rrprime)\frac{\partial\Phi(\rrprime)}{\partial \boldsymbol{\text{n}}'}d^2\rrprime  - \frac{\epsilon_0}{e}\int\limits_\mathscr{M} \Phi(\rrprime)\frac{\partial G^\mathrm{M}_\mathscr{S}(\rr, \rrprime) }{\partial \boldsymbol{\text{n}}'}d^2\rrprime.
\label{Phi_mixed}		
\end{equation}
The superscript $\mathrm{M}$ stands for mixed boundary conditions. Formally, we can define 
\begin{equation}
\gamma_\mathscr{S}(\rr,\rrprime)\equiv\left.
\begin{cases}	
( \frac{\epsilon_0}{e} G^\mathrm{M}_\mathscr{S}(\rr, \rrprime) & \forall \,\rrprime \in \mathscr{D})\\
(-\frac{\epsilon_0}{e}\frac{\partial G^\mathrm{M}_\mathscr{S}(\rr, \rrprime) }{\partial \boldsymbol{\text{n}}'}& \forall \, \rrprime \in \mathscr{M}).
\end{cases}\right\}
\label{most_general_gamma}		
\end{equation}
Then, we can write Eq.~\ref{Phi_mixed} compactly as 
\begin{equation}
\Phi(\rr)= \langle\Phi\rangle_\mathscr{D} + \oint\limits_\mathscr{S} \gamma^\mathrm{M}_\mathscr{S}(\rr,\rrprime)\bigg[\frac{\partial\Phi(\rrprime)}{\partial \boldsymbol{\text{n}}'}\theta(\rrprime) + \Phi(\rrprime)(1-\theta(\rrprime))\bigg]d^2\rrprime,
\label{most_general_Phi}		
\end{equation}
where we have defined 
\begin{equation}
\theta(\rrprime)\equiv\left.
\begin{cases}			
1 & \quad \text{if} \,\, \rrprime \in  \mathscr{D}\\
0 & \quad \text{if} \,\,  \rrprime \in  \mathscr{M}
\end{cases}\right\}
\label{theta}		
\end{equation}
The average in Eq.~\ref{most_general_Phi} is carried out over the surface potential on all dielectric surface parts $\mathscr{D}$. This equation has the form of Eq.~\ref{inv_I_O}, if we consider $\Phi(\rr)- \langle\Phi\rangle_\mathscr{D}$ as the image function. Its inversion would provide us, at least conceptually, with the object function of SQDM, which turns out to be a surface potential on the metallic part of the surface and a charge density on the dielectric part. The constant $\langle\Phi\rangle_\mathscr{D}$ appearing in the object function is the average potential on the dielectric due to the polarization at its surface, resulting from the Neumann boundary conditions.  

While Eq.~\ref{most_general_Phi} establishes, for the most general sample, a formal relationship between the SQDM image and object functions, i.e.~corresponds to Eq.~\ref{inv_I_O}, its practicality is hampered by the following: 
\begin{enumerate}
	\item the partitioning of the surface into metallic and dielectric parts is not known a priori,
	\item $\gamma^\mathrm{M}_\mathscr{S}(\rr,\rrprime)$ is difficult to determine, and 
	\item when we adjust the bias voltage applied to the sample to compensate local surface potential variations (Section~\ref{primaryMeasurands}), we execute a boundary value problem Eq.~\ref{boundary_problem_conceptual} on the \textit{metallic} surface $\surf'$. For this reason, our primary measurands \Vpm embody information on $\surf'$, not on \surf. As a consequence, when inverting Eq.~\ref{inv_I_O} to get Eq.~\ref{I_O}, we obtain information on $\surf'$, not \surf.
\end{enumerate}

For these reasons, we will now introduce an approximation regarding the type of boundary conditions and boundary shape. We will use this assumption throughout the rest of the paper for the interpretation of SQDM images. Rather than pursuing Eq.~\ref{most_general_Phi}, we go back to Dirichlet boundary conditions on the metallic surface $\surf'$ in Fig.~\ref{Fig_SI_B} and regard the dielectric parts between $\mathscr{D}$ and $\surf'$ as a perturbation.

\subsection{Dielectric topography}
\label{sec:Approximativesolution}

\begin{figure}[!ht]
	\centering
	\includegraphics[width=6cm]{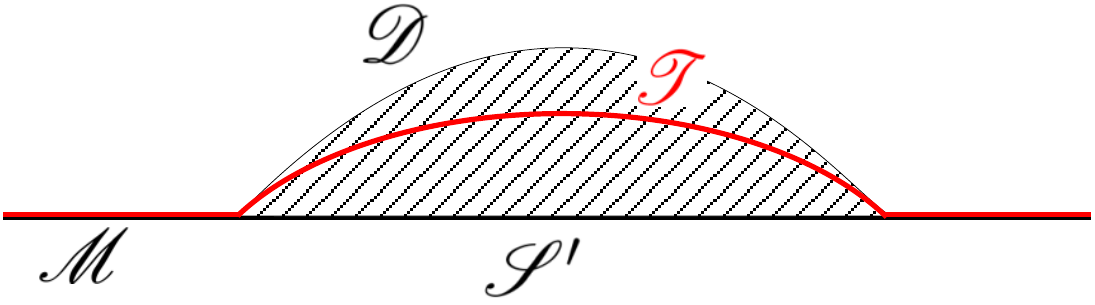}
	\sffamily
	\caption{\footnotesize  \textbf{Effective dielectric topography.} For the two extreme cases of zero or infinite permittivity of the dielectric, $\mathscr{T}$ would either coincide with $\mathscr{S}'$ or with $\mathscr{D}$ respectively, while it naturally coincides with \surf on all metallic parts of the surface. \normalsize}
	\normalfont 
	\label{Fig_SI_C} 
\end{figure}

\noindent From the solution to the Dirichlet boundary value problem in Section~\ref{sec:Dirichlet} it is clear that the topography of the metal surface influences the $\frac{\partial \GSD(\rr, \rrprime) }{\partial \boldsymbol{\text{n}}'}$ function, and hence also determines the effect of $\Phis$ and \Vb on the potential in \vol. Since any dielectric above the metal additionally influences how $\Phis$ and \Vb on the metal surface affect the potential in \vol, the influence of the dielectric can be lumped together with the metallic topography into an effective $\frac{\partial \GTD(\rr, \rrprime) }{\partial \boldsymbol{\text{n}}'}$ function on the surface \surfT (Fig.~\ref{Fig_SI_C}). We refer to this effective surface \surfT as \textit{dielectric topography}. $\mathscr{T}$ is defined such that the gating efficiency resulting from the hypothetical metallic surface \surfT is identical to the original situation of a mixed surface $\mathscr{S}$, consisting of dielectric surface  $\mathscr{D}$ and metal surface $\mathscr{M}$. Note that generally $\surfT \neq \surf$. 

When determining $\Vstar$ in the experiment by executing the boundary value problem Eq.~\ref{boundary_problem_conceptual} (i.e.~nulling the effect of $\Phis$ on the potential at the QD, as in Eq.~\ref{boundary_problem_conceptual_constant}) we cannot distinguish between the metallic topography and the contribution of the dielectric. Hence, the inversion of Eq.~\ref{integralalpha1} will necessarily yield an effective object function which combines metallic topography and dielectric contributions. Moreover, the object function \Phis (Eq.~\ref{equation3}) is given on the dielectric topography \surfT (see Fig.~\ref{Fig_SI_C}), and not on $\surf'$ directly.

To illustrate the influence of the dielectric on the potential at the QD, we consider the case of a parallel plate capacitor that is partially filled with a dielectric. Specifically, we calculate how the partial filling with dielectric influences the gating efficiency $\alpha$ (Eq.~\ref{EqAlpha1}) which is the integral over $\rrprime$ of the effective $\frac{\partial \GTD(\rr, \rrprime) }{\partial \boldsymbol{\text{n}}'}$ as we will show later. Splitting the capacitor of width \zt into two capacitors in series, with widths $t$ and $\zt-t$, and filling the capacitor $t$ by a dielectric with permittivity $\epsilon_r$, we obtain for the total capacity
\begin{equation}
  \frac{1}{C}=\frac{t}{\epsilon_0 \epsilon_r A}+\frac{\zt-t}{\epsilon_0 A}
  \label{}
\end{equation}
or 
\begin{equation}
C=\frac{\epsilon_0 A}{t(\epsilon_r^{-1}-1)+ \zt}.
  \label{}
\end{equation}
The gating efficiency is given by \cite{Temirov2018}
\begin{equation}
 \alpha=\frac{C}{C+C_\mathrm{QD/tip}},
   \label{}
\end{equation}
such that 
\begin{equation}
 \alpha=\frac{1}{1+\frac{C_\mathrm{QD/tip}}{\epsilon_0 A}[t(\epsilon_r^{-1}-1)+\zt]},
   \label{}
\end{equation}
Noting that $C^0_\mathrm{QD/sample}=\frac{\epsilon_0 A}{\zt}$, we obtain
\begin{equation}
 \alpha=\frac{1}{1+\frac{C_\mathrm{QD/tip}}{C^0_\mathrm{QD/sample}}\bigg[\frac{t(\epsilon_r^{-1}-1)+\zt}{\zt}\bigg]}.
   \label{}
\end{equation}
Since $C^0_\mathrm{QD/sample} \ll C_\mathrm{QD/tip}$ \cite{Temirov2018}, we can expand the above equation
\begin{equation}
 \alpha \approx \frac{C^0_\mathrm{QD/sample}}{C_\mathrm{QD/tip}\big[1+\frac{t}{\zt}(\epsilon_r^{-1}-1)\big]}.
   \label{}
\end{equation}
With $\alpha^{\epsilon_r=1}\equiv \frac{C^0_\mathrm{QD/sample}}{C_\mathrm{QD/tip}+C^0_\mathrm{QD/sample}}\approx \frac{C^0_\mathrm{QD/sample}}{C_\mathrm{QD/tip}} $ this becomes
\begin{equation}
 \alpha \approx \alpha^{\epsilon_r=1} \Big[1-\frac{t}{\zt}(\epsilon_r^{-1}-1)\Big]= \alpha^{\epsilon_r=1} \Big[1+\frac{t}{\zt}|\epsilon_r^{-1}-1|\Big].
   \label{}
\end{equation}
This shows that the dielectric will increase the gating efficiency, in lowest order linearly with its thickness $t$. Since $t$ and $|\epsilon_r^{-1}-1|$ enter as a product, a given change in gating efficiency can be explained by any combination of (increasing) thickness and (decreasing) permittivity. A metal with $\epsilon_r=\infty$ yields a minimum thickness. When we apply Dirichlet boundary conditions to the mixed surface \surf in Fig.~\ref{Fig_SI_C}, we implicitly assume an effective topography $\mathscr{T}$ of metallic nature. Therefore, in the dielectric regions $\mathscr{T}$ will lie below \surf. 

Since we will use the assumption of Dirichlet boundary conditions on an effective surface \surfT throughout the rest of the paper, we drop the indices $\mathrm{D}$ and \surfT from now on. The respective Green's function and its gradient will hence be simply denoted as $\mathrm{G}$ and \gammaplain but should be interpreted in the context of this assumption.

\subsection{Dirichlet solution at the surface given by the dielectric topography}
\label{sec:Dirichlet_SQDM}

\begin{figure}[!ht]
	\centering
	\includegraphics[width=10cm]{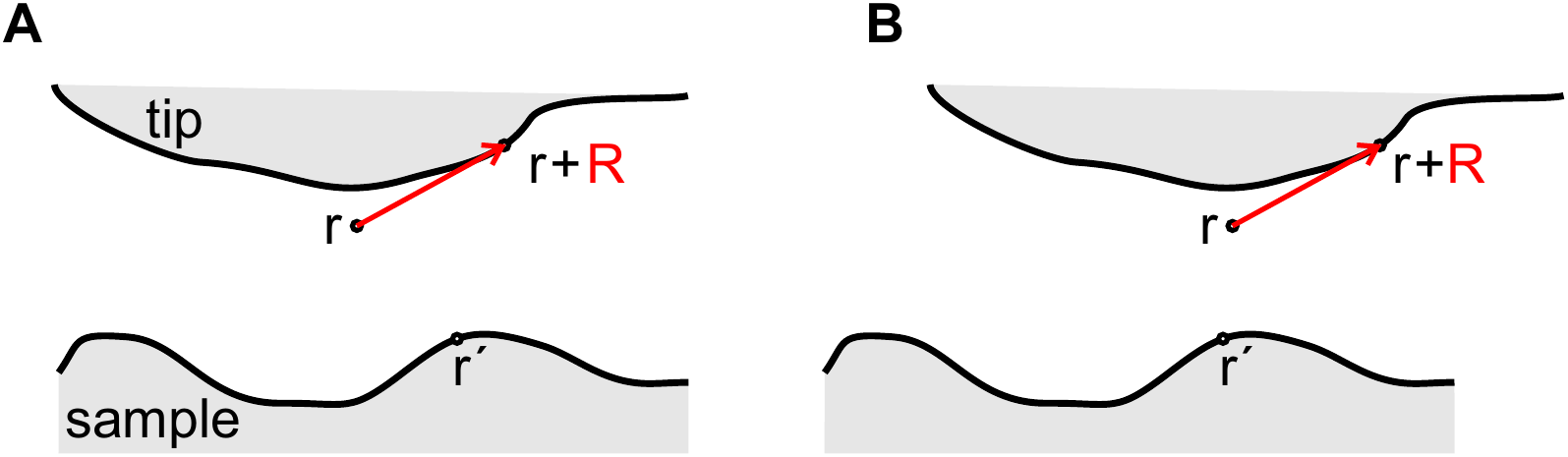}
	\sffamily
	\caption{\footnotesize  \textbf{Spatial relation between QD and points on the tip surface.} \textbf{A} The spatial relation between the QD at \rr and points on the surface of the tip at $\rr+\RR$ (red arrow) are fixed and independent of \rr. Hence, any point on the tip surface can be addressed by its specific \RR. \textbf{B} In contrast, the relative position of the QD to any feature on the sample surface changes during scanning.   \normalsize}
	\normalfont 
	\label{Fig_TipInfluence} 
\end{figure}

\noindent In SQDM we essentially measure the electric potential $\Phi=\PhiQD$ in \vol, the vacuum space in the NC-AFM junction where $\rho=0$. The surface \surfT consists of the tip and sample surfaces, which are metallic (tip) or effectively metallic (sample) and which we assume connected infinitely far away from the tip apex. Thus \surfT is metallic and closed, and we can apply Eq.~\ref{PhiGF} to the SQDM situation. Specifically, the boundary values of the potential are $\Phit(\RR)$  on the surface of the grounded tip (Fig.~\ref{Fig_TipInfluence}), and $\Phis(\rrprime)+\Vb$ on the surface of the sample. Local variations of \Phit and \Phis arise from the local atomic structure of the tip and nanostructures on the surface. \Vb is the bias voltage applied to sample. The integration in Eq.~\ref{PhiGF} is over the tip and sample surfaces such that and we obtain

\begin{equation}
\begin{aligned}
\PhiQD(\rr) =
&&- \iint\limits_\text{sample surface}[\Phis(\rrprime)+\Vb]\frac{\epsilon_0}{e}\frac{\partial G(\rr, \rrprime)}{\partial \boldsymbol{\text{n}}'}d^2\rrprime \\
&&- \iint\limits_\text{tip surface}\Phit(\RR)\frac{\epsilon_0}{e}\frac{\partial G(\RR)}{\partial \boldsymbol{\text{n}}'}d^2\RR
\end{aligned}
\label{PhiGF2WithTip}.
\end{equation}
We assume that tip and sample are sufficiently well separated such that nanostructures on the latter do not influence the Green's function on the former and vice versa. Then, the integral over the tip surface yields a constant contribution \PhiTip which is an integral over the \textit{relative} positions \RR between all points on the tip surface and the QD (Fig.~\ref{Fig_TipInfluence}), and is thus independent of the lateral or vertical positions of the sensor. Thus we obtain
\begin{equation}
\PhiQD(\rr)
=  - \iint\limits_\text{surface}[\Phis(\rrprime)+\Vb]\frac{\epsilon_0}{e}\frac{\partial G(\rr, \rrprime)}{\partial \boldsymbol{\text{n}}'}d^2\rrprime+\PhiTip
\label{PhiGF2}.
\end{equation}
Here, "surface" refers to the sample part of \surfT.
Defining 
\begin{equation}
\gammaplain(\rr,\rrprime)
\equiv  - \frac{\epsilon_0}{e}\frac{\partial G(\rrprime, \rr)}{\partial \boldsymbol{\text{n}}'},
\label{defgamma}
\end{equation}
Eq.~\ref{PhiGF2} becomes 
\begin{equation}
\PhiQD(\rr)
= \iint\limits_\text{surface}\gammaplain(\rr,\rrprime)
[\Phis(\rrprime)+\Vb]d^2\rrprime+\PhiTip.
\label{PhiGF3}
\end{equation}
With this, we have derived, in the context of a boundary value problem via the Green's function formalism, the fundamental equation of SQDM (Eq.~\ref{EqPhiQD2} of Section~\ref{SecImagingFormalism}).

We briefly analyze the sign of \gammaplain. Since $\boldsymbol{\text{n}}'$ points outward from \surfT, i.e.~away from \vol, and since $G(\rr, \rrprime)$, as the potential at $\rrprime$ of the test charge $+e$ at \rr (see Eq.~\ref{Laplace_delta}) with $G(\rr, \rrprime)=0$ for $\rrprime \in \surfT$, becomes less positive towards the surface, $\frac{G(\rr, \rrprime)}{\partial \boldsymbol{\text{n}}'}$ must be negative. $\gammaplain(\rr,\rrprime)$, given by Eq.~\ref{defgamma}, is therefore positive. For a negative test charge $-e$ at \rr the sign on the right hand sides of Eqs.~\ref{Laplace_delta}, \ref{PhiGF} and \ref{defgamma} changes, and thus, because also $\frac{G(\rr, \rrprime)}{\partial \boldsymbol{\text{n}}'}$ changes sign, $\gammaplain(\rr,\rrprime)$ remains positive. $\gammaplain(\rr,\rrprime)$ is therefore always positive. 

A further notable property is the decay of \gammaplain with distance $|\rr-\rrprime|$ which is generally expected because of the $1/|\rr-\rrprime|$ term in Eq.~\ref{G_one_over_R_plus_F}, notwithstanding the contribution of $F$. In fact, as we will see below, Dirichlet boundary conditions and the associated $F$ in Eq.~\ref{G_one_over_R_plus_F} lead to an even faster (exponential) decay of $\gammaplain(\rr,\rrprime)$ with the lateral distance between \rr and \rrprime. This effectively limits the area of the surface around \rr for which $\gammaplain(\rr,\rrprime)$ is nonzero to a few hundred nm$^2$. For example, if we consider a flat surface, the flatness of the surface needs to be fulfilled only within a $~\approx \unit[10]{nm}$ radius around the lateral QD position, since the integrand in Eq.~\ref{EqVstar2} is practically zero outside this region. 

We now return to Eq.~\ref{EqPhiQD1},
\begin{equation}
\PhiQD=\alpha \Vb+ \Phistar+\PhiTip.
\label{equation1}
\end{equation}
Subtracting Eq.~\ref{PhiGF3} for the case of $\Phis = 0$ from Eq.~\ref{equation1}, we find
\begin{equation}
0=\Phistar - \Vb \iint\limits_\text{surface} \gammaplain(\rr,\rrprime) d^2\rrprime + \alpha \Vb.
\label{equation1b}
\end{equation}
Since this has to be valid for any \Vb, we can conclude that $\Phistar=0$ and
\begin{equation}
\alpha=\alpha(\rr)
= \iint\limits_\text{surface} \gammaplain(\rr,\rrprime)d^2\rrprime,
\label{integralalpha}
\end{equation}
which, with the definition of \alpharel (Eq.~\ref{EqAlphaRel1}) is identical to Eq.~\ref{integralalpha1}.
Thus, the electrode geometry \surfT that determines the boundary value problem also determines the gating efficiency $\alpha(\rr)$ of the quantum dot, through the common integral kernel \gammaplain.  Eq.~\ref{integralalpha} allows the definition of a kernel \gammastar 
\begin{equation}
\gammastar(\rr,\rrprime)\equiv \frac{1}{\alpha(\rr)}\gammaplain(\rr,\rrprime),
\label{definitiongammastar}
\end{equation}
which is normalized if integrated over \rrprime. 
When $\Vb=0$, the comparison of Eq.~\ref{EqPhiQD1} with Eq.~\ref{PhiGF3} shows that 
\begin{equation}
\Phistar(\rr) = \iint\limits_\text{surface} \gammaplain(\rr,\rrprime) \Phis(\rrprime)d^2\rrprime
\label{Phistar}.
\end{equation}
Dividing this by $\alpha(\rr)$, we obtain, because of Eq.~\ref{definitiongammastar},
\begin{equation}
\frac{\Phistar(\rr)}{\alpha(\rr)}
= \iint\limits_\text{surface} \gammastar(\rr,\rrprime) \Phis(\rrprime)d^2\rrprime
\label{vstarphis}.
\end{equation}
We define 
\begin{equation}
\Vstar(\rr)\equiv\frac{\Phistar(\rr)}{\alpha(\rr)}
\label{vstarphistar}
\end{equation}
and thus finally arrive at
\begin{equation}
\Vstar(\rr) = \iint\limits_\text{surface} \gammastar(\rr,\rrprime) \Phis(\rrprime)d^2\rrprime
\label{EqVstar2}.
\end{equation}
This is Eq.~\ref{equation3}. Applying $-\Vstar(\rr)$ to the sample exactly cancels the effect of the surface potential \Phis on the QD as can be seen by the following argument: We assume that at the sensor position \rr, the potential is $\Phistar(\rr)$. If then we apply a bias voltage $\Vb = \Vpm_0 - \Vstar(\rr)$ (given by Eq.~\ref{EqVstar2}) we obtain
\begin{equation}
\begin{aligned}
\PhiQD(\rr)&=\alpha(\rr)\Vb+ \Phistar(\rr) + \PhiTip\\
&=\alpha(\rr) [V_0^\pm - \Vstar(\rr)]+ \Phistar(\rr) + \PhiTip\\
&=\alpha(\rr) V_0^\pm - \alpha(\rr)\Vstar(\rr) + \Phistar(\rr) + \PhiTip\\
&= \alpha(\rr) V_0^\pm -\alpha(\rr) \frac{\Phistar(\rr)}{\alpha(\rr)} + \Phistar(\rr) + \PhiTip\\
&=\alpha(\rr) V_0^\pm + \PhiTip.
\end{aligned}
\label{cancellation}
\end{equation}
In  other words, additionally applying $-\Vstar$ keeps the potential at the QD at exactly the same charging threshold $\PhiQD^\pm(\rr)=\alpha(\rr) V_0^\pm$ as if \Phistar was not present, irrespective of the local gating efficiency $\alpha(\rr)$. Eq.~\ref{cancellation} thus demonstrates that \Vstar, as defined in Eq.~\ref{vstarphistar} (or \ref{equation3}), fulfills Eq.~\ref{boundary_problem_conceptual_constant}and can thus be identified with the experimentally determined image function $I(\rr)$ of SQDM. Since, for translationally invariant (i.e. flat) samples $\gammastar(\rr,\rrprime) = \gammastar(|\rr-\rrprime|)$, the inversion of Eq.~\ref{EqVstar2} to determine the object function $\Phis(\rr)$ becomes a deconvolution problem.

\newpage

\section{From image to object function}
\label{sec:SpecificSolutions}

\subsection{The role of non-local screening}
\label{sec:GeneralProblem}

In Sec.~\ref{primaryMeasurands} we found that SQDM provides two independent secondary measurands, \Vstar and \alpharel. Within the formalism outlined in Sec.~\ref{SecImagingFormalism}, these two quantities can be interpreted as image functions. In Sec.~\ref{sec:Dirichlet_SQDM} we have identified the surface potential \Phis as the object function related to \Vstar and derived a way to obtain the former via inversion of Eq.~\ref{EqVstar2}. While we have stated in Sec.~\ref{SecImagingFormalism} that the topography of the sample surface is the object function of the second image function \alpharel, we have not formally derived the respective relation yet. In fact, this relation is not straightforward. The reason is non-local screening of the potential by the sample topography, as we will show now.

In Sec.~\ref{sec:Dirichlet}, we have seen that with Dirichlet boundary conditions the Green's function $\GTD(\rr,\rrprime)$ is the sum of a point charge potential and the function $\FT(\rr,\rrprime)$ (Eq.~\ref{G_one_over_R_plus_F}). Here we have replaced \surf by \surfT, following the discussion in Sec.~\ref{sec:Approximativesolution}. \FT can be understood as the potential of a charge distribution just outside \surfT which screens the influence of the charge at \rrprime on the potential at \rr. Evidently, this charge distribution and therefore \FT must depend on the shape of \surfT, because the screening charges are distributed over the entire region just outside the surface. In other words, the shape of \surfT non-locally determines \FT and therefore, via Eqs.~\ref{G_one_over_R_plus_F} and \ref{defgamma}, also \gammaplain. Consequently, \gammaplain and via Eqs.~\ref{integralalpha1} and \ref{integralalpha}, also \alpharel is therefore a functional of the surface topography \surfT. This non-locality poses a problem for retrieving \surfT from \alpharel. To approach this problem, we first express the shape of the surface \surfT by a scalar function \teff which then becomes the object function associated to \alpharel. Then, we derive relations between \gammaplain and \teff for a series of approximations. These relations will also be important for the recovery of \Phis, since knowledge of \gammastar is required for the inversion of Eq.~\ref{EqVstar2}.

\subsection{Infinite planes}  

In the simplest case for which we can solve the boundary value problem explicitly and thus obtain $\gammaplain \equiv \partial G/\partial \boldsymbol{\text{n}}'$, tip and sample are approximated by infinitely extended parallel planes. For the sample surface, this will become our general assumption for the rest of the paper, because we will treat topographic features and thus the object function \teff as perturbations of this plane. This simplification allows us to recast the spatial coordinates \rr and \rrprime, separating them in vertical and horizontal components. We first define a positive $z'$ axis vertical to the sample plane which has its origin at the sample and points towards the tip. Details of the dielectric surface topography can then be described via scalar heights $\teff(\rparprime)$ along $z'$ at corresponding lateral positions \rparprime, such that $\rrprime \equiv \colvec{\rparprime}{\teff}$. We denote the height of the QD above the sample plane as $z$, such that $\rr \equiv \colvec{\rpar}{z}$.

\subsection{Surface potential reconstruction: From \Vstar to \Phis}
\label{Sec_PhisRecovery}

\subsubsection{Parallel planar surfaces}
\label{sec:PPapprox}

\begin{figure}[!ht]
	\centering
	\includegraphics[width=10cm]{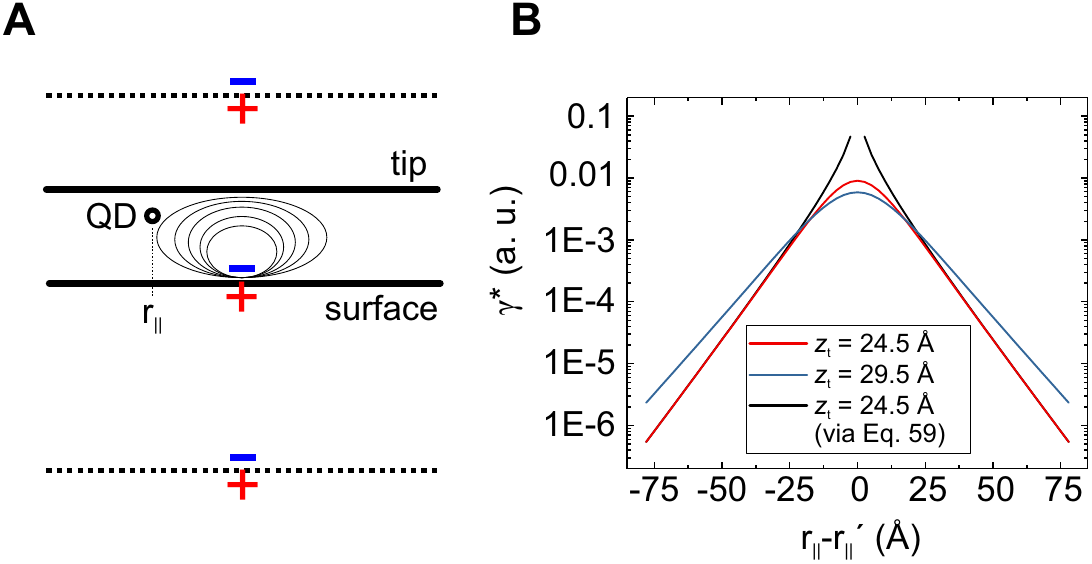}
	\sffamily
	\caption{\footnotesize  \textbf{Green's function between two grounded planes} \textbf{A} Illustration of the image charge method used to calculate the potential of a point charge between two conductive planes. Since the charge is placed next to the bottom plane, effectively, a charge-image-charge dipole is created which produces a minimal deformation of the surface potential. The indicated equipotential lines show that the potential is zero on both planes.  \textbf{B} Comparison between the potential from panel A calculated with the image charge method for $d=\unit[7]{\text{\AA}}$ at two different plate separations \zt (red, blue), and the asymptotic expression Eq.~\ref{EqPumplin} for one of the separations (black). \normalsize}
	\normalfont 
	\label{Fig_ImageCharges} 
\end{figure}

\noindent As the simplest possible approximation we assume a planar tip and a sample topography with $\teff(\rparprime) = 0\ \forall \rparprime$. An important consequence of this approximation is a simplification in the functional dependency of \gammaPP (the index pp stands for parallel plates) on \rpar and \rparprime, because the Green's function becomes invariant under lateral translations: 
\begin{equation}
\gammaplain(\rpar,z,\rparprime) = \gammaplain(|\rpar-\rparprime|,z).
\label{EqGamma2}
\end{equation}
Here we have also assumed an isotropic surface, i.e., neglected the atomic lattice structure. Eq.~\ref{EqGamma2} makes Eq.~\ref{EqVstar2} a straightforward 2D convolution with an axially symmetric kernel function. In this case, the kernel $\gammaPP^{\ast}$ can be easily calculated. To this end we place a point charge just above the surface at $(\rparprime,z_\mathrm{c})$. This point charge produces an image charge just below the surface. Charge and image charge together produce a minimal deformation $\Phis(\rpar'')=\phi\delta(\rpar''-\rparprime)$ of the surface potential. According to Eq.~\ref{EqVstar2}, $\Vstar(\rpar,z)=\phi\gammaPP^{\ast}(|\rpar-\rparprime|,z)$. Since $\Vstar=\Phistar/\alpha$, we may calculate the potential at $(\rpar, z=\zt-d)$, where \zt denotes the tip height and $d$ the tip-QD distance, to obtain, after suitable normalization, the function $\gammaPP^{\ast}(|\rpar-\rparprime|,z)$. The potential can be calculated via an infinite series of image dipoles (Fig.~\ref{Fig_ImageCharges}A). Fig.~\ref{Fig_ImageCharges}B displays numerically calculated potentials $\Phistar = \PhiQD$ of this dipole for two \zt values, together with an asymptotic expression for large $|\rpar-\rparprime|$ which clearly reveals a (faster than) exponential decay \cite{Pumplin1969},
\begin{equation}
\PhiQD(\rpar,z) =  \alpha \phi\gammaPP^{\ast}(|\rpar-\rparprime|,z) \propto\sqrt{\frac{8}{|\rpar-\rparprime| \zt}}\sin\left(\frac{z}{\zt}\pi\right) \sin\left(\frac{z_\text{c}}{\zt}\pi\right)e^{-\frac{\pi}{\zt}|\rpar-\rparprime|}.
\label{EqPumplin}
\end{equation}
Note that the symmetry of the Green's function with respect to the positions of test charge and QD which we mentioned in Sec.~\ref{sec:Dirichlet} is also present in Eq.~\ref{EqPumplin}, since it contains a product of the two sine functions with interchangeable $z$ and $z_\text{c}$.

The numerical calculation (Fig.~\ref{Fig_ImageCharges}B) and its asymptotic behaviour (Eq.~\ref{EqPumplin}) provides us with the sought-after $\gammaPP^{\ast}(|\rpar-\rparprime|,z)$. This is, in fact, the point spread function (PSF) of SQDM (for the planar electrode configuration). Although we sense long-range electrostatic fields in order to measure the object function, the joint screening of tip and sample leads to an exponential decay of the PSF and thus puts SQDM in line with scanning tunneling microscopy, where the tunneling probability also decays exponentially with distance. In both cases, the result is a superior lateral resolution, because the influence of objects which are not right beneath the probe is strongly attenuated. 

Equation~\ref{EqPumplin} shows that there is a considerable influence of the tip-surface separation \zt on the shape of \gammaPP. However, this is not a problem, because the experimental tip height is precisely measured while we attach the molecular QD to the tip by molecular manipulation \cite{Wagner2015}. Hence, we can  calculate the respective \gammaPP function for each \Vstar image.

\subsubsection{Beyond parallel planar surfaces}
\label{SecBeyondPP}

\begin{figure}[!ht]
	\centering
	\includegraphics[width=12cm]{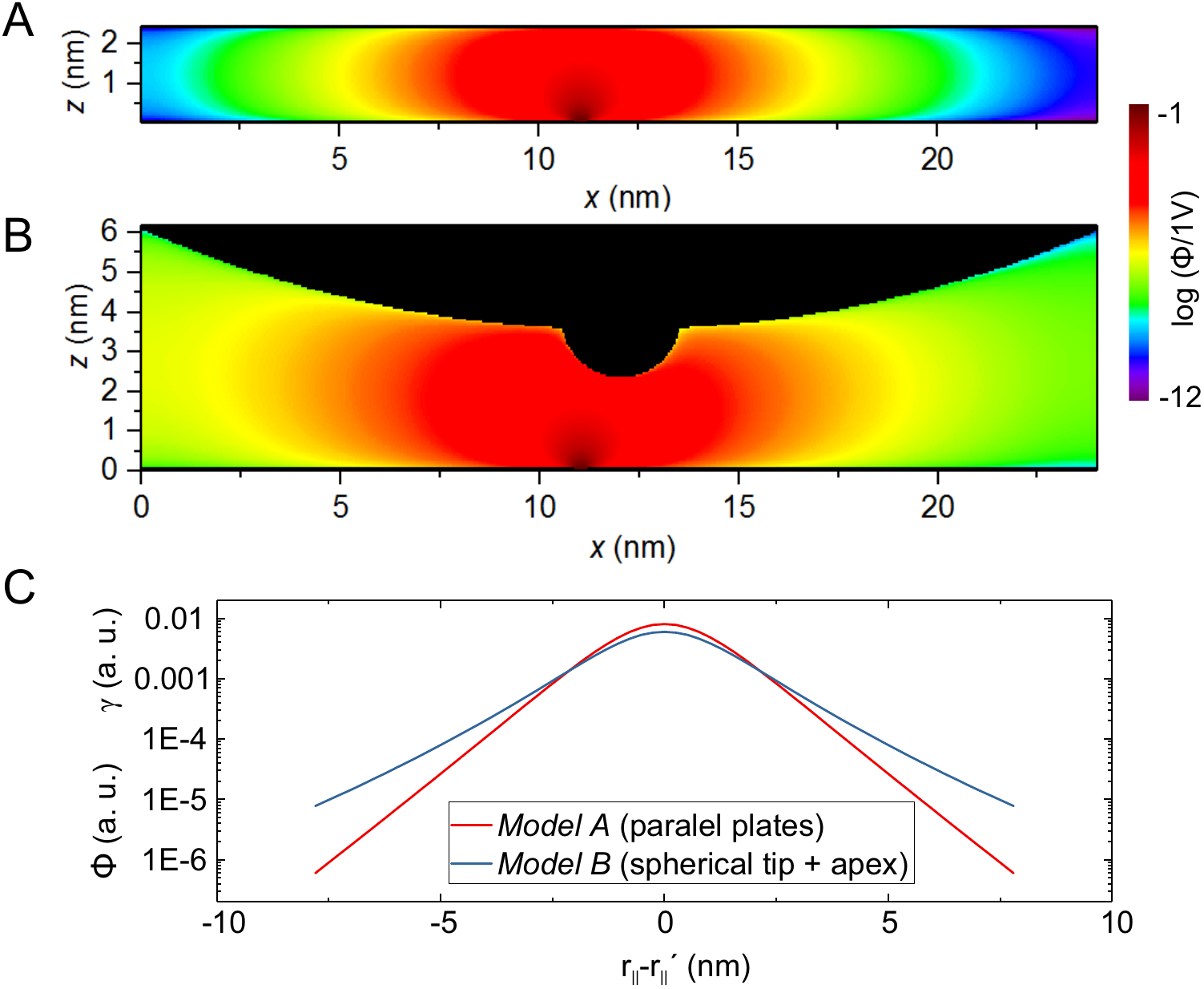}
	\sffamily
	\caption{\footnotesize  \textbf{Potential distribution in the tip-surface junction.} Shown on a log scale is a cut through the simulated potential of an on-surface nanostructure of $3 \times 3 $ \AA$^2$ with a potential of \unit[0.09]{V} in the presence of a grounded tip and surface. We compare two different tip models: A parallel plane (panel \textbf{A}) and a tip apex modeled with two spheres with \unit[300]{\AA} and \unit[15]{\AA} radius respectively (panel \textbf{B}). In both cases the nominal tip height is $\zt = \unit[24]{\text{\AA}}$. \textbf{C} Comparison of the corresponding $\gammaplain(\rpar-\rparprime)$ profiles from panels A and B measured $\unit[7]{\text{\AA}}$ below the tip apex.\normalsize}
	\normalfont
	\label{Fig_PSF1} 
\end{figure}

\noindent There exists a hierarchy of approximations for the boundary shape which add increasing levels of complexity beyond the model of parallel planes for tip and surface. First, we will discuss the implications of a non-planar tip. If we initially retain the restriction of an axially symmetric tip, Eq.~\ref{EqGamma2} is left intact. The only difference to the planar tip is that $\gammaAx(|\rpar-\rparprime|,z)$ will have a different shape than \gammaPP. \gammaAx, which denotes an entire class of functions depending on the precise tip shape, can in principle be obtained for any given axially symmetric tip surface by a finite element simulation. We have done finite element simulations for two tip models.  The results are displayed in Fig.~\ref{Fig_PSF1}. A sharper tip results in a weaker screening of the point dipole at \rparprime (Fig.~\ref{Fig_ImageCharges}A) and thus yields a weaker decay of $\gamma_\text{axial}^*$ with $|\rpar-\rparprime|$. In contrast, a hypothetical tip with a concave apex into which the QD is embedded could enhance the decay compared to \gammaPP even  further and thus increase the lateral resolution of SQDM beyond Eq.~\ref{EqPumplin}.

If we drop the assumption of an axially symmetric tip, Eq.~\ref{EqGamma2} looses its validity and the dependency of \gammaplain on $|\rpar-\rparprime|$ is replaced by a dependency on $\rpar-\rparprime$ which still implies translational symmetry,
\begin{equation}
\gammaplain(\rpar,z,\rparprime) = \gammaplain(\rpar-\rparprime,z).
\label{EqGamma3}
\end{equation}

Without axial symmetry of \gammaplain, \Vstar images of a point-like object would loose their axial symmetry. However, on a larger scale the exponential decay of \gammaplain would still dominate. In fact, we have never observed any significant distortions of \Vstar images of circular objects such as single adatoms which could have been attributed to an irregular tip shape. Nevertheless, distortions in an image of a highly symmetric object could, in principle, be used to gain some information about \gammaplain in Eq.~\ref{EqGamma3}. 

With Eq.~\ref{EqGamma3}, which still retains translational symmetry, the inversion of Eq.~\ref{EqVstar2} becomes a deconvolution with an non-symmetric kernel function which could, for example, be obtained from a finite element simulation of the non-symmetric tip. As we will see later, Eq.~\ref{EqGamma3} still allows obtaining surface dipole moments $p$ of nanostructures from integration of \Vstar images (i.e. without prior deconvolution and thus without the necessity of knowing the function \gammaplain). Hence, $p$ is independent of the tip shape, which explains our high reproducibility in the experimentally measured values of this quantity.

The most general situation is reached if we drop the assumption of a planar surface. Then, irrespective of the tip shape, neither Eq.~\ref{EqGamma2} nor Eq.~\ref{EqGamma3} are valid anymore and \gammaplain maintains its full separate dependency on \rpar and \rparprime. In this case, two situations must be distinguished: 

\begin{enumerate}
\item
A surface which is \textit{locally} flat, i.e., has flat terraces with a radius which is at least as large as the decay length of \gammaplain, can still be considered as completely flat. Due to the rapid decay of \gammaplain with $|\rpar-\rparprime|$ local flatness is already fulfilled if the terrace radius exceeds $\approx \unit[10]{nm}$. In this case the inversion of Eq.~\ref{EqVstar2} can still be achieved by deconvolution, one just needs to substitute (example for a symmetric tip) $\gammaplain_{\text{pp/axial}}(|\rpar-\rparprime|,z)$ by $\gammaplain_{\text{pp/axial}}(|\rpar-\rparprime|,z-\teff(\rpar))$, where $\teff(\rpar)$ is the dielectric topographic height of the locally flat region below the QD. Therefore, all that is required is to determine \gammaAx functions for an entire set of surface-QD distances $z-\teff$ (by the methods discussed above) and then select the specific \gammaAx which has been calculated for the respective $z-\teff$ value when deconvolving \Vstar on a locally flat sample region of height \teff. For example, this procedure allows for a correct deconvolution of \Vstar for nanostructures on different terraces of a stepped sample surface, where \teff varies from terrace to terrace.
\item
Under the assumption that \teff varies substantially on length scales \textit{smaller} than $\approx \unit[10]{nm}$, both axial and translational symmetry are broken and the determination of the Green's function is far from straightforward. In principle, a finite element simulation could solve this problem if the topography is known. In this context, we note that SQDM, in principle, contains information on the dielectric topography in the measurand \alpharel (see Sec.~\ref{sec:GeneralProblem}). Therefore, we will now discuss the inversion of Eq.~\ref{integralalpha} with the goal of estimating both, \gammaplain and \teff, consistently from experimental \alpharel data. While \teff reveals the dielectric topography, the knowledge of \gammaplain potentially improves our recovery of \Phis, via inversion of Eq.~\ref{EqVstar2}, in cases of non-trivial topographies.  
\end{enumerate}

\subsection{Dielectric topography: From \alpharel to \teff}
\label{sec:Dielectrictopography}

\subsubsection{Statement of the problem}

\noindent While the approximation of a planar surface allows us to obtain $\gammaplain(\rpar,z,\rparprime)$ (even if the surface is not strictly flat) and thus recover \Phis, in contrast any interpretation of lateral variations in \alpharel necessarily requires us to go beyond the assumption of a flat surface, because a for a truly flat surface \alpharel is unity everywhere and therefore contains no information. 

We can again, as in Sec.~\ref{SecBeyondPP}, separately discuss two situations: Firstly, locally flat surfaces where \alpharel and thus \teff are constant in a sufficiently large area, and secondly, surfaces where this is not the case. Both cases can be rationalized as different limits of treating \gammaplain within the framework of high-dimensional model representation (HDMR). In the following section (\ref{sec:HDMR}), we will first introduce HDMR and then discuss various approximations in sections~\ref{sec:ZerothOrderApprox} and \ref{SecFirstOrderApprox}.

\subsubsection{High-dimensional model representation of \gammaplain}
\label{sec:HDMR}

\begin{figure}[!ht]
	\centering
	\includegraphics[width=10cm]{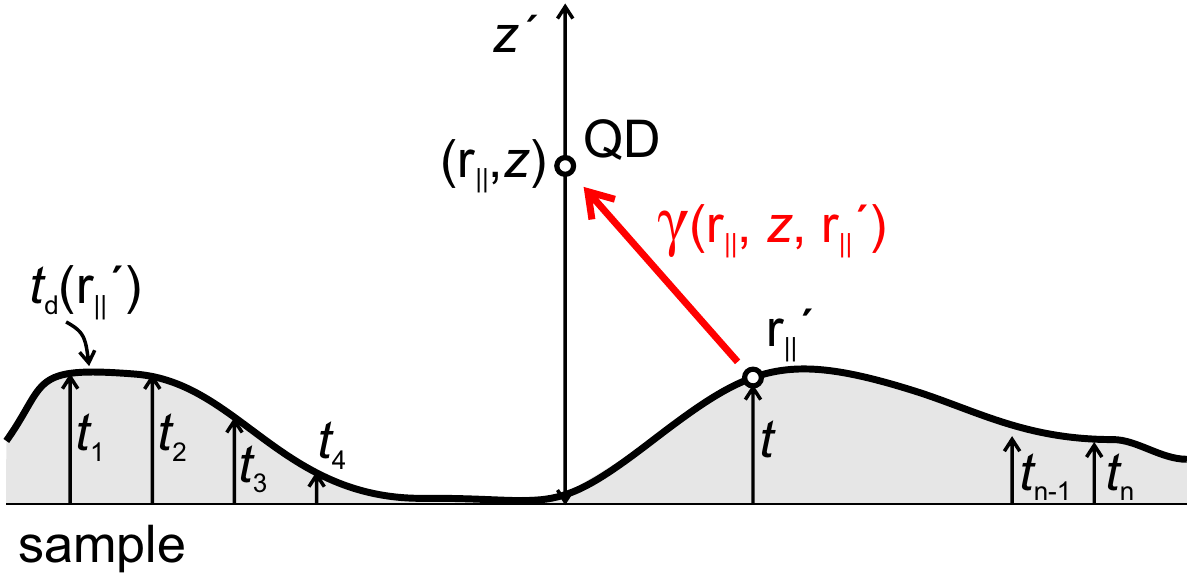}
	\sffamily
	\caption{\footnotesize  \textbf{High dimensional model representation of the sample surface.} The sample surface with dielectric topography $\teff(\rparprime)$ is discretized in 2D in $n$ patches at positions \rparidprime with respective heights $t_i$. \gammaplain is visualized by an arrow for one pair $(\rpar,\rparprime)$.   \normalsize}
	\normalfont 
	\label{Fig_HDMR} 
\end{figure}
\noindent As a consequence of the non-local relation between \teff and \gammaplain, the latter is a functional \hatGamma of the dielectric topography, i.e.,
\begin{equation}
  \gammaplain(\rpar,z,\rparprime,\teff(\rparprime))=\hatGamma(\rpar,z,\rparprime,\teff(\rparprime))[\teff(\rpardprime)].
  \label{gamma_as_functional}
\end{equation}
For simplicity we define $\teff \equiv \teff(\rparprime)$.  With this, Eq.~\ref{integralalpha1} becomes  
\begin{equation}
	\alpharel(\rpar,z)= \frac{1}{\alpha_0(z)}\iint\limits_\text{surface} \hatGamma(\rpar,z,\rparprime,\teff)[\teff(\rpardprime)]d^2\rparprime.
\label{EqAlphaRelFromFunctional}
\end{equation}
The task of recovering the surface topography means to invert $\alpharel(\rpar,z)$ for \teff. This is a special case of a widespread class of problems, namely finding the relationship between a high-dimensional input $(t_1, ...\,t_n)$ and the output $f(t_1, ...\,t_n)$. In high-dimensional model representation (HDMR), the output $f(t_1, ...\,t_n)$ is represented as a hierarchical correlated function expansion in terms of the input variables $(t_1, ...\,t_n)$ \cite{Rabitz1999}:   
\begin{align*}
	 f(t_1, ...\,t_n)=f_0 & +\sum\limits_{i=1}^n f_i(t_i)+\sum\limits_{1\leq i < j \leq n} f_{ij}(t_i, t_j) + \sum\limits_{1\leq i < j <k \leq n} f_{ijk}(t_i, t_j, t_k)\\
	 & + \dots + f_{12\dots n}(t_1, t_2, \dots t_n) \numberthis \label{EqHDMR3}
\end{align*}
The general advantage of HDMR is the high convergence. This is achieved by regrouping the standard Taylor expansion \cite{Rabitz1999}.

Within the HDMR framework we can write the functional \hatGamma as
\begin{align*}
\hatGamma(\rpar,\zz,\rparprime,\teff)\left[\teff(\rpardprime)\right]  & = f_0(\rpar,\zz,\rparprime,\teff) \\ &
+ \sideset{}{'}\sum_{i=1}^{n} f_i(\rpar,\zz,\rparprime,\teff,t_i) +\sideset{}{'}\sum_{1\leq i\le j \leq n}  f_{ij}(\rpar,\zz,\rparprime,\teff,t_i,t_j) + \ldots \numberthis \label{EqHDMR2}
\end{align*}
Here, we have discretized $\teff(\rpardprime)$ on the entire surface to $n$ values $t_i$ at the respective positions \rparidprime, i.e., $t_i \equiv \teff(\rparidprime)$. Note that we do not need to provide the positions $\rparidprime$ as arguments in Eq.~\ref{EqHDMR2}, since the different positions of the discrete $t_i$ are encoded in their respective functions $f_i$, $f_{ij}$ etc. Moreover, since a topographic feature cannot screen itself, the cases of $\rparidprime = \rparprime$, $\boldsymbol{\text{r}}_{||j}'' = \rparprime$, $\ldots$ have to be excluded from the sums. This is indicated by a prime at the sums in Eq.~\ref{EqHDMR2}.

\begin{figure}[!ht]
	\centering
	\includegraphics[width=11cm]{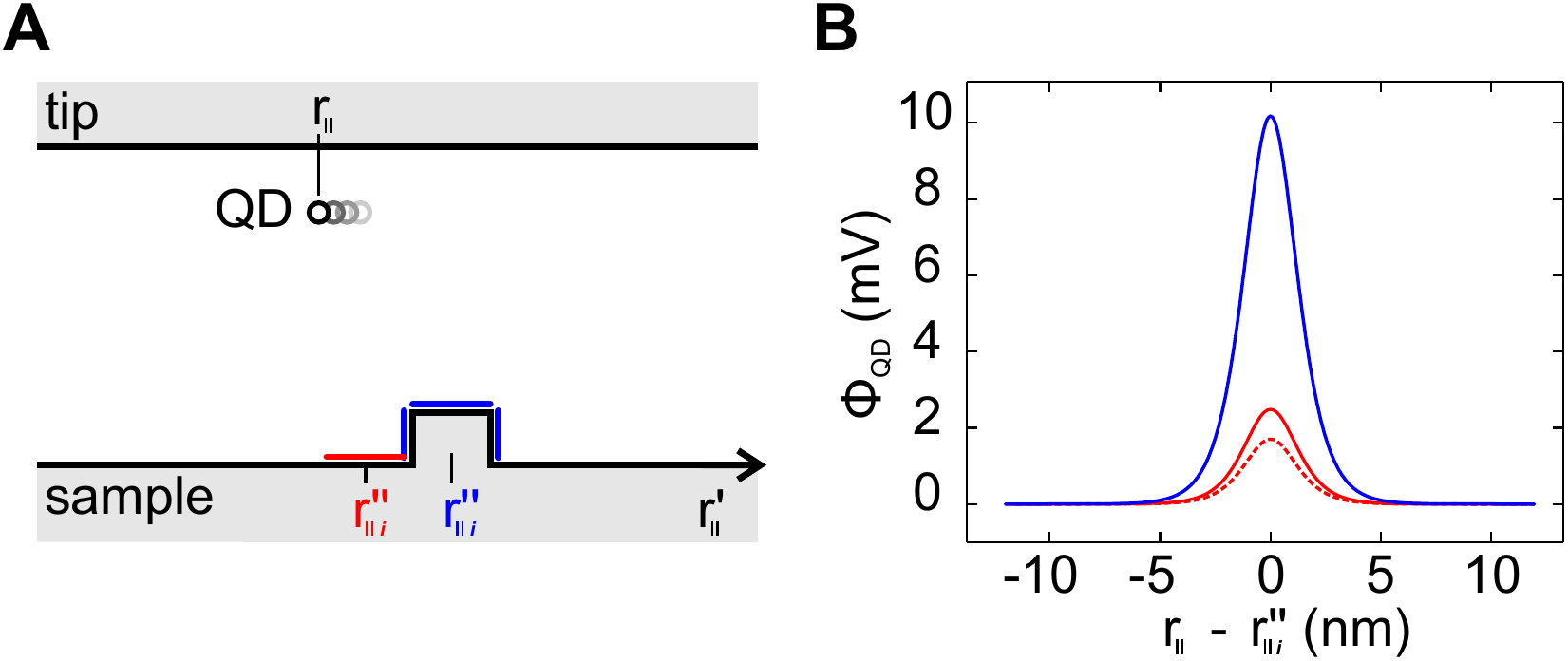}
	\sffamily
	\caption{\footnotesize  \textbf{Effect of the sample topography on the Green's function. A finite element simulation.} \textbf{A}~Illustration of the finite element (FE) simulation setup: A small cuboid-shaped nanostructure ($\unit[3.5 \times 3.5 \times 2]{\text{\AA}^3}$) on an otherwise flat surface. Tip and sample ($z=\unit[23]{\text{\AA}},d=\unit[7]{\text{\AA}}$) are kept at zero potential except for a potential of $\unit[1]{V}$ which is \textit{either} applied to the patch marked with a red line \textit{or} to the cuboid (blue lines) in the FE simulation. \textbf{B}~Simulated potential $\PhiQD(\rpar)\propto \gammaplain(\rpar,z,\rparidprime)$ as the QD is scanned across the cuboid-shaped nanostructure in panel A. For the blue (red) curve, the nanostructure (patch) at \rparidprime is biased. The solid red curve is simulated in the \textit{absence} of the blue cuboid protrusion, i.e., disregarding its screening influence, while the dashed red curve is obtained \textit{with} the cuboid (at zero potential) included in the FE simulation. This exemplifies how nanostructures at the surface screen or enhance the Green's function (cf. Fig.~\ref{Fig_SI_A}). \normalsize}
	\normalfont 
	\label{Fig_FirstOrderApprox1} 
\end{figure}

An intuitive physical interpretation of the terms in Eq.~\ref{EqHDMR2} is possible if we define the (flat) part of the sample surface in a reference area (in which $\alpha_0$ has been defined) as the origin of the $z$ axis and all heights \teff accordingly. Then, $f_0$ is the contribution of the topographic object of height \teff at \rparprime on an otherwise completely flat surface with $t_i=0\ \forall\ \rparidprime \neq \rparprime$ because
\begin{equation}
f_i(\rpar,\zz,\rparprime,\teff,t_i = 0) = 0,
\label{Eq_66}
\end{equation}
\begin{equation}
f_{ij}(\rpar,\zz,\rparprime,\teff,t_i = 0,t_j = 0) = 0,
\end{equation}
and so on.
The second term in Eq.~\ref{EqHDMR2} includes first order screening effects where the contribution of the feature at \rparprime is modified by the presence of each of the non-zero topography elements $t_i$ individually. Likewise, the third term describes how two topographic elements $t_i$ and $t_j$ affect each other in the screening of the feature at \rparprime. A quantitative example of first order screening is shown in Fig.~\ref{Fig_FirstOrderApprox1}.

The aspect of non-locality means that all terms beyond $f_0$ are, in principle, needed for the correct description of \hatGamma, but at the same time we expect their contributions to decrease with increasing number of parameters. Therefore, we ignore terms beyond $f_i$ and proceed in two steps. First, we introduce the zeroth-order approximation in which we only include $f_0$. Then, we introduce the first-order approximation to include non-local screening to lowest order into the formalism.

\subsubsection{Zeroth-order approximation}
\label{sec:ZerothOrderApprox}

Considering a single zero-dimensional nanostructure in an otherwise flat region of the surface, we apply the zeroth-order approximation. This simplifies Eq.~\ref{EqHDMR2} to
\begin{equation}
\hatGamma(\rpar,z,\rparprime,\teff)\left[\teff(\rpardprime)\right] = f_0(\rpar,z,\rparprime,\teff).
\label{EqHDMRZeroOrder}
\end{equation}
Inserting this into Eq.~\ref{EqAlphaRelFromFunctional} one obtains a local description of the gating efficiency 
\begin{equation}
\alpharel(\rpar,z) = \frac{1}{\alpha_0(z)}\iint\limits_\text{surface} f_0\left(\rpar,z,\rparprime,\teff\right)d^2\rparprime.
\label{EqZeroOrder1}
\end{equation}
Note that even in this simple case of a zero-dimensional nanostructure, the use of the zeroth-order approximation is questionable as we will discuss at the end of Sec.~\ref{SecFirstOrderApprox}.

To linearize the integrand of Eq.~\ref{EqZeroOrder1} in \teff, we perform a Taylor expansion of $f_0$ around $\teff=0$ and obtain 
\begin{equation}
f_0(\rpar,z,\rparprime,\teff) \approx f_0(\rpar,z,\rparprime,0) + \frac{\partial f_0(\rpar,z,\rparprime,\teff)}{\partial \teff}\Bigr|_{\teff=0} \teff.
\label{EqTaylorExp}
\end{equation}
If we insert this result into Eq.~\ref{EqZeroOrder1}, we get
\begin{equation}
\alpharel(\rpar,z) = \frac{1}{\alpha_0(z)}\iint\limits_\text{surface} f_0(\rpar,z,\rparprime,0) d^2\rparprime + \frac{1}{\alpha_0(z)}\iint\limits_\text{surface} \frac{\partial f_0(\rpar,z,\rparprime,\teff)}{\partial \teff}\Bigr|_{\teff=0} \teff(\rparprime) d^2\rparprime.
\label{EqZeroOrder2}
\end{equation}
By assuming a completely planar surface with $\teff = 0$ everywhere, the second integral in Eq.~\ref{EqZeroOrder2} vanishes. A comparison of Eq.~\ref{EqZeroOrder2} (also assuming an axially symmetric tip) with Eq.~\ref{integralalpha1} then yields
\begin{equation}
f_0(\rpar,z,\rparprime,0) = \gammaAx(|\rpar-\rparprime|,z).
\label{f0_flat}
\end{equation}
Since $f_0(\rpar,z,\rparprime,0)$ is independent of \teff, Eq.~\ref{f0_flat} is also valid for surfaces with $\teff \neq 0$.
Because of 
\begin{equation}
\alpha_0(z) = \iint\limits_\text{surface} \gammaAx(|\rpar-\rparprime|,z)  d^2\rparprime = \iint\limits_\text{surface} f_0(\rpar,z,\rparprime,0)  d^2\rparprime,
\label{EqAlphaZero}
\end{equation}
(Eq.~\ref{integralalpha} and definition of $\alpha_0$) we obtain the zeroth-order approximation for \alpharel
\begin{equation}
\alpharel(\rpar,z) = 1 + \frac{1}{\alpha_0(z)}\iint\limits_\text{surface} \frac{\partial f_0(\rpar,z,\rparprime,\teff)}{\partial \teff}\Bigr|_{\teff=0} \teff(\rparprime)  d^2\rparprime.
\label{EqTaylorZeroOrder}
\end{equation}

If Eq.~\ref{EqTaylorZeroOrder} is to be employed for determining $\teff(\rparprime)$ for a given \alpharel, the integral kernel 
\begin{equation}
\gammaTopo\left(\rpar,z,\rparprime,\teff\right) \equiv \frac{\partial f_0(\rpar,z,\rparprime,\teff)}{\partial \teff}\Bigr|_{\teff=0}
\label{def_gamma_topo}
\end{equation}
needs to be known.
We discuss shape and norm of this kernel separately. In the zeroth-order approximation we can write, in analogy to Eq.~\ref{f0_flat},
\begin{equation}
\frac{\partial f_0(\rpar,z,\rparprime,\teff)}{\partial \teff} = \frac{\partial f_0(|\rpar-\rparprime|,z,\teff)}{\partial \teff}.
\label{EqRelativeDistance}
\end{equation}
We obtain the shape of \gammaTopo from the consideration that a zero-dimensional topographic feature is a polarizable object. In the homogeneous gating field above the otherwise flat sample (as assumed in the zeroth-order approximation) it therefore behaves as a local dipole. This dipole is analogous to a point-like deformation of the surface potential (see Sec.~\ref{sec:PPapprox}) $\delta\Phis(\rparprime) = \beta \Vb \delta(\rparprime-\rparprimeZero)$ on a homogeneous background \Vb, where $\phi = \beta \Vb$ is the potential due to the local dipole on a flat surface. Inserting this into Eq.~\ref{PhiGF3}, we obtain
\begin{equation}
\PhiQD(\rpar,z) = \iint\limits_\text{surface} \gammaAx(|\rpar-\rparprime|,z)(\Vb + \Vb \beta \delta(\rparprime-\rparprimeZero)) d^2\rparprime +\PhiTip,
\label{EqShape1}
\end{equation}
where we have moreover used Eq.~\ref{EqGamma2} which is valid because we have abstracted the zero-dimensional topographic feature as a pure dipole potential on a flat surface. The integral over the second term in the integrand (Eq.~\ref{EqShape1}) collapses at $\rparprime=\rparprimeZero$ such that we obtain
\begin{equation}
\PhiQD(\rpar,z) = \alpha_0(z)\Vb  + \Vb\beta \gammaAx(|\rpar-\rparprimeZero|,z) +\PhiTip.
\label{EqShape}
\end{equation}
If we instead consider the original case of a bias \Vb homogeneously applied to a surface with a zero-dimensional topographic feature and $\Phis = 0$ everywhere, we find (Eq.~\ref{PhiGF3})
\begin{equation}
\PhiQD(\rpar,z) =  \Vb\iint\limits_\text{surface} \gammaplain(\rpar,z,\rparprime,\teff) d^2\rparprime +\PhiTip = \alpha(z) \Vb + \PhiTip.
\label{EqShape2}
\end{equation}
Since (Eqs.~\ref{EqTaylorExp}, \ref{f0_flat}, \ref{def_gamma_topo}, \ref{EqRelativeDistance})
\begin{equation}
\gammaplain = \gammaAx(|\rpar-\rparprime|,z) + \gammaTopo(|\rpar-\rparprime|,z)\teff\delta(\rparprime-\rparprimeZero),
\label{EqShape3}
\end{equation}
we obtain from Eq.~\ref{EqShape2} with Eq.~\ref{EqAlphaZero}
\begin{equation}
\PhiQD(\rpar,z) =  \alpha_0(z)\Vb + \Vb \teff \gammaTopo(|\rpar-\rparprimeZero|,z,\teff) +\PhiTip.
\label{EqShape4}
\end{equation}
A comparison of Eqs.~\ref{EqShape} and Eq.~\ref{EqShape4} reveals that the integral kernel \gammaTopo which determines the influence of a topographic feature on $\alpha$ is proportional to \gammaAx, i.e., proportional to the kernel that determines the influence of a local deformation of the potential on \PhiQD,
\begin{equation}
\gammaTopo(|\rpar-\rparprime|,z) \propto \gammaAx(|\rpar-\rparprime|,z).
\label{EqZeroOrderTopoKernel}
\end{equation}
Thus, in the present approximation also \gammaTopo has axial symmetry.

\begin{figure}[!ht]
	\centering
	\includegraphics[width=11cm]{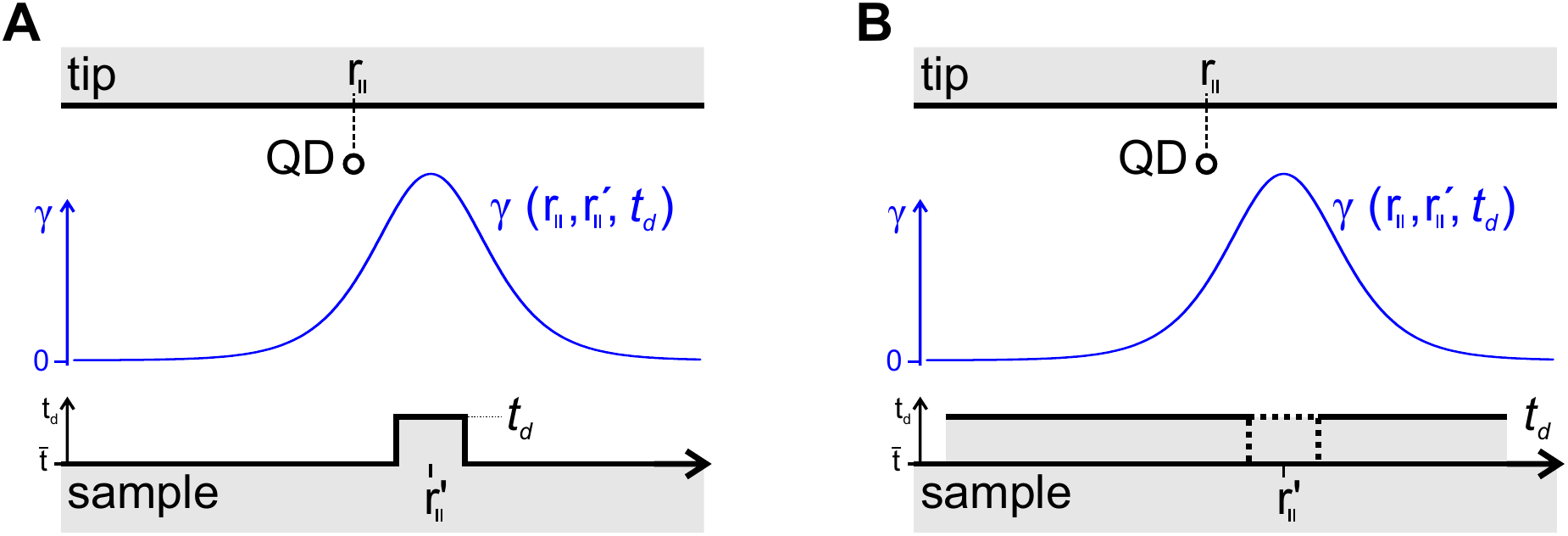}
	\sffamily
	\caption{\footnotesize  \textbf{Norm of \gammaTopo in the zeroth-order approximation.} The norm of \gammaTopo and thus also \gammaplain itself for a nanostructure of height \teff is the same, no matter whether the nanostructure is isolated on the surface (panel \textbf{A}) or part of an extended terrace (panel \textbf{B}). \normalsize}
	\normalfont 
	\label{Fig_ZerothOrderApprox} 
\end{figure}

The norm of \gammaTopo in the zeroth-order approximation quantifies the influence on \alpharel of a single zero-dimensional feature on a flat surface. Several methods of determining the norm of \gammaTopo are conceivable. For example, one could perform a finite element (FE) analysis. In fact, such an analysis with $z = \unit[24]{\text{\AA}}$, $d=\unit[7]{\text{\AA}}$ and a cuboid-shaped nanostructure with a lateral diameter of $\unit[3.5]{\text{\AA}}$ (Fig.~\ref{Fig_FirstOrderApprox1}) fully confirms the result Eq.~\ref{EqZeroOrderTopoKernel} regarding the shape of the kernel function \gammaTopo. However, due to the simplifications inherent in the zeroth-order approximation (Sec.~\ref{sec:HDMR}), using the norm of \gammaTopo that is correct for the case of a single zero-dimensional nanostructure will under practically no circumstances yield correct deconvolution results for \teff for realistic surfaces. 

Therefore, we propose the use of a different norm for \gammaTopo, constructed such that the correct height \teff of extended terraces is recovered from the deconvolution of \alpharel according to Eq.~\ref{EqTaylorZeroOrder}. This implies the assumption that a single isolated nanostructure has the same influence on the QD as the "nanostructure" within a matrix of material (Fig.~\ref{Fig_ZerothOrderApprox}). We note that this approximation constitutes an inconsistency in the zeroth-order approximation (see below). To calculate the norm of \gammaTopo in this approximation, we consider \alpharel on two extended surface terraces and denote the \alpharel functions on the two terraces as $\alpharel^\text{upper}$ and $\alpharel^\text{lower}$, respectively. The height difference between the two terraces is \teff. Then we take advantage of the fact that 
\begin{equation}
\alpharel^\text{upper}(\rpar,z_0+\teff) = \alpharel^\text{lower}(\rpar,z_0).
\label{arel_t_1}
\end{equation}
This equation is valid because the constant offset \teff of the complete dielectric topography from the lower to the upper terrace can be compensated by a corresponding offset of the vertical position $z$ of the sensor.  
We expand
\begin{equation}
\alpharel^\text{upper}(\rpar,z) \approx \alpharel^\text{upper}(\rpar,z_0)-g\times(z-z_0),
\label{arel_t_2}
\end{equation}
where 
\begin{equation}
g \equiv-\frac{\partial \alpharel^\text{upper}(\rpar,z)}{\partial z}\Big|_{z_0}=-\frac{\partial \alpharel^\text{lower}(\rpar,z)}{\partial z}\Big|_{z_0-\teff}.
\label{arel_t_2a}
\end{equation}
Because $\alpha$ and thus also $\alpharel$ increases with decreasing tip-sample separation, $g>0$. 
Setting $z=z_0+\teff$ in Eq.~\ref{arel_t_2}, we obtain
\begin{equation}
\alpharel^\text{upper}(\rpar,z_0+\teff)  \approx \alpharel^\text{upper}(\rpar,z_0)-g\teff.
\label{arel_t_3}
\end{equation}
With Eq.~\ref{arel_t_1} this becomes
\begin{equation}
\alpharel^\text{lower}(\rpar,z_0) \approx \alpharel^\text{upper}(\rpar,z_0)-g\teff.
\label{arel_t_4}
\end{equation}
If we assume without loss of generality that $\alpha_0$ is defined on the lower terrace, then by definition $\alpharel^\text{lower}(z_0)=\frac{\alpha^{\text{lower}}(z_0)}{\alpha_0(z_0)}=1$ and Eq.~\ref{arel_t_4} becomes
\begin{equation}
\alpharel^\text{upper}(z_0)-1 \approx g\teff.
\label{arel_t_6}
\end{equation}
On the other hand, according to Eq.~\ref{EqTaylorZeroOrder}, the gating efficiency of a flat terrace of empty surface \teff above the reference terrace is
\begin{equation}
\alpharel^\text{upper}(z)-1 = \frac{\teff}{\alpha_0(z)}\iint\limits_\text{surface} \frac{\partial f_0(|\rpar-\rparprime|,z,\teff)}{\partial \teff}\Bigr|_{\teff=0}  d^2\rparprime.
\label{arel_t_7}
\end{equation}
Comparing eqs.~\ref{arel_t_6} and \ref{arel_t_7}, and using Eqs.~\ref{def_gamma_topo} and \ref{EqRelativeDistance}, we find in the zeroth-order approximation
\begin{equation}
\iint\limits_\text{surface} \gammaTopo(|\rpar-\rparprime|,z_0)d^2\rparprime \approx  g \alpha_0(z_0).
\label{EqLinearDeconvolute3}
\end{equation}
In other words, in the zeroth-order approximation the norm of the integral kernel \gammaTopo is equal to $g \alpha_0(z_0)$, where according to its definition in Eq.~\ref{arel_t_2a}, $g$ can be measured by a calibration experiment above the empty surface. Note that $g$ varies from tip to tip \cite{Temirov2018} and thus needs to be determined for each tip and $z_0$ separately. Typical experimental $g$ values are in the range of $\unit[0.03]{\text{\AA}^{-1}}$. We thus have with Eqs.~\ref{EqZeroOrderTopoKernel} and \ref{EqLinearDeconvolute3}
\begin{equation}
\gammaTopo(|\rpar-\rparprime|,z_0) \approx  g \gammaAx(|\rpar-\rparprime|,z_0).
\label{EqLinearDeconvolute4}
\end{equation}

It was our goal to derive an expression for \gammaplain which we have achieved now in the zeroth-order approximation. With Eqs.~\ref{gamma_as_functional}, \ref{EqHDMRZeroOrder}, \ref{EqTaylorExp}, \ref{f0_flat}, \ref{EqZeroOrderTopoKernel}, and \ref{EqLinearDeconvolute4} we arrive at
\begin{equation}
\begin{aligned}
\gammaplain(\rpar,z,\rparprime,\teff)&\overset{\text{(\ref{gamma_as_functional})}}{=}\hatGamma(\rpar,z,\rparprime,\teff)\left[\teff(\rpardprime)\right] \\ 
&\overset{\text{(\ref{EqHDMRZeroOrder})}}{=} f_0(\rpar,z,\rparprime,\teff)\\
&\overset{\text{(\ref{EqTaylorExp},\ref{f0_flat})}}{=} \gammaAx(|\rpar-\rparprime|,z) + \gammaTopo(|\rpar-\rparprime|,z)\teff(\rparprime)\\ 
&\overset{\text{(\ref{EqZeroOrderTopoKernel},\ref{EqLinearDeconvolute4})}}{\approx} \gammaAx(|\rpar-\rparprime|,z) + g\gammaAx(|\rpar-\rparprime|,z)\teff(\rparprime)\\ 
&= (1+g\teff(\rparprime))\gammaAx(|\rpar-\rparprime|,z).
\end{aligned}
\label{ZeroOrderApproxFinalResult}
\end{equation}
With Eq.~\ref{EqAlphaRelFromFunctional}, this result leads to a straightforward relationship between \alpharel and \teff, namely 
\begin{equation}
\alpharel(\rpar,z) = \frac{1}{\alpha_0(z)}\iint\limits_\text{surface}(1+g\teff(\rparprime))\gammaAx(|\rpar-\rparprime|,z) d^2\rparprime,
\label{ZeroOrderApproxAlphaResult}
\end{equation}
which allows for the deconvolution of \alpharel images of obtain $\teff(\rparprime)$ maps. However, as stated above, the way in which the norm of \gammaTopo was determined in this section is in fact inconsistent. In the following section we will therefore develop one possible first-order approximation of \hatGamma which removes this inconsistency.

\subsubsection{First-order mean field approximation}
\label{SecFirstOrderApprox}

\noindent For the first-order approximation we truncate the HDMR expression Eq.~\ref{EqHDMR2} after the second term. Hence, we need to specify the functions $f_i$. While it is not impossible to calculate the $f_i(\rpar,z,\rparprime,\teff,t_i)$ and thus \gammaplain for a large number of $\rparidprime$ and $t_i$ values via a set of FE simulations, this is cumbersome.

To simplify the $f_i$, we introduce an effective height $\tBar(\rparprime)$, taking a weighted average over the $t_i$, with weights that decrease with increasing lateral distance $|\rparidprime-\rparprime|$. Because $t_i = \tBar \ \forall i$, this approach resembles a mean field approximation. The sum over all $f_i$ can be expressed as a single function $\bar{f}$, 
\begin{equation}
\bar{f}\left(\rpar,z,\rparprime,\teff,\tBar\right) \equiv \sideset{}{'}\sum_{i=1}^{n}f_i(\rpar,z,\rparprime,\teff,t_i),
\label{EqDerjaguinApprox}
\end{equation}
and thus 
\begin{equation}
\hatGamma(\rpar,z,\rparprime,\teff)\left[\teff(\rpardprime)\right] = f_0(\rpar,z,\rparprime,\teff) + \bar{f}(\rpar,z,\rparprime,\teff,\tBar).
\end{equation}
With the introduction of a flat effective topography \tBar we have practically eliminated the difference in complexity between $f_0$ and $\bar{f}$. The term $f_0$ describes the case $\tBar=0$ ($\bar{f}(\tBar=0)=0$), while $\bar{f}$ expresses all changes that occur when \tBar takes on other values. Hence, we can combine both terms into a single function $\fm(\rpar,z,\rparprime,t,\tBar)$ which reduces to $f_0$ for $\tBar=0$:
\begin{equation}
\hatGamma(\rpar,z,\rparprime,\teff)\left[\teff(\rpardprime)\right] = \fm(\rpar,z,\rparprime,\teff,\tBar)
\label{gammaHatFirstOrder}
\end{equation}
Note that both \teff and \tBar are functions of \rparprime. Here, \fm (m stands for mean field) describes \gammaplain for a local topographic feature of height \teff at \rparprime on an infinite plane of height \tBar.

We expand Eq.~\ref{gammaHatFirstOrder}  around $\teff=0$ and $\tBar=0$
\begin{equation}
\fm(\rpar,\zt,\rparprime,\teff,\tBar) \approx \fm(\rpar,\zt,\rparprime,0,0) + \frac{\partial \fm}{\partial \teff}\Bigr|_{\substack{\teff=0\\ \tBar=0}} \teff + \frac{\partial \fm}{\partial \tBar}\Bigr|_{\substack{\teff=0\\ \tBar=0}} \tBar.
\label{EqTaylorExpFirstOrder}
\end{equation}
Since the first two terms of this expansion are identical to Eq.~\ref{EqTaylorExp}, the third term must contain the non-local screening effects. With the following considerations we bring this term into a more intuitive form. First, we express the derivatives as differences for an infinitely small $\epsilon$,
\begin{equation}
\frac{\partial \fm}{\partial \teff}\Bigr|_{\substack{\teff=0\\\tBar=0}} = \frac{\fm(\rpar,z,\rparprime,\epsilon,0)-\fm(\rpar,z,\rparprime,-\epsilon,0)}{2\epsilon} 
\label{EqEps1}
\end{equation}
and
\begin{equation}
\frac{\partial \fm}{\partial \tBar}\Bigr|_{\substack{\teff=0\\\tBar=0}}=\frac{\fm(\rpar,z,\rparprime,0,\epsilon)-\fm(\rpar,z,\rparprime,0,-\epsilon)}{2\epsilon}.
\label{EqEps2}
\end{equation}
Then, we realize that the two cases $\teff=\epsilon,\tBar=0$ and $\teff=0,\tBar=-\epsilon$ describe the same topography with the sole difference that the tip-QD separation has increased to $z = z+\epsilon$ in the latter case. Hence we can write
\begin{equation}
\fm(\rpar,z,\rparprime,0,\epsilon) = \fm(\rpar,z-\epsilon,\rparprime,-\epsilon,0) =\fm(\rpar,z,\rparprime,-\epsilon,0) - \frac{\partial \fm}{\partial z}\epsilon
\label{EqEps3}
\end{equation}
and
\begin{equation}
\fm(\rpar,z,\rparprime,0,-\epsilon) = \fm(\rpar,z+\epsilon,\rparprime,\epsilon,0)= \fm(\rpar,z,\rparprime,\epsilon,0) + \frac{\partial \fm}{\partial z}\epsilon
\label{EqEps4}
\end{equation}
If we insert Eqs.~\ref{EqEps3} and \ref{EqEps4} into Eq.~\ref{EqEps2} we obtain
\begin{equation}
\frac{\partial \fm}{\partial \tBar}\Bigr|_{\substack{\teff=0\\ \tBar=0}}=\frac{\fm(\rpar,z,\rparprime,-\epsilon,0)-\fm(\rpar,z,\rparprime,\epsilon,0)}{2\epsilon}-\frac{\partial \fm}{\partial z}.
\end{equation}
A comparison with Eq.~\ref{EqEps1} reveals that the first term on the right hand side is simply $-\partial \fm / \partial \teff$, such that we get
\begin{equation}
\frac{\partial \fm}{\partial \tBar}\Bigr|_{\substack{\teff=0\\\tBar=0}}=-\frac{\partial \fm}{\partial \teff}\Bigr|_{\substack{\teff=0\\\tBar=0}}-\frac{\partial \fm}{\partial z}\Bigr|_{\substack{\teff=0\\\tBar=0}}.
\end{equation}
Inserting this into Eq.~\ref{EqTaylorExpFirstOrder} turns the Taylor expansion into
\begin{equation}
\fm(\rpar,\zt,\rparprime,\teff,\tBar) \approx  \fm(\rpar,z,\rparprime,0,0) + \frac{\partial \fm}{\partial \teff}\Bigr|_{\substack{\teff=0\\ \tBar=0}} (\teff-\tBar) - \frac{\partial \fm}{\partial z}\Bigr|_{\substack{\teff=0\\ \tBar=0}} \tBar.
\label{GammaFirstOrderApprox}
\end{equation}

The third term in Eq.~\ref{GammaFirstOrderApprox} describes the influence of an extended terrace (for which the second term vanishes because $\tBar = \teff$) on \gammaplain , while the second term describes the effect of isolated protrusions located on a terrace where $\tBar = 0$. Eq.~\ref{GammaFirstOrderApprox} therefore eliminates the inconsistency present in the zeroth-order approximation (Eq.~\ref{ZeroOrderApproxAlphaResult}), where a single term was used to describe both, isolated topographic features and terraces. For situations of several nanostructures which screen each other, Eq.~\ref{GammaFirstOrderApprox} interpolates between the two limits of a single nanostructure and a terrace. In this respect, the first-order approximation is the simplest possible consistent formulation for \gammaplain.

The first and the third terms of the sum in Eq.~\ref{GammaFirstOrderApprox}, in combination, are simply \gammaAx as calculated for the surface-QD distance $z-\tBar$ instead of $z$, such that we obtain
\begin{equation}
\gammaplain(\rpar,z,\rparprime,\teff) = \gammaAx(|\rpar-\rparprime|,z-\tBar(\rparprime))  + \frac{\partial \fm}{\partial \teff}\Bigr|_{\substack{\teff=0\\ \tBar=0}} (\teff-\tBar).
\label{GammaFirstOrderApprox3}
\end{equation} 
The second term in Eq.~\ref{GammaFirstOrderApprox3} describes a single protrusion (or depression) of height $\teff-\tBar$ on a flat surface. Hence, by analogy with Eq.~\ref{def_gamma_topo} $\partial \fm / \partial \teff$ has the shape of \gammaTopo (and \gammaAx, Eq.~\ref{EqZeroOrderTopoKernel}), while its amplitude in the first-order mean field approximation remains to be determined (Eq.~\ref{EqLinearDeconvolute3} has been derived in the zeroth-order approximation and is not valid here). Therefore, Eq.~\ref{GammaFirstOrderApprox3} simplifies to
\begin{equation}
\gammaplain(\rpar,z,\rparprime,\teff(\rparprime)) = \gammaAx(|\rpar-\rparprime|,z-\tBar(\rparprime))  + \gammaTopo(|\rpar-\rparprime|,z) (\teff(\rparprime)-\tBar(\rparprime)).
\label{GammaFirstOrderApprox2}
\end{equation}
For clarity, in this equation we have made the dependence of both \teff and \tBar on \rparprime explicit.
Inserting this into Eq.~\ref{EqAlphaRelFromFunctional}, we obtain a description of the gating efficiency in the first-order mean field approximation,
\begin{equation}
\alpharel(\rpar,z)= \frac{1}{\alpha_0(z)}\iint\limits_\text{surface} \gammaAx(|\rpar-\rparprime|,z-\tBar(\rparprime)) d^2\rparprime + \frac{1}{\alpha_0(z)}\iint\limits_\text{surface}  \gammaTopo(|\rpar-\rparprime|,z) (\teff(\rparprime)-\tBar(\rparprime))d^2\rparprime.
\label{EqAlphaRelFromFunctional1}
\end{equation}

While in the second integral the dependence on $\tBar(\rparprime)$ is explicit, in the first term it is still implicit in the dependence of \gammaAx on $z-\tBar(\rparprime)$. For a constant \tBar,  the  first integral, i.e. the norm of \gammaAx, is given by  (Eqs.~\ref{arel_t_3}, \ref{arel_t_4}, \ref{arel_t_6})
\begin{equation}
\alpha(\rpar,z-\tBar) = (1+g\tBar)\alpha_0(z).
\label{EqAlphaRelFromFunctional1d}
\end{equation}
but for an arbitrary function $\tBar(\rparprime)$ the value of the first integral in Eq.~\ref{EqAlphaRelFromFunctional1} is unclear. We introduce a function $\gammaAx^0$ the norm of which is always $\alpha_0(z)$, i.e.,
\begin{equation}
\alpha_0(z) = \iint\limits_\text{surface} \gammaAx^0(|\rpar-\rparprime|,z-\tBar(\rparprime)) d^2\rparprime
\label{EqAlphaRelFromFunctional2}
\end{equation}
irrespective of the function $\tBar(\rparprime)$, and write in analogy to Eqs.~\ref{EqAlphaRelFromFunctional1d} and \ref{ZeroOrderApproxFinalResult}
\begin{equation}
\gammaAx(|\rpar-\rparprime|,z-\tBar(\rparprime))=(1+g\tBar(\rparprime)) \gammaAx^0(|\rpar-\rparprime|,z-\tBar(\rparprime))
\label{EqGamma_anlogous_to_alpha}
\end{equation}
Clearly, for constant \tBar, Eq.~\ref{EqAlphaRelFromFunctional1d} follows from Eq.~\ref{EqGamma_anlogous_to_alpha}, and therefore Eq.~\ref{EqGamma_anlogous_to_alpha} yields the correct limit. Note that the argument $z-\tBar(\rparprime)$ in $\gammaAx^0$ is still important (unlike in Eq.~\ref{ZeroOrderApproxFinalResult}), since it influences the shape of $\gammaAx^0$. Specifically, $\gammaAx^0$ becomes narrower for larger \tBar values.  Using Eq.~\ref{EqGamma_anlogous_to_alpha} in Eq.~\ref{EqAlphaRelFromFunctional1}, we obtain 
\begin{equation}
\begin{aligned}
\alpharel(\rpar,z) &= \frac{1}{\alpha_0(z)}\iint\limits_\text{surface} (1+g\tBar(\rparprime)) \gammaAx^0(|\rpar-\rparprime|,z-\tBar(\rparprime)) d^2\rparprime \\
& + \frac{1}{\alpha_0(z)}\iint\limits_\text{surface} \gammaTopo(|\rpar-\rparprime|,z) (\teff(\rparprime)-\tBar(\rparprime))d^2\rparprime.
\end{aligned}
\label{EqTaylorFirstOrderSimple}
\end{equation}
This equation connects to object function $\teff(\rparprime)$ to the image function $\alpharel(\rpar)$ in the first-order mean field approximation. It replaces Eq.~\ref{ZeroOrderApproxAlphaResult} which is valid in the zeroth-order approximation. Note that, not surprisingly,  the first term in Eq.~\ref{EqTaylorFirstOrderSimple}, describing the contribution of the flat effective topography, is analogous to Eq.~\ref{ZeroOrderApproxAlphaResult}, while the second term quantifies the contribution of local protrusions (or depressions) around the flat effective topography.

It is important to note that the introduction of the first-order approximation breaks the axial and translational symmetry, as was already noted in Sec.~\ref{SecBeyondPP}. These symmetries are expressed by the fact that \gammaplain depends on $(|\rpar-\rparprime|,z)$ instead of $(\rpar,z,\rparprime,\teff)$  (Eq.~\ref{EqGamma2}), which implies 
\begin{equation}
\iint\limits_\text{surface} \gammaAx(|\rpar-\rparprime|,z)d^2\rparprime = \iint\limits_\text{imaging plane} \gammaAx(|\rpar-\rparprime|,z)d^2\rpar.
\label{EqTranslationalInvariance}
\end{equation}
Because axial and translational symmetry are broken, this equation is not valid for the \gammaplain in Eq.~\ref{GammaFirstOrderApprox2}.

\begin{figure}[!ht]
	\centering
	\includegraphics[width=11cm]{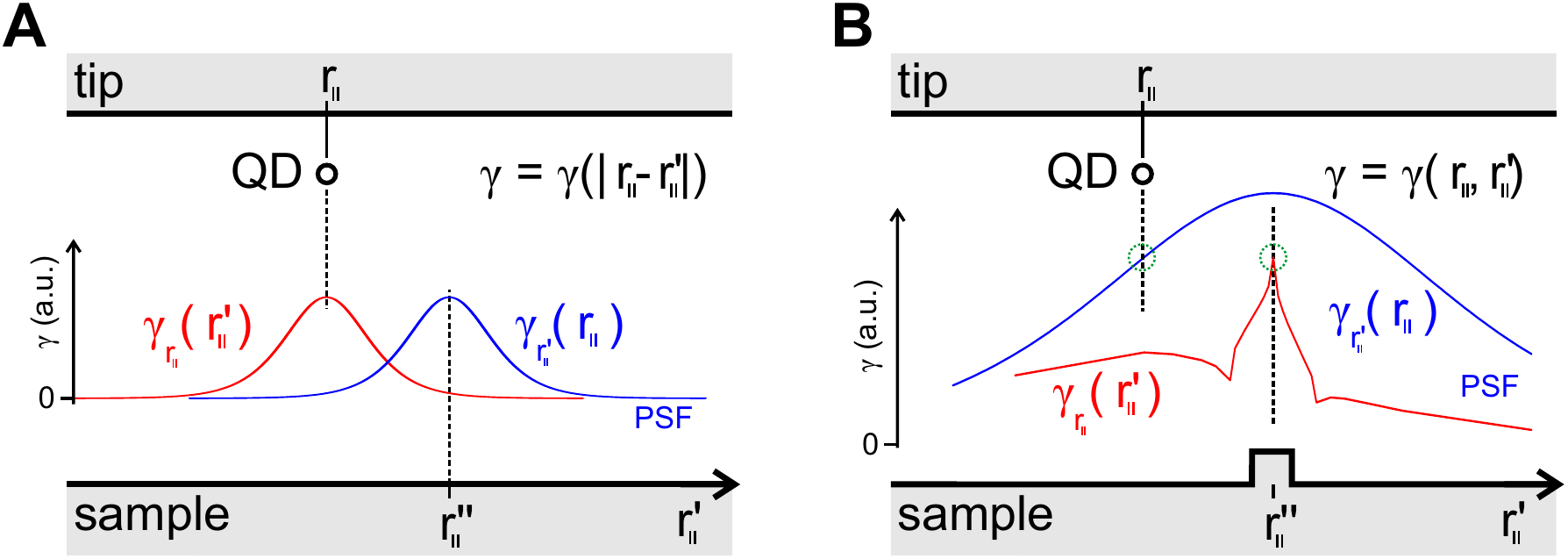}
	\sffamily
	\caption{\footnotesize \textbf{Symmetries of \gammaplain} \textbf{A} For a flat surface, \gammaplain depends only on the distance $|\rpar-\rparprime|$ and has therefore translational and axial symmetry with respect to \rpar and \rparprime. Its norm is uniquely defined (see text). \textbf{B} If a nanostructure at \rpardprime breaks the symmetry of the surface, also \gammaplain loses its symmetries. Shown are two cuts through $\gammaplain(\rpar,\rparprime)$ where either \rpar or \rparprime is held constant (FE simulated curves; scaling different from panel A). Encircled in green are the points where the two cuts intersect and which have hence identical \gammaplain values. The blue curve illustrates that \gammaplain retains its axial symmetry for $\rparprime=\rpardprime$ and in the absence of other nanostructures. The red curve illustrates how the local surface potential is screened or enhanced by a nanostructure (cf. Fig.~\ref{Fig_FirstOrderApprox1}.) \normalsize}
	\normalfont 
	\label{Fig_TranslationalInvariance} 
\end{figure}

To aid further discussion we introduce the terminology $\gammaplain_{\rpar}(\rparprime) \equiv \gammaplain(\rpar=\text{const},\rparprime)$ and $\gammaplain_{\rparprime}(\rpar) \equiv \gammaplain(\rpar,\rparprime=\text{const})$. In the case of axial and translational symmetry (Fig.~\ref{Fig_TranslationalInvariance}A), the integral of $\gammaplain_{\rpar}(\rparprime)$ (red) over the entire surface, i.e, over all \rparprime, which determines $\alpha(\rpar)$, is equal to the integral of the PSF $\gammaplain_{\rparprime}(\rpar)$ (blue) over \rpar. In the first-order mean field approximation and for a given sensor position \rpar, $\gammaplain(\rpar,z,\rparprime)$  is an explicit function of $\teff(\rparprime)$ and $\tBar(\rparprime)$ and has therefore no inherent symmetries for a surface with an arbitrary dielectric topography (red curve in Fig.~\ref{Fig_TranslationalInvariance}B). The PSF, i.e, $\gammaplain_{\rparprime}(\rpar)$ for any given point \rparprime on the surface (blue curve in Fig.~\ref{Fig_TranslationalInvariance}B), however, maintains its full axial symmetry, because \rpar enters \gammaplain only via the distance $|\rpar-\rparprime|$ (Eq.~\ref{GammaFirstOrderApprox2}). We therefore may determine the amplitude of \gammaTopo in Eqs.~\ref{GammaFirstOrderApprox2} and \ref{EqTaylorFirstOrderSimple} via an integration of $\gammaplain_{\rparprime}(\rpar)$ over \rpar, in spite of the fact that translational symmetry is broken and therefore
\begin{equation}
\iint\limits_\text{surface} \gammaplain(\rpar,z,\rparprime,\teff(\rparprime))d^2\rparprime \neq \iint\limits_\text{imaging plane} \gammaplain(\rpar,z,\rparprime,\teff(\rparprime))d^2\rpar.
\label{EqTranslationalInvariance2}
\end{equation}
As a consequence of Eq.~\ref{EqTranslationalInvariance2}, while \gammaTopo in the zeroth-order approximation has a uniquely defined norm (see Eq.~\ref{EqTranslationalInvariance}), this is not true in the first-order mean field approximation. Therefore, we calculate its integral over \rpar 
\begin{equation}
A = \iint\limits_\text{imaging plane} \gammaTopo(|\rpar-\rparprime|,z) d^2\rpar
\label{EqAmplitudeA}
\end{equation}
and refer to it as amplitude $A$.

\begin{figure}[!ht]
	\centering
	\includegraphics[width=12cm]{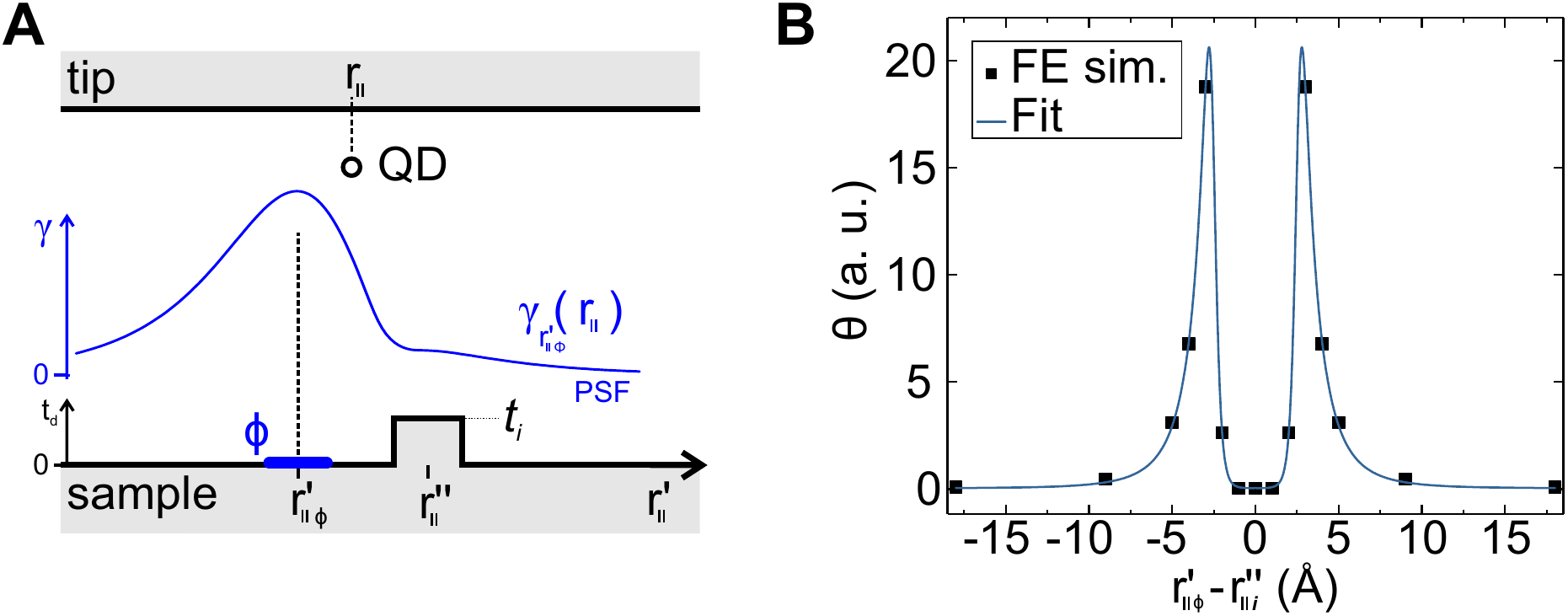}
	\sffamily
	\caption{\footnotesize  \textbf{The weighting function $\theta$.} \textbf{A} Scheme employed for the calculation of the weighting function $\theta$, including a plot of $\rparprimePhi(\rpar)$ for one specific distance $\rparprimePhi-\rpardprime$ between the location \rparprimePhi of the minimal deformation of the surface potential and the screening nanostructure. To create a visible effect, a large height of $t_i = \unit[10]{\text{\AA}}$ is chosen in the FE simulation. \textbf{B} Plot of $\theta(\rparprime-\rparidprime)$ (not normalized) as obtained from the integrals $\iint \Delta\PhiQD(\rpar)d^2\rpar$ for a series of FE simulations (black squares). The fit function is given by Eq.~\ref{EqScreeningKernel}. \normalsize}
	\normalfont 
	\label{Fig_ScreeningKernel} 
\end{figure}

Before Eq.~\ref{EqTaylorFirstOrderSimple} can be inverted to obtain the object function \teff from the image function \alpharel, we need to find expressions for $\tBar(\rparprime)$ and the amplitude $A$ of \gammaTopo. \tBar expresses the collective screening action of all elements $(\rparidprime,t_i)$ on $\gammaplain(\rpar,z,\rparprime,\teff)$. We use the sum
\begin{equation}
\tBar(\rparprime) = \sideset{}{'}\sum\limits_{i=1}^n \theta(|\rparprime-\rparidprime|) t_i(\rparidprime)
\label{tDKernel}
\end{equation}
to determine \tBar. The weighting function $\theta$ is given by an analytical formula fitted to a series of FE simulations (Fig.~\ref{Fig_ScreeningKernel}). Since Eqs.~\ref{EqDerjaguinApprox} and \ref{tDKernel} are sums of terms each of which contains one $t_i$ only, we consider each topography element $(\rparidprime,t_i)$ in a separate FE simulation and add the terms according to Eq.~\ref{tDKernel}. To this end, we place a single cuboid-shaped nanostructure ($\unit[2 \times 2 \times 1.3]{\text{\AA}^3}$) at a series of distances $|\rr_{||\phi}'-\rparidprime|$ from $\rr_{||\phi}'$ (Fig.~\ref{Fig_ScreeningKernel}A). To obtain the influence of this protrusion on $\gammaplain(\rpar,z,\rparprime,\teff)$, we introduce a minimal deformation $\Phis(\rparprime)=\phi\delta(\rparprime-\rr_{||\phi}')$ of the surface potential at $\rr_{||\phi}'$ while keeping $\Phis = 0$ everywhere else. Then, the integral in Eq.~\ref{Phistar} collapses to $\Phistar(\rpar) = \phi \gammaplain(\rpar,z,\rr_{||\phi}',0) \propto \gammaplain_{\rr_{||\phi}'}(\rpar)$. Since we have defined \tBar as a sample property, i.e., independent of the position \rpar of the sensor, the weight $\theta$ has to take into account the effect of the protrusion $(\rparidprime,t_i)$ on $\gammaplain_{\rr_{||\phi}'}(\rpar)$ at all \rpar. Therefore, we calculate $\theta$ by averaging over \rpar as
\begin{equation}
\theta(|\rr_{||\phi}'-\rparidprime|) \propto -\iint (\Phistar(\rpar)-\Phistar_0(\rpar)) d^2\rpar = -\iint \Delta\Phistar(\rpar) d^2\rpar,
\label{EqIntDeltaPhistar}
\end{equation}
where the potential $\Phistar(\rpar)$ is calculated for the situation shown in Fig.~\ref{Fig_ScreeningKernel}A and $\Phistar_0(\rpar)$ is calculated in the absence of the protrusion, i.e., describes the situation without screening. The minus sign in Eq.~\ref{EqIntDeltaPhistar} is needed because $\Delta \Phistar<0$. In the FE simulation we realize $\Phis(\rparprime)=\phi\delta(\rparprime-\rr_{||\phi}')$ via a small surface patch at a potential of $\unit[1]{V}$ (Fig.~\ref{Fig_ScreeningKernel}A). The results of the simulation are shown as black dots in Fig.~\ref{Fig_ScreeningKernel}B. We employ the empirical function
\begin{equation}
\theta(|\rparprime-\rparidprime|) = \frac{b} {\left(|\rparprime-\rparidprime|\frac{1}{\text{\AA}} + 0.5\right)^{4.3}\times \left(1 + e^{6\left(-|\rparprime-\rparidprime|\frac{1}{\text{\AA}} + 2.6\right)}\right)}
\label{EqScreeningKernel}
\end{equation}
to interpolate the FE simulations. Here $b$ is chosen such that the norm of $\theta$ is 1; this is required to yield $\tBar = t_i$ for a large area where $t_i=\text{const}\ \forall i$. In order to emulate the primed sum in Eq.~\ref{tDKernel}, which originates from the fact that a topographic feature cannot screen itself and thus the case of $\rparidprime = \rparprime$ has to be excluded from the sum, the center of $\theta$ is cut out by a Fermi-type damping function. The size of the cut-out region is chosen empirically such that it represents roughly the size of a single atom. 

Equation~\ref{EqScreeningKernel} can be used in conjunction with Eq.~\ref{EqTaylorFirstOrderSimple} to obtain the object function \teff from the image function \alpharel. Note, however, that a variety of alternative procedures for the determination of \tBar exist and that, in principle, the entire HDMR series in Eq.~\ref{EqHDMR2} can be emulated by an elaborate method to compute \tBar from the $t_i$. This consideration could be the starting point for alternative formulations of the first-order approximation and beyond.

\begin{figure}[!ht]
	\centering
	\includegraphics[width=14cm]{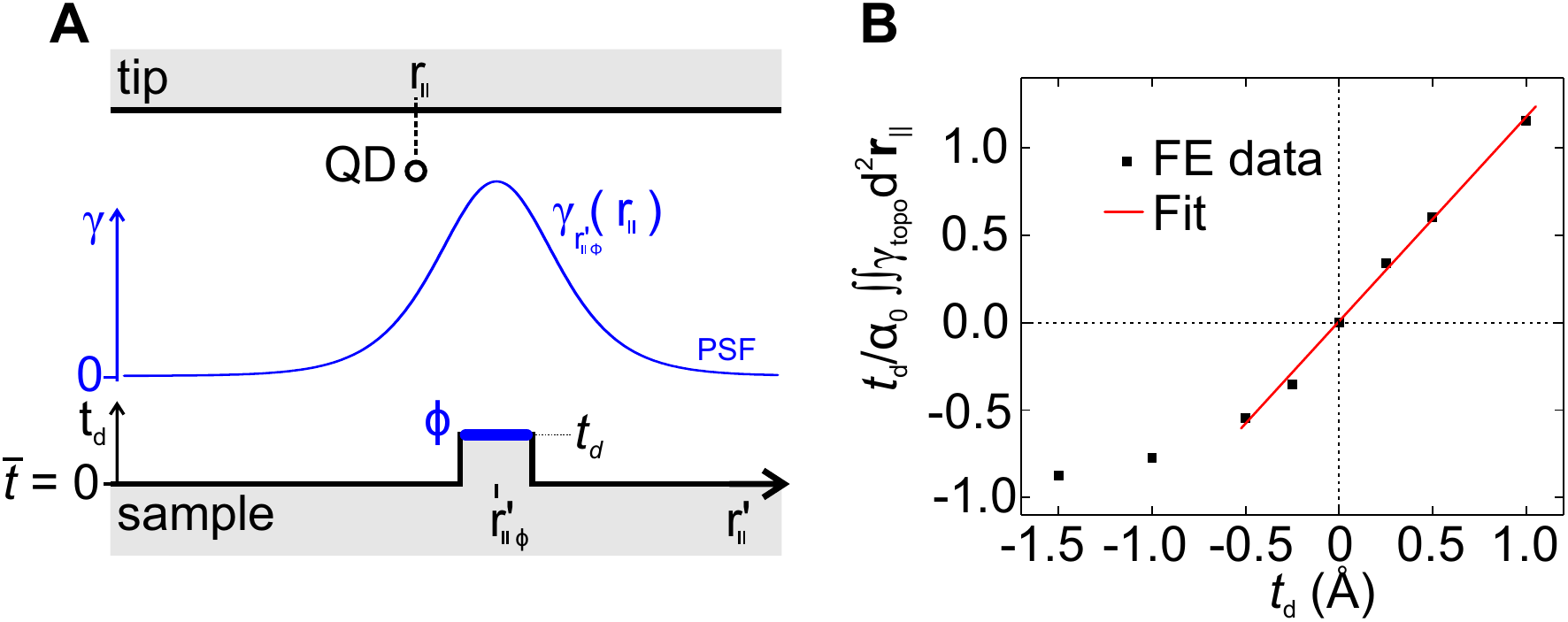}
	\sffamily
	\caption{\footnotesize  \textbf{Calculation of the norm of \gammaTopo.} \textbf{A} Scheme employed for the calculation of the amplitude $A$ of \gammaTopo. A minimal deformation of the surface potential is placed on top of a protrusion (shown) or into a depression (not shown) in the otherwise flat surface and a FE simulation of the resulting potential distribution (blue) is performed, which is proportional to \gammaplain. Then, the left hand side of Eq.~\ref{EqScreeningKernel6} is calculated and used to obtain the right hand side. \textbf{B} Plot of the quantity on the left hand side of Eq.~\ref{EqScreeningKernel6} as obtained from FE simulations. The fit used to determine the amplitude of \gammaTopo is performed only in a limited range of \teff values and thus invalid for deep depressions. \normalsize}
	\normalfont 
	\label{Fig_ttBarNorm} 
\end{figure}

To determine the amplitude of \gammaTopo in the first-order mean field approximation, we use Eq.~\ref{GammaFirstOrderApprox2} and obtain
\begin{equation}
\gammaTopo(|\rpar-\rparprime|,z) (\teff(\rparprime)-\tBar(\rparprime)) = \gammaplain(\rpar,z,\rparprime,\teff) - \gammaAx(|\rpar-\rparprime|,z-\tBar(\rparprime)).
\label{EqScreeningKernel2}
\end{equation}
Integrating Eq.~\ref{EqScreeningKernel2} over \rpar yields
\begin{equation}
\begin{aligned}
&\iint\limits_\text{imaging plane} \gammaTopo(|\rpar-\rparprime|,z) (\teff(\rparprime)-\tBar(\rparprime)) d^2\rpar = \\
&\iint\limits_\text{imaging plane} \gammaplain(\rpar,z,\rparprime,\teff) d^2\rpar - \iint\limits_\text{imaging plane} \gammaAx(|\rpar-\rparprime|,z-\tBar(\rparprime)) d^2\rpar.
\end{aligned}
\label{EqScreeningKernel3}
\end{equation}
We now consider a zero-dimensional topographic feature of height \teff at \rparprimePhi on an otherwise flat surface ($\tBar(\rparprimePhi) = 0$), see Fig.~\ref{Fig_ttBarNorm}A, and take the integrals in Eq.~\ref{EqScreeningKernel3} at $\rparprime = \rparprimePhi$, such that we get with Eqs.~\ref{EqAlphaZero} and \ref{EqTranslationalInvariance}
\begin{equation}
\frac{\teff}{\alpha_0} \iint\limits_\text{imaging plane} \gammaTopo(|\rpar-\rparprimePhi|,z) d^2\rpar = 
\frac{1}{\alpha_0} \iint\limits_\text{imaging plane} \gammaplain(\rpar,z,\rparprimePhi,\teff) d^2\rpar - 1.
\label{EqScreeningKernel6}
\end{equation}
Here we also have divided by $\alpha_0$. Using Eq.~\ref{EqScreeningKernel6}, we can calculate the amplitude $A = \iint \gammaTopo(|\rpar-\rparprimeZero|,z) d^2\rpar$ (Eq.~\ref{EqAmplitudeA}). To this end, we calculate the integral on the right hand side of Eq.~\ref{EqScreeningKernel6} from a FE simulation in the presence of the nanostructure for several \teff values (Fig.~\ref{Fig_ttBarNorm}) and divide by $\alpha_0$, which can be calculated as the integral over \gammaAx (i.e. in the absence of the nanostructure, Eq.~\ref{EqAlphaZero}). In the FE simulation we assume a minimal deformation $\Phis(\rparprime)=\phi\delta(\rparprime-\rr_{||\phi}')$ of the surface potential at \rparprimePhi while keeping $\Phis = 0$ everywhere else (Fig.~\ref{Fig_ttBarNorm}A). Since the amplitude of \gammaTopo depends on the size of the nanostructure we use the correct limit of a single atom.  The result is plotted in Fig.~\ref{Fig_ttBarNorm}B. From a fit to the FE data (red) we obtain its slope $c$. The amplitude of \gammaTopo is therefore
\begin{equation}
A = \iint\limits_\text{imaging plane} \gammaTopo(|\rpar-\rparprimePhi|,z) d^2\rpar = c\alpha_0.
\label{EqScreeningKernel7}
\end{equation}
Together with Eq.~\ref{EqTaylorFirstOrderSimple} and \ref{tDKernel} this constitutes the relation between image function \alpharel and object function \teff in the first-order mean field approximation. 
We note the similarity of Eq.~\ref{EqScreeningKernel7} with Eq.~\ref{EqLinearDeconvolute3}: The amplitude of \gammaTopo in the first-order mean field approximation is proportional to $\alpha_0$, as is the norm of \gammaTopo in the zeroth-order approximation. The difference is, however, that the norm in Eq.~\ref{EqLinearDeconvolute3} has been derived for a nanostructure within an extended terrace (which is equal to a flat surface!), i.e., in the presence of maximal screening, even though the zeroth-order approximation disregards screening by construction. This creates the inconsistency that was mentioned in section \ref{sec:ZerothOrderApprox}. In the first-order mean field approximation this inconsistency is removed, which is also evidenced by the fact that $c \gg g$. We obtain $c=\unit[1.17]{\text{\AA}^{-1}}$, whereas typical $g$ values are in the range of $\unit[0.03]{\text{\AA}^{-1}}$. This comparably high value of $c$ is, however, compensated to a large extent when calculating $\alpha$ in the first order mean field approximation (Fig.~\ref{Fig_ExplainGammaShape}), as becomes obvious in the next section when we discuss the special case of a single isolated nanostructure in the zeroth and first-order mean field approximations.

It is important to note that we perform the linear fit in Fig.~\ref{Fig_ttBarNorm}B over a limited $\teff-\tBar$ interval where the FE results are almost linear. The reason for the saturation to -1 at large negative $\teff-\tBar$ values is the strong screening of the potential at the bottom of a deep depression in the topography (Fig.~\ref{Fig_SI_A}). The value of the integral on the right hand side of Eq.~\ref{EqScreeningKernel6} converges to $\iint\limits \gammaplain(\rpar,z,\rparprimeZero,\teff \longrightarrow -\infty) d^2\rpar \longrightarrow 0$. When using the first-order mean field  approximation, one therefore has to keep in mind that it produces unphysical behavior for large negative $\teff-\tBar$ values.

\subsubsection{The example of a single isolated nanostructure}
\label{Sec_FirstOrderNano}

\begin{figure}[!ht]
	\centering
	\includegraphics[width=16cm]{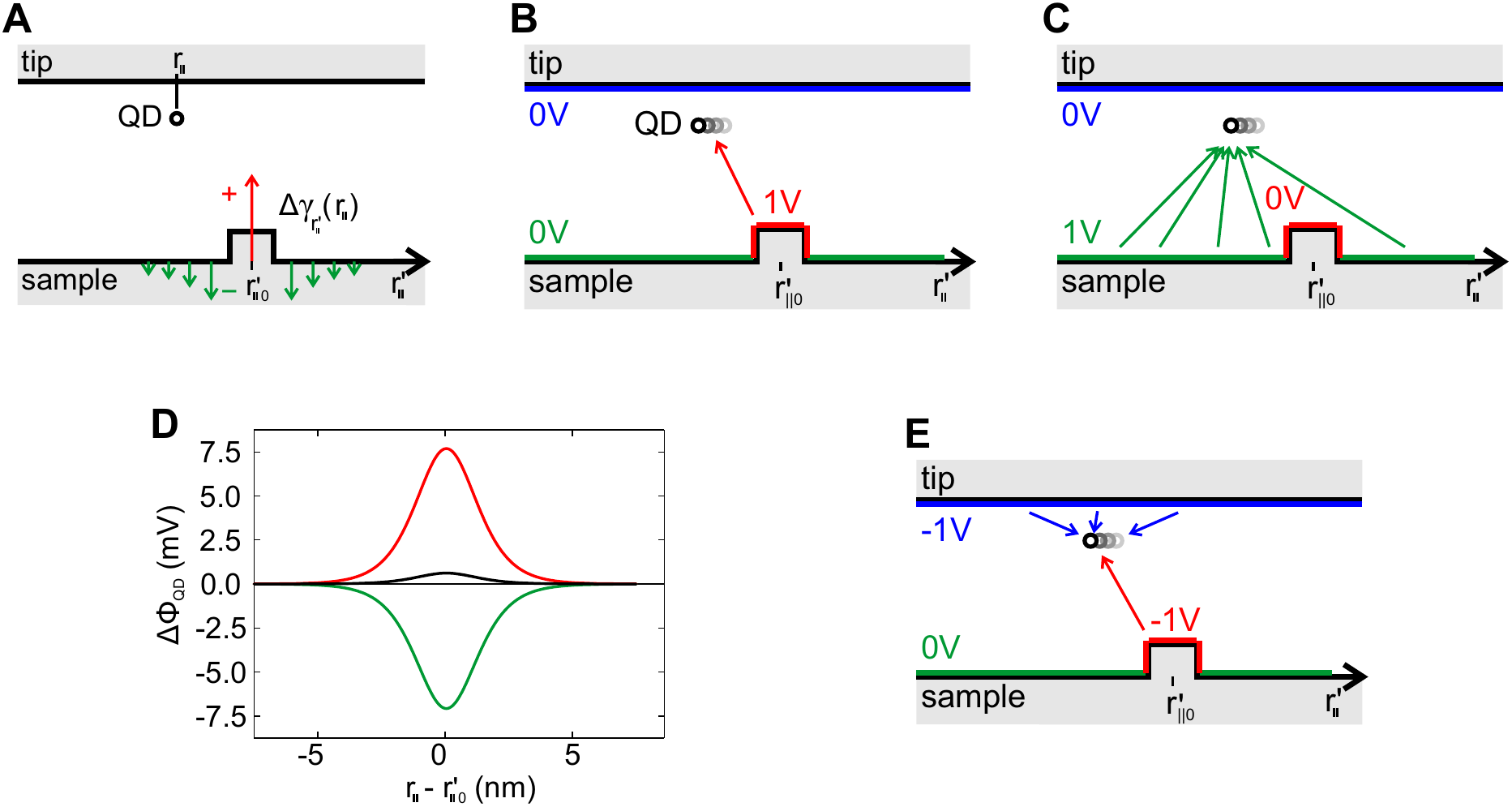}
	\sffamily
	\caption{\footnotesize  \textbf{Quantification of first-order screening effects on \gammaplain.} \textbf{A}~Qualitative illustration of the change in $\gammaplain_{\rparprime}(\rpar)$ upon the introduction of a nanostructure at \rparprimeZero on the otherwise flat surface. \textbf{B}, ~\textbf{C}~The quantitative changes in $\gammaplain_{\rparprime}(\rpar)$ can be obtained from a FE simulation in which the surface potential either at the nanostructure (panel B) or everywhere else on the flat surface (panel C) is set to a finite value (here: $\unit[1]{V}$). The red and green arrows indicate which parts of the surface contribute to \PhiQD. \textbf{D}~Plot of the potential $\Delta \PhiQD(\rpar) \propto \gammaplain_{\rparprime}(\rpar)$ for the situations in panels B and C. The case of an entirely flat surface (no nanostructure) but with voltages applied as in B and C is used as a reference and subtracted to obtain the respective $\Delta \PhiQD$ curves.The red curve shows the increase in \PhiQD (Eq.~\ref{EqAlphaRelNanostruct2}) due to the increase in $\gammaplain_{\rparprimeZero}(\rpar)$, the green curve the decrease of \PhiQD due to the decrease of $\gammaplain_{\rparprime}(\rpar)$ for all \rparprime around the nanostructure (i.e., the screening by the nanostructure (Eqs.~\ref{EqAlphaRelNanostruct3} and \ref{EqAlphaRelNanostruct4})). The black curve shows the real net effect which results if the entire surface (nanostructure and flat part) is biased with \Vb. As expected from the considerations in the text (and panel E), all three curves have exactly the same shape. \textbf{E} Mere redefinition of the potential from panel C by subtraction of $\unit[1]{V}$ does not change the shape of $\Delta \PhiQD$ but leads to a situation where instead of the surface around the nanostructure now the nanostructure itself and the structureless tip contribute to $\Delta \PhiQD$, which is a situation comparable to panel B. This proves that the $\gammaplain_{\rparprime}(\rpar)$ curves in panels B and C have similar shape.
		\normalsize}
	\normalfont 
	\label{Fig_ExplainGammaShape} 
\end{figure}

To all orders, the effect of a single isolated nanostructure at \rparprimeZero on the image function \alpharel can be expressed as (Eq.~\ref{integralalpha})
\begin{equation}
\alpharel(\rpar,z) = \frac{1}{\alpha_0(z)}\left[ \gamma(\rpar,z,\rparprimeZero,\teff) + \int_{\setminus \rparprimeZero} \gamma(\rpar,z,\rparprime,0) d^2\rparprime \right],
\label{EqAlphaRelNanostruct1}
\end{equation}
where we have separated the contribution of the nanostructure at \rparprimeZero from the contribution of all other points on the surface. The splitting in Eq.~\ref{EqAlphaRelNanostruct1} can be realized in a thought experiment in which we apply the bias voltage $\Vb$ (Eq.~\ref{EqPhiQD2}) either exclusively to the nanostructure (first case, Fig.~\ref{Fig_ExplainGammaShape}B) or only to rest of the surface (second case, Fig.~\ref{Fig_ExplainGammaShape}C). The hypothesis, which we are going to prove now, is that the two terms in Eq.~\ref{EqAlphaRelNanostruct1}, corresponding to the two cases of the thought experiment, have the same shape as a function of the distance $|\rpar-\rparprimeZero|$. 

If $\Phis = 0$, as we assume in Fig.~\ref{Fig_ExplainGammaShape}B-C, the fundamental equation of SQDM (Eq.~\ref{PhiGF3}) yields $\PhiQD = \Vb\iint\gammaplain(\rr,\rrprime) d^2\rrprime+\PhiTip$. Without loss of generality we assume the tip to be at zero potential ($\PhiTip=0$). The first term in Eq.~\ref{EqAlphaRelNanostruct1} thus generates a potential
\begin{equation}
\Phi_{\text{QD\_1}}(\rpar) = \Vb \gamma(\rpar,z,\rparprimeZero,\teff) = \Vb \gamma(|\rpar-\rparprimeZero|,z,\teff)
\label{EqAlphaRelNanostruct2}
\end{equation}
at the quantum dot (Fig.~\ref{Fig_ExplainGammaShape}B), while the second term yields (Fig.~\ref{Fig_ExplainGammaShape}C)
\begin{equation}
\Phi_{\text{QD\_2}}(\rpar) = \Vb \int_{\setminus \rparprimeZero} \gamma(\rpar,z,\rparprime,0) d^2\rparprime.
\label{EqAlphaRelNanostruct3}
\end{equation}
Evidently, both $\Phi_{\text{QD\_1}}(\rpar)$ and $\Phi_{\text{QD\_2}}(\rpar)$ exhibit axial symmetry with respect to the point \rparprimeZero, while in Eq.~\ref{EqAlphaRelNanostruct3} the \gammaplain function for individual \rparprime does not have this symmetry because of the presence of the nanostructure (Fig.~\ref{Fig_ExplainGammaShape}C). This is an important difference to the respective functions in the first order mean field approximation, where because of the mean-field \tBar approximation (Eq.~\ref{EqDerjaguinApprox}, Fig.~\ref{Fig_Performance}) \textit{each} \gammaplain exhibits axial symmetry with respect to \rparprime (Eq.~\ref{GammaFirstOrderApprox2}). 
Next, we apply a bias offset of $-\Vb$ to the entire tip+surface boundary in the second case (Eq.~\ref{EqAlphaRelNanostruct3}, Fig.~\ref{Fig_ExplainGammaShape}E). This corresponds to a redefinition of the potential without loss of generality. Because the potential of the entire surface excluding the nanostructure is now zero, we obtain
\begin{equation}
\Phi_{\text{QD\_2}}(\rpar) = -\Vb \gamma(|\rpar-\rparprimeZero|,z, \teff) + \PhiTip.
\label{EqAlphaRelNanostruct4}
\end{equation}
Since the offset \PhiTip has no dependency on \rpar, this immediately proves our hypothesis, namely that the two terms in Eq.~\ref{EqAlphaRelNanostruct1} do indeed have the same shape as function of the distance $|\rpar-\rparprimeZero|$. In other words, the increase in the SQDM image function $\alpharel(\rpar)$ around $\rpar=\rparprimeZero$, caused by the increase of $\gammaplain_{\rparprimeZero}(\rpar)$ due to the nanostructure at \rparprimeZero, has the same shape but opposite sign as the \textit{sum} of the decreases in $\alpharel(\rpar)$ that originate from the screening action of the nanostructure on $\gammaplain_{\rparprime}(\rpar)$ at all $\rparprime \neq \rparprimeZero$. This is indicated by the red and green arrows in Fig.~\ref{Fig_ExplainGammaShape}A. However, due to the different potential of the tip electrode, the two cases B and C do not cancel precisely to zero, as the FE simulation in Fig.~\ref{Fig_ExplainGammaShape}D shows. The strong reduction of red to the black curve in Fig.~\ref{Fig_ExplainGammaShape}D explains why the large ratio $c/g$ in Sec.~\ref{SecFirstOrderApprox} does not lead to a similarly strong enhancement of \alpharel in the first-order mean field approximation.

\subsubsection{Performance of the first-order mean-field approximation}

\begin{figure}[!ht]
	\centering
	\includegraphics[width=15cm]{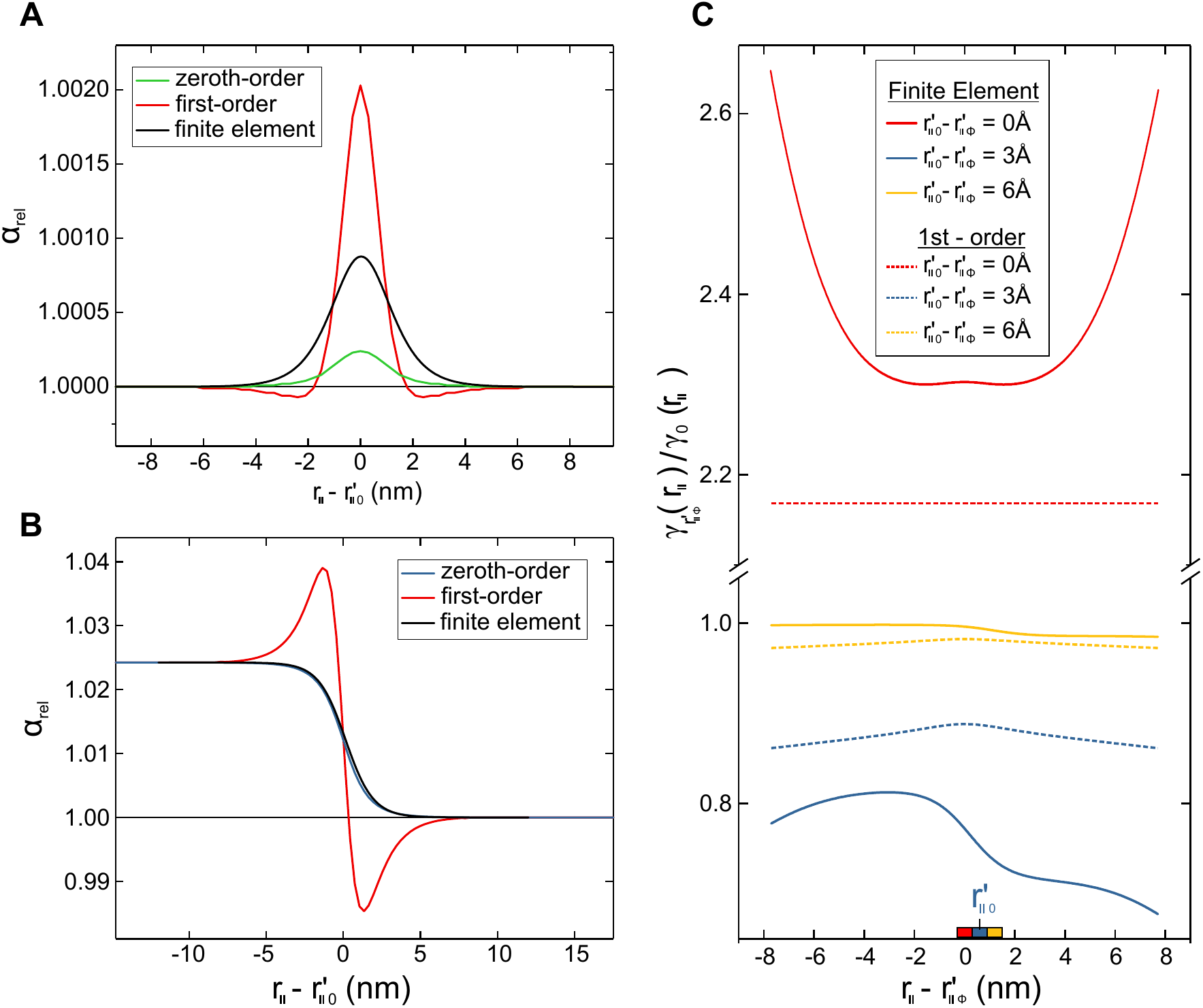}
	\sffamily
	\caption{\footnotesize  \textbf{Comparison of zeroth- and first-order approximation with FE simulations.} \textbf{A} Relative gating efficiency \alpharel above a zero dimensional nanostructure ($3\times 3 \times 1 \ \text{\AA}^3$) at \rparprimeZero on an otherwise flat surface as calculated with the three different methods. The image resolution for the zeroth- and first-order approximation is $\unit[1]{\text{pixel/\AA}}$, for the FE simulation it is $\unit[3]{\text{voxel/\AA}}$. \textbf{B} Profile of \alpharel perpendicular to a step edge (height $\unit[1]{\text{\AA}}$). While the zeroth-order approximation comes very close to the FE result, the \alpharel curve from first-order approximation is characterized by strong overshooting in the vicinity of the step edge. \textbf{C} Plots of the relative enhancement or damping of $\gammaplain_{\rparprimePhi}(\rpar)$ with $\rparprimePhi=0$ as induced by the zero-dimensional nanostructure from panel A placed at different positions \rparprimeZero and calculated with the first-order approximation and FE simulations. The reference curve $\gamma_0(\rpar)$ is calculated in the absence of the nanostructure.
	\normalsize}
	\normalfont 
	\label{Fig_Performance} 
\end{figure}

\noindent We demonstrate the use of the first-order mean field approximation and its deviation from the zeroth-order approximation for the example of a single nanostructure of height \teff at \rparprimeZero on an otherwise flat surface. By definition, all individual $f_i$ in Eq.~\ref{EqHDMR2} are zero in the zeroth-order approximation and consequently the $\gammaplain_{\rparprime}(\rpar)$ for $\rparprime \neq \rparprimeZero$ are unaffected by the presence of the nanostructure. In contrast, in the first-order mean field approximation, the sum 
\begin{equation}
\sideset{}{'}\sum_{i=1}^{n}f_i(\rpar,z,\rparprime,\teff,t_i)
\label{EqFirstOrderTerm1}
\end{equation}
is nonzero, unless $\rparprime = \rparprimeZero$  whence $t_i =0\ \forall\, \rparidprime \neq \rparprimeZero$ (Eq.~\ref{Eq_66}, note that $\rparidprime = \rparprimeZero$ is excluded from the sum).

A major difference between the full description of the potential around a nanostructure as captured in FE simulations on the one hand and the first-order approximation in combination with the mean field \tBar approximation on the other hand is the shape of the individual \gammaplain functions in Eq.~\ref{EqAlphaRelNanostruct3}. In the \tBar approximation axial symmetry around \rparprime  is enforced for every individual \gammaplain function. We highlight the consequences of this provision by two exemplary scenarios, (1) a zero-dimensional nanostructure (Fig.~\ref{Fig_Performance}A) and (2) a surface step (Fig.~\ref{Fig_Performance}B). We compare $\alpharel(\rpar-\rparprimeZero)$ as obtained from the zeroth-order and the first-order mean field approximations with FE simulations for the two scenarios. Fig.~\ref{Fig_Performance}A shows that the zeroth-order approximation (green curve) substantially underestimates the strength of the influence of the nanostructure on \alpharel. The obvious reason is that the factor $g$ which determined the norm of \gammaplain has been defined for the closed layer with maximal screening. The first-order approximation in principle corrects this deficiency as it increases \alpharel (red curve in Fig.~\ref{Fig_Performance}A). However, in combination with the mean field \tBar approximation it also results in an unrealistic "sombrero" shape of $\alpharel(|\rpar-\rparprimeZero|)$. 

The origin of the sombrero shape can be understood when comparing the individual axially symmetric $\gammaplain_{\rparprime}(\rpar)$ functions from the first-order mean field approximation with their FE-simulated counterparts (Fig.~\ref{Fig_Performance}C). To facilitate this comparison, we plot the relative change $\gammaplain_{\rparprimePhi}(\rpar)/\gamma_0(\rpar)$, where $\gamma_0(\rpar)$ is the $\gammaplain_{\rparprimePhi}(\rpar)$ function in the absence of the nanostructure. As before, the potential is applied at $\rparprimePhi = 0$.

When the nanostructure is placed at zero where also the potential is applied ($\rparprimeZero = \rparprimePhi = 0$), a scenario with axial symmetry is created as reflected by the red curves in Fig.~\ref{Fig_Performance}C which are both symmetric around $\rpar=0$. Since $\tBar(0)=0$ (no self-screening), $\gammaplain_{\rparprimePhi}$ has the same shape as $\gamma_0$ but an amplitude that is larger by a factor $1+c\teff$. In our example, we have chosen $\teff = \unit[1]{\text{\AA}}$ which leads to the dashed red curve. The fitting procedure used for the determination of $c$ (Fig.~\ref{Fig_ttBarNorm}) results in a good correspondence between the amplitudes of the FE simulation (solid red) and of the first-order approximation. 

When the nanostructure is placed beside the point $\rparprimePhi = 0$ where the potential is applied, at $\rparprimeZero = \unit[3]{\text{\AA}}$ (blue) and $\unit[6]{\text{\AA}}$ (yellow), it partially screens the potential at $\rparprimePhi$, thereby damping the respective $\gammaplain_{\rparprimePhi}$. Since the axial symmetry is now broken, the FE simulation curves are asymmetric. The FE simulations reveal that the screening predominantly attenuates \gammaplain in the region above and behind the nanostructure ($\rparprimeZero \lesssim \rpar$), whereas \gammaplain is less affected in the opposite direction ($\rparprimeZero \gtrsim \rpar$). However, since for large offsets $|\rpar-\rparprimePhi|$ both, $\gammaplain_{\rparprimePhi}(\rpar)$ and $\gamma_0(\rpar)$, approach zero very quickly, the maximal \textit{absolute} damping of $\gammaplain_{\rparprimePhi}$ occurs in the immediate vicinity of the nanostructure ($\rpar \approx \rparprimeZero$). This is also where the absolute enhancement of $\gammaplain_{\rparprimeZero}(\rpar)$ due to the nanostructure is maximal (for $\rparprimePhi = \rparprimeZero$). Since both effects are strongest at the same position, namely at $\rpar = \rparprimeZero$, both contributions cancel to a large extent and the shape of \alpharel upon introduction of the nanostructure is barely changed which leads to the results in Figs.~\ref{Fig_ExplainGammaShape}D and \ref{Fig_Performance}A. This is in marked contrast to the first-order approximation which leads to the sombrero shape.

In the first-order approximation the axial symmetry of the $\gammaplain_{\rparprime}$ curves is retained due to the \tBar mean-field approximation. Since $\tBar > 0$ for the dashed blue and yellow curves, the shapes of $\gammaplain_{\rparprime}$ and $\gammaplain_0$ are slightly different which explains the (weak) symmetric \rpar dependency of the two curves in Fig.~\ref{Fig_Performance}C. The relative amplitude $\gammaplain_{\rparprimePhi}/\gammaplain_0$ is now determined by $1+g\tBar+c(\teff-\tBar)$ (Eqs.~\ref{EqTaylorFirstOrderSimple} and \ref{EqScreeningKernel7}) which is smaller than 1 since $\tBar(\rparprimePhi) > 0$ whereas $\teff(\rparprimePhi) = 0$, and $g \ll c$. The precise value of the relative amplitude depends on the weighting function $\theta$ used to calculate \tBar around the nanostructure (Fig.~\ref{Fig_ScreeningKernel}B) and rapidly approaches 1 as $|\rparprimeZero-\rparprimePhi|$ increases (dashed yellow curve).

Since the screening factor is applied uniformly to $\gammaplain_{\rparprimePhi}$ (independent of \rpar) in the first-order mean field approximation,  too little screening occurs at the position \rparprimeZero of the nanostructure, whereas too much screening is applied in the region around the nanostructure. This creates the regions of $\alpharel < 1$ in the red curve in Fig.~\ref{Fig_Performance}A and, in general, its sombrero shape. It is interesting to note that, for a single nanostructure, the second term in Eq.~\ref{EqTaylorFirstOrderSimple} which depends on $\teff - \tBar$ is zero if integrated over the entire \rpar imaging plane, since $\theta$ (Eq.~\ref{EqScreeningKernel}) is normalized to 1.

As a second example to assess the properties of the zeroth and first-order approximation we consider a step edge. Figure~\ref{Fig_Performance}B shows that in this case the \tBar approximation and the resulting enforced axial symmetry of $\gammaplain_{\rparprime}(\rpar)$ has much stronger consequences, namely an over-damping and corresponding overshooting at the lower and higher terrace side of the step respectively. The zeroth-order approximation on the other hand reproduces the FE simulated \alpharel curve almost perfectly. 

In conclusion, we find that of the two approximations evaluated here, the zeroth-order approximation is the better choice for an actual SQDM image deconvolution since it is devoid of overshooting artifacts. Its major shortcoming, a general underestimation of \alpharel for isolated objects, can be overcome by the first-order approximation, but the additional mean field \tBar approximation as employed here leads to artifacts. As an outlook, the first-order approximation could, however, become the better choice if the \tBar approximation and the associated axial symmetry of \gammaplain is replaced by a more sophisticated approach. Such an approach could, for example, be based on a set of FE simulations as the ones in Fig.~\ref{Fig_Performance}C.

\newpage

\section{Surface dipoles of nanoscale objects}

\subsection{Surface potential and dipole density}
\label{sec:surfacepotentialdipoledensity}

Quite generally, the surface potential $\Phis$ can also be understood as a local surface dipole densities $\piPerp \equiv \Pperp/A$ relative to the empty surface. This can be seen as follows: If a charge density $\sigma_\mathrm{c}(\rrprime)$ is applied at height $z_\mathrm{c}$ to the metallic sample surface \surf, this creates an image charge density $\sigma_\mathrm{i}(\rrprime)$ below the surface. Together, they form a capacitor. Because the lower "plate" with charge density $\sigma_\mathrm{i}(\rrprime)$ is in the metal, it is grounded, i.e.~$\Phi_\mathrm{i}(\rrprime)\equiv 0$. For the purpose of the present argument, we fix the zero of the coordinate $z'$ normal to the surface at this plate. Furthermore, we assume that locally we can approximate the capacitor at $\rrprime$ as a parallel plate capacitor. Then, according to Gauss's law the electric potential due to the image charge plate a $z$ is 
\begin{equation}
	\Phi_\mathrm{i}(\rrprime,z)=-\frac{1}{2\epsilon}\sigma_\mathrm{i}(\rrprime)z,
\end{equation}
while the potential of the upper plate is given by 
\begin{equation}
	\Phi_\mathrm{c}(\rrprime,z)=\frac{1}{2\epsilon}\sigma_\mathrm{c}(\rrprime)(z-z_\mathrm{c})+\frac{1}{2\epsilon}\sigma_\mathrm{c}(\rrprime)z_\mathrm{c}.
\end{equation}
Since $\sigma_\mathrm{i}(\rrprime)=-\sigma_\mathrm{c}(\rrprime)$, we obtain
\begin{equation}
	\Phis(\rrprime)=\Phi_\mathrm{c}(\rrprime,z_\mathrm{c})+\Phi_\mathrm{i}(\rrprime,z_\mathrm{c})= \frac{1}{\epsilon}\sigma_\mathrm{c}(\rrprime) z_\mathrm{c}.
\end{equation}
If we define 
\begin{equation}
	\piPerp (\rrprime) =\sigma_\mathrm{c}(\rrprime)  z_\mathrm{c},
\end{equation}
we obtain the Helmholtz equation
\begin{equation}
\Phis(\rrprime)= \frac{1}{\epsilon}\piPerp (\rrprime).
\label{dipoledensity}
\end{equation}
In other words, the surface potential is directly proportional to a perpendicular dipole density. For simplicity, we may set $\epsilon=\epsilon_0$. Eq.~\ref{dipoledensity} accounts properly for the orientation of the dipole density. If $\sigma_\mathrm{c} > 0$ ($\sigma_\mathrm{c} < 0$), then also $\piPerp > 0$ ($\piPerp < 0$), and the dipole density points outward (inward), always towards the positive charge, as it must. The work function change $\Delta W$ associated with the surface potential $\Phis$ is simply $\Delta W=e\Phis$, where $e$ is the charge of the electron (with sign).

\subsection{Dipole moments of nanostructures from 2D integration}
\label{sec:dipolemomentsfromintegration}

While the dipole moment density \piPerp is a meaningful quantity for extended uniform surface nanostructures on the sample surface, it is of little meaning at Angstrom-sized inhomogeneous structures, for which the surface dipole moment $P_\perp$ itself gives a more meaningful description. The latter can be obtained from the former by integration over the area $\mathscr{N}$ of the nanostructure where $\piPerp \neq 0$, 

\begin{equation}
\Pperp = \iint\limits_{\mathscr{N}}\pi(\rrprime)d^2\rrprime = \epsilon_0\iint\limits_{\mathscr{N}}\Phis(\rrprime)d^2\rrprime.
\label{EqP}
\end{equation}

Therefore, surface dipoles $\Pperp$ of individual nanostructures can be obtained by SQDM if the respective structure is located on the empty surface such that $\Phis=0$ for the entire border of the integration area $\mathscr{N}$. While we stick to the  $\Phis=0$ case in the following, in fact, relative dipole moments can also be obtained if \Phis provides a constant, non-zero background over the entire integration area. An example for the latter case would be the dipole of a vacancy inside an otherwise closed adsorbate layer.

\begin{figure}[!ht]
	\centering
	\includegraphics[width=4cm]{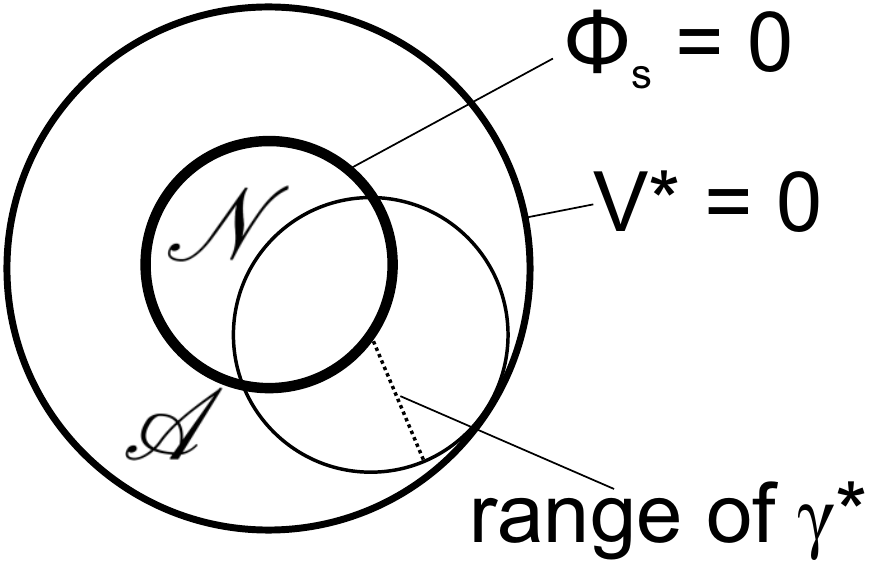}
	\sffamily
	\caption{\footnotesize  \textbf{Integration areas required to obtain surface dipole moments.} To be able to obtain \Pperp from an integration of \Vstar the following conditions have to hold: $\Phis \neq 0$ in $\mathscr{N}$ and $\Phis = 0$ outside. $\Vstar \neq 0$ in $\mathscr{A}$ and $\Vstar = 0$ outside. Consequently, the minimal size of $\mathscr{A}$ is given by the range of \gammastar. \normalsize}
	\normalfont 
	\label{Fig_SI_F} 
\end{figure} 

We note that under certain assumptions we find 
\begin{equation}
\Pperp= \iint\limits_{\mathscr{N}}\Phis(\rrprime)d^2\rrprime= \iint\limits_{\mathscr{A}}\Vstar(\rr)d^2\rr,
\label{intPhiequalsintvstar}
\end{equation}
Here, we have considered a situation where the condition $\Phis=0$ holds even in an area $\mathscr{A}$ around the nanostructure which is defined such that $\gammastar(\rpar,z,\rparprime) \approx 0$ for all pairs $(\rpar,\rparprime)$ where \rpar lies on the border of $\mathscr{A}$ and $\rparprime \in \mathscr{N}$ (Fig.~\ref{Fig_SI_F}). In this case we can write
\begin{equation}
\begin{aligned}
\iint\limits_{\mathscr{A}} \Vstar(\rpar,z)d^2\rpar &= \iint\limits_{\mathscr{A}}\left( \iint\limits_{\mathscr{N}}\Phis(\rparprime)\gammastar(\rpar,z,\rparprime)d^2\rparprime \right) d^2\rpar\\
&= \iint\limits_{\mathscr{N}}\left( \iint\limits_{\mathscr{A}}\Phis(\rparprime)\gammastar(\rpar,z,\rparprime)d^2\rpar \right) d^2\rparprime
\end{aligned}
\end{equation}

Here we have made use of Fubini's theorem to reverse the order of integration. Since $\Phis(\rparprime)$ does not depend on \rpar, we find
\begin{equation}
\iint\limits_{\mathscr{A}} \Vstar(\rpar,z)d^2\rpar = \iint\limits_{\mathscr{N}}\left(\Phis(\rparprime) \iint\limits_{\mathscr{A}}\gammastar(\rpar,z,\rparprime)d^2\rpar \right) d^2\rparprime
\end{equation}

The inner integral of \gammastar over $\mathscr{A}$ yields 1 for any given \rpar by the definition of \gammastar. Note that this is only valid if there is a translational symmetry on the surface. In general the integrals $\iint \gammaplain(\rpar,z,\rparprime)d^2\rpar \neq \iint \gammaplain(\rpar,z,\rparprime)d^2\rparprime$ (Eq.~\ref{EqTranslationalInvariance2}). This means that the normalization criterion of Eq.~\ref{integralalpha} does not hold if there is no translational invariance in \gammaplain.
Thus we finally arrive at
\begin{equation}
\begin{aligned}
\iint\limits_{\mathscr{A}} \Vstar(\rpar,z)d^2\rpar &= \iint\limits_{\mathscr{N}}\Phis(\rparprime)d^2\rparprime\\
&= \frac{1}{\epsilon_0}\iint\limits_{\mathscr{N}}\piPerp(\rparprime)d^2\rparprime\\
&= \frac{1}{\epsilon_0} \Pperp.
\end{aligned}
\end{equation}

This relation proves that dipole moments of nanostructures can potentially be inferred directly from the measured \Vstar image without the need for the deconvolution procedure.

\section{Conclusion}
\subsection{Summary}

SQDM is a powerful imaging technique which can reveal the rich electrostatic landscape at the nanoscale with high resolution. The working principle of SQDM is based on the gating of a QD which is mechanically strongly but electronically weakly coupled to the conductive tip of a non-contact atomic force microscope. The potential difference between the QD and the tip is influenced by the potential and the shape of the surfaces of tip and sample. SQDM can hence be interpreted in the framework of a boundary-value problem of electrostatics. In this paper, we have derived the respective general formalism for metallic as well as dielectric boundaries. 

In SQDM we measure the constantly changing bias voltage which needs to be applied to the sample during scanning to continually compensate the influence of the locally varying static surface potential on the QD. These are the primary measurands \Vpm. The secondary measurands \Vstar and \alpharel are calculated from \Vpm. The relative gating efficiency \alpharel quantifies the contribution of the surface topography on the QD potential \PhiQD, while \Vstar quantifies the influence of \Phis on \PhiQD \textit{for a given topography}. Consequently, we have always have to define a topography before attempting to recover the surface potential from \Vstar. 

Since changes in \alpharel during scanning cannot be uniquely attributed to either metallic or dielectric topographic features, we have introduced the concept of a \textit{dielectric topography} \teff which combines variations in topography \textit{and} in dielectric properties of the sample surface into a single quantity. The dielectric topography of the sample is the hypothetical metallic topography which would yield the same relative gating efficiency \alpharel as the actual surface topography. With this definition, we have two unknowns \Phis and \teff in the \textit{object plane} of the sample surface which need to be recovered from the two secondary measurands \Vstar and \alpharel measured in the \textit{imaging plane} (at the height $z$ of the QD). As the central part of this paper we have outlined a strategy to achieve this recovery. This is an inverse problem in which the point spread function \gammaplain plays a central role. \gammaplain is related to the gradient of the Green's function on the sample surface (Eq.~\ref{defgamma}) which, in turn, encodes \teff. However, this encoding is non-local (Sec.~\ref{sec:GeneralProblem}) such that \gammaplain becomes a functional of \teff (Eq.~\ref{gamma_as_functional}). There exists no analytical expression for \gammaplain, even for the simplest case in which tip and surface are approximated as parallel planes. Hence, we have presented a series of approximative solutions and discussed their implications. 

Initially, we limited our considerations to a flat sample surface and discussed the cases of a flat tip (Sec.~\ref{sec:PPapprox}), an axially symmetric tip (Sec.~\ref{SecBeyondPP}), and an arbitrarily shaped tip (Sec.~\ref{SecBeyondPP}). All these approximations are convenient since they yield a single \gammaplain function for all points in an image (i.e. \gammaplain is translationally invariant), which is important for the recovery of surface dipole moments directly from \Vstar images (Sec.~\ref{sec:dipolemomentsfromintegration}). However, since a flat surface implies that $\teff = 0$ and $\alpharel = \text{const}$, these approximations cannot be used to interpret \alpharel images in terms of \teff. To go beyond flat surfaces, we have approximated the functional which relates \gammaplain and \teff (Eq.~\ref{gamma_as_functional}) using the mathematical approach of high-dimensional model representation (Sec.~\ref{sec:HDMR}) and truncating the resulting sum (Eq.~\ref{EqHDMR2}) after the first or second term. In the \textit{zeroth-order approximation} (Sec.~\ref{sec:ZerothOrderApprox}), which only includes the first term of HDMR, the non-local nature of the relation between \teff and \gammaplain is completely neglected (Eq.~\ref{EqHDMRZeroOrder}). In the course of this derivation we define \gammaTopo as a measure for the local difference in \gammaplain between a flat surface (\gammaplain = \gammaAx) and a surface with $\teff \neq 0$ (Eq.~\ref{EqShape3}). We find that \gammaTopo is proportional to \gammaAx such that in the zeroth-order approximation \gammaplain differs from its flat-surface pendant \gammaAx only insofar as its amplitude now increases or decreases with the local value of \teff (Eq.~\ref{ZeroOrderApproxFinalResult}). This yields a straightforward relation between \gammaplain and \alpharel (Eq.~\ref{ZeroOrderApproxAlphaResult}) which can be used to recover the object function \teff via deconvolution.

In the \textit{first-order approximation} (Sec.~\ref{SecFirstOrderApprox}) we include non-local effects into our description of the relation between \gammaplain and \teff. To reduce the complexity of the problem, this is done via a mean field approximation in which the entire topography around each location \rparprime on the surface is summed into a single number \tBar via a kernel summation (Eq.~\ref{tDKernel}). Similar to the zeroth-order approximation, also in the first-order approximation, \gammaTopo is proportional to \gammaAx. However, with respect to the zeroth-order approximation there are two notable differences: First, \gammaAx is calculated for a tip-QD distance $z-\tBar$ instead of $z$ and second, \gammaTopo is scales with $\teff - \tBar$ instead of \teff (Eq.~\ref{GammaFirstOrderApprox2}). 

In the first order approximation a protrusion at \rparprime causes an increase in \gammaplain at \rparprime (where \teff-\tBar is positive) but simultaneously a decrease in the region around \rparprime (where \teff-\tBar is negative). These two effects compensate each other to a large extent (Fig.~\ref{Fig_ExplainGammaShape}). In the zeroth-order approximation, on the other hand, there is no influence of the protrusion on its surrounding. Thus it is not surprising that the amplitude of \gammaTopo varies much stronger with the height of the protrusion \teff in the first-order approximation compared to the zeroth-order approximation. We derive a ratio of $c/g \approx 40 : 1$ for the strength of this scaling from finite element simulations (Fig.~\ref{Fig_ttBarNorm}) and experiments. With this information, we can use Eq.~\ref{EqTaylorFirstOrderSimple} which gives the relation between \alpharel and \teff in the first-order approximation to recover the object function \teff via deconvolution.

In the last part of the paper, we compare the performance of the zeroth and first-order approximation to finite-element simulations for two prototypical surface structures: An atom-sized protrusion and a step edge (Fig.~\ref{Fig_Performance}). The benchmark is made with the help of simulated \alpharel profiles for a given \teff(\rparprime) function. We find that the zeroth-order approximation reproduces the shape of \alpharel very well in both cases. This can be explained because an isolated nanostructure is a case for which the integral over all $\gammaplain(\rpar,\rparprime)$ with respect to \rparprime has the same shape as $\gammaAx(\rpar)$ (Fig.~\ref{Fig_ExplainGammaShape}). However, the amplitude predicted by the zeroth-order approximation for the isolated protrusion is too small because the factor $c$ which determines the scaling with \teff is chosen to yield correct results for an extended layer which is characterized by maximal screening. This also explains the generally good correspondence in the prediction of \alpharel for the step edge. 

The first-order approximation accounts for the scaling with \teff in a better way which is directly reflected in the higher amplitude of \alpharel predicted for the isolated protrusion. However, the mean field approximation which we used in the first-order approximation causes a sombrero-like prediction for \alpharel, since it is not capable of locating the source of screening at a certain topographic element but instead produces a smeared-out isotropic screening effect (Fig.~\ref{Fig_Performance}c). Consequently, the predicted screening at the position of the protrusion is too weak, whereas it is too strong in the area around the protrusion. The same reason explains the strong overshooting visible in the \alpharel profile across the step edge. These results suggest that at the current level of theoretical description of \gammaplain as reported in this paper, the zeroth-order approximation should be preferred over the first-order approximation for the analysis of SQDM images and the recovery of \teff and \Phis images. The convincing results of such a recovery which are reported in Ref.~\cite{Wagner2019} confirm this conclusion.

\subsection{Outlook}
Our work establishes the theoretical description of SQDM on the firm basis of the boundary value formalism and connects the basic Green's function formalism to practically applicable recipes for the recovery of dielectric topography and surface potential images from secondary measurands of SQDM. Our theoretical analysis and benchmarking also clearly indicates where the analysis can be improved further. Most importantly, we have linked the poor overall performance of the first-order approximation to the inability of the mean-field approximation to describe local screening effects. Here we see the biggest chance for a future improvement of the methodology: A more flexible approximation fitted to a series of finite element simulations could overcome this limitation and further improve our ability to recover dielectric topographies and the surface potentials from SQDM images. Other data-driven approaches based on simulations seem also feasible. While it seems implausible at first to simulate all possible topographic structures of a surface and use it in, for example, a machine learning approach, the situation is severely simplified by the short-sightedness of SQDM. Because \gammaplain drops exponentially with $|\rpar-\rparprime|$, any topographic structure outside a region with a radius of about $\unit[10]{nm}$ becomes irrelevant. A second aspect of SQDM image analysis which is worth further research efforts is the determination of \gammaplain functions from experimental data on model systems. Here, high quality ground truth data for \Phis (and possible \teff) is required which can be obtained by state-of-the-art ab-initio methods.

\section*{Acknowledgements}
C.W. acknowledges funding through the European Research Council (ERC-StG 757634 "CM3").

\bibliographystyle{naturemag}

\end{document}